\documentclass{article}
\pdfoutput=1

\usepackage{natbib}
\usepackage{epubtk}
\usepackage{amsmath}
\usepackage{amssymb}
\usepackage{graphicx}
\usepackage{textcomp}
\usepackage{xspace}

\showlistoftablesfalse 

\newcommand{\kms}{km~s\super{-1}\xspace}
\newcommand\e{\mathrm{e}}
\newcommand\p{\mathrm{p}}

\begin{document}

\title{Solar Modulation of Cosmic Rays}

\author{%
\epubtkAuthorData{Marius S.\ Potgieter}{%
  Centre for Space Research, \\
  North-West University, \\
  2520 Potchefstroom, \\
  South Africa}{%
  Marius.Potgieter@nwu.ac.za}{%
  }
}

\date{}
\maketitle

\begin{abstract}
This is an overview of the solar modulation of cosmic rays in the heliosphere. 
It is a broad topic with numerous intriguing aspects so that a 
research framework has to be chosen to concentrate on. The review focuses on 
the basic paradigms and departure points without presenting advanced 
theoretical or observational details for which there exists a large number 
of comprehensive reviews. Instead, emphasis is placed on numerical modeling 
which has played an increasingly significant role as computational resources 
have become more abundant. A main theme is the progress that has been made 
over the years. The emphasis is on the global features of CR modulation and 
on the causes of the observed 11-year and 22-year cycles and charge-sign 
dependent modulation. Illustrative examples of some of the theoretical
and observational milestones are presented, without attempting to
review all details or every contribution made in this field of
research. Controversial aspects are discussed where appropriate, with
accompanying challenges and future prospects. The year 2012 was the
centennial celebration of the discovery of cosmic rays so that several
general reviews were dedicated to historical aspects so that such
developments are briefly presented only in a few cases.
\end{abstract}

\epubtkKeywords{Cosmic rays, heliosphere, solar modulation, solar
  cycles, solar activity}

\newpage
\tableofcontents


\newpage 
\section{Introduction}

Galactic cosmic rays encounter a turbulent solar wind with an embedded 
heliospheric magnetic field (HMF) when entering the heliosphere. This leads 
to significant global and temporal variations in their intensity and in 
their energy as a function of position inside the heliosphere. This process 
is identified as the solar modulation of cosmic rays (CRs). For this review, 
CRs are considered to have energies above 1~MeV/nuc and come mainly from 
outside the heliosphere, with the exception of the anomalous component of 
cosmic rays (ACRs) which originates inside the heliosphere.

The purpose of this overview is to explain and discuss progress of this 
process with its intriguing facets. The modulation of CRs is considered to 
happen from below $\sim$~30~GeV/nuc. The review also includes some aspects 
of ACRs but solar energetic particles, compositional abundances and 
isotopes are not included. The fascinating origin of CRs and acceleration in 
galactic space and beyond are not discussed. The emphasis is on the global 
features of CR modulation and on the causes of the observed 11-year and 
22-year cycles and phenomena such as charge-sign dependent modulation. 
Shorter-term CR variations, on scales shorter than one solar rotation, are 
not part of this review. Space weather and related issues are reviewed by 
\citet{SheaSmart2012}, amongst others.

The spotlight is first on the global features of the heliosphere and how it 
responds to solar activity. It is after all this extensive volume in which 
solar modulation takes place, mainly determined by what happens on and with 
the Sun. Transport and modulation theory is explained and the recurrent 
behaviour that CRs exhibit in the heliosphere is discussed within this 
context. Major predictions and accomplishments based on numerical modeling 
and some observational highlights are given. This overview is meant to be 
informative and didactic in nature.

\section{The Global Heliosphere and its Main Features}

\subsection{Physical boundaries}

The heliosphere moves through the interstellar medium so that an interface 
is formed. In the process the solar wind undergoes transitions with the main 
constituents its termination shock (TS), the heliopause (HP), and a bow wave 
(BW), with the regions between the TS, the HP, and the BW defined as the 
inner and outer heliosheath, respectively. An important goal of the 
heliospheric exploration by the two Voyager spacecraft has always been to 
observe this TS and HP. The TS is predicted to be highly dynamic in its 
location and structure, both globally and locally, and is as such confirmed 
observationally by Voyager~1 to be at 94~AU in December 2004 and at 84~AU by 
Voyager~2 in August 2007 \citep{Stone-etal2005, Stone-etal2008}. 
Observing the TS in situ was a true milestone. 
According to Voyager~2 plasma observations, the low solar
wind dynamic pressure beyond the TS lead to an inward movement of the
TS of about 10~AU to an assumed minimum position of 73~AU in 2010
\citep{RichardsonWang2011}. By the end of 2013, Voyager~1 will be at
126.4~AU while Voyager~2 will be at 103.6~AU, both moving into the
nose region of the heliosphere but at a totally different
heliolatitude, about 60\textdegree\ apart in terms of heliocentric
polar angles, and with a significant azimuthal difference (see
{\urlstyle{same}\url{http://voyager.gsfc.nasa.gov}}).

\citet{WebberMcDonald2013} reported that Voyager~1 
might have crossed the HP (or something that behaves like it) at the end of 
August 2012, a conclusion based on various CR observations. This is another 
major milestone for the Voyager mission. Since their launch in 1977, the two 
spacecraft have explored heliospace from Earth to the HP over almost four decades and 
will next explore a region probably very close to the pristine local interstellar 
medium. The passage of the Voyager~1 and 2 spacecraft through the inner 
heliosheath has revealed a region somewhat unlike than what was observed 
up-stream of the TS (towards the Sun). The question arises if the
modulation of CRs in this region is actually different from the rest
of the heliosphere?

Based on magnetohydrodynamic (MHD) modeling, it is generally accepted that 
the heliospheric structure is asymmetric in terms of a nose-tail (azimuthal) 
direction, yielding a ratio of $\sim$~1:2 for the upwind-to-downwind TS 
distance from the Sun. This asymmetry is pronounced during solar 
minimum conditions because the TS propagates toward or away from the Sun 
with changing solar activity. The exact dependence is still unknown.
Encounters with big transients in the HMF may 
also cause the TS position to change and even oscillate locally with 
interesting effects on CRs. The HP has always been considered as the 
heliospheric boundary from a CR modulation point of view because it 
supposedly separates the solar and interstellar media. Ideally, the solar 
wind should not propagate beyond this boundary. According to MHD models the 
HP is properly demarcated in the direction that the heliosphere is moving 
but not so in the tail direction. Some instabilities can be anticipated at 
the HP that may modify this picture
\citep[e.g.,][]{Zank-etal2009}. It is expected that these aspects will
be studied with models in greater detail in future. The general
features of the heliospheric geometry are shown in Figure~\ref{fig1}. For
illustrations of these features obtained with magnetohydrodynamic
(MHD) models, see, e.g., \citet{Opher-etal2009a} and \citet{Pogorelov-etal2009a}.

\epubtkImage{Figure1.png}{%
\begin{figure}[htbp]
\centerline{\includegraphics[width=0.8\textwidth]{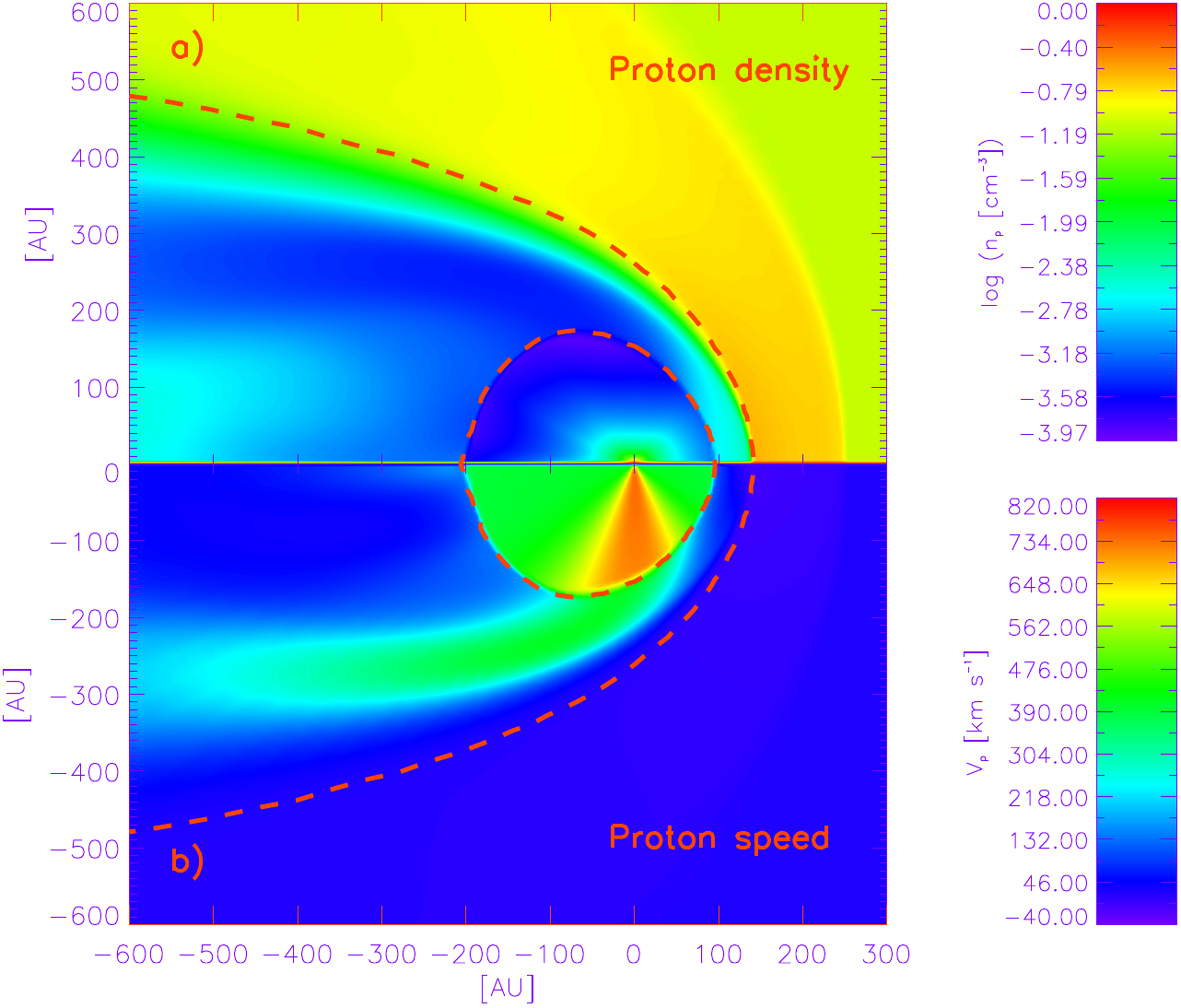}}
  \caption{The basic features of the global heliospheric geometry
    according to the hydrodynamic (HD) models of
    \citet{FerreiraScherer2004} in terms of the solar wind density
    (upper panel) and solar wind speed (lower panel). The heliosphere
    is moving through the interstellar medium to the right. Typical
    solar minimum conditions are assumed so that the solar wind speed
    has a strong latitudinal dependence.}
  \label{fig1}
\end{figure}}

The last decade witnessed the development of CR transport models based
on improved HD and MHD models of the heliosphere that provide
realistic geometries and detailed backgrounds (solar wind flow and
corresponding magnetic field lines) to global transport models, called
hybrid models. It is known that CRs exert pressure and, therefore,
also modify the heliosphere \citep[e.g.,][]{Fahr2004}. These MHD
models also predict an asymmetry in a north-south (meridional or
polar) direction, making it most likely that the heliosheath is wider
in the direction that Voyager~1 is moving than in the Voyager~2
direction. The local interstellar magnetic field causes the
heliosphere to become tilted as featured already in earlier global
simulations of the heliosphere \citep[e.g.,][]{Ratkiewicz-etal1998,
  Linde-etal1998}. Although this asymmetry seems somewhat
controversial from a MHD point of view, energetic neutral particle
(ENAs) observations from the IBEX mission sustain this view of the
heliosphere \citep{McComas-etal2012a}. This mission also established
that the boundary where the heliosphere begins to disturb the
interstellar medium, because it is moving with respect to this medium,
should not be seen as a bow shock but rather as a bow wave
\citep{McComas-etal2012b}. The relative motion of the Sun with respect
to the interstellar medium seems slower and also in a slightly
different direction than previously thought. See also
\citet{Zank-etal2013}. A schematic presentation of this new view of
the geometrical shape of the heliosphere is shown in
Figure~\ref{fig2}. Models of the heliosphere based on HD and MHD
approaches have become very sophisticated over the past decade and
many of the predicted features still have to be incorporated in the
hybrid modeling approach. Whether these detailed features are
contributing more than higher order effects to the global solar
modulation of CRs is to be determined. For reviews on MHD modeling,
see, e.g., \citet{Opher-etal2009b} and \citet{Pogorelov-etal2009b}.

\epubtkImage{Figure2.jpg}{%
\begin{figure}[htbp]
\centerline{\includegraphics[width=0.8\textwidth]{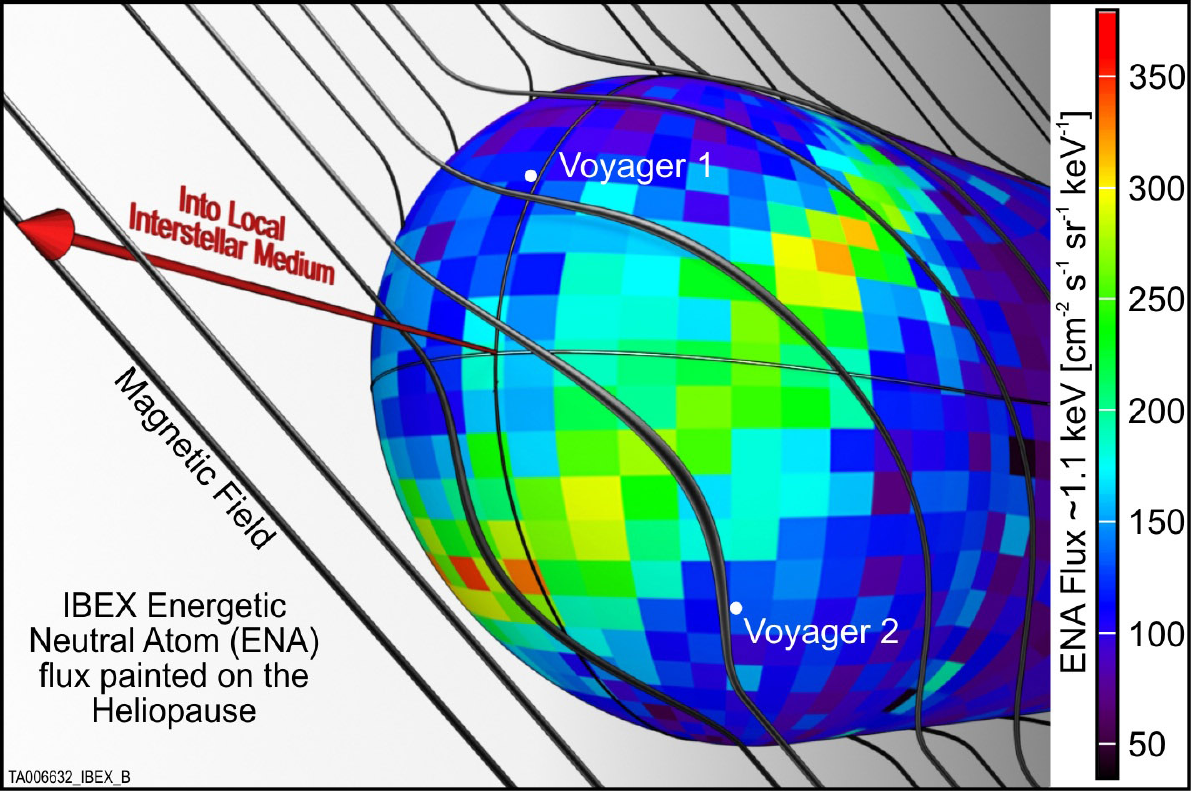}}
  \caption{A schematic view of an asymmetric heliosphere together with
    the directions of the interstellar magnetic field lines. The
    measured ENA flux at $\sim$~1.1~keV is superposed on the
    heliopause with the bright ENA ribbon appears to correlate with
    where the field is most strongly curved around it. (From the
    Interstellar Boundary Explorer, IBEX spacecraft's first all-sky
    maps of the interstellar interaction at the edge of the
    heliosphere.) See \citet{McComas-etal2009} for details. Image
    credit: Adler Planetarium/Southwest Research Institute.}
  \label{fig2}
\end{figure}}

It is to be determined if the outer heliosheath, the region beyond the HP, 
has any effect on CRs when they enter this region from the interstellar 
medium. \citet{Scherer-etal2011} presented arguments that this may be the case 
followed by \citet{Strauss-etal2013} who presented numerical modeling that 
produces at 100~MeV a small radial intensity gradient between 0.2\% to 
0.4\% per AU, depending on solar activity, and if the BW is assumed 
at 250~AU. It may also be that CRs do not enter the heliosphere completely 
isotropic as always been assumed \citep[e.g.,][]{NgobeniPotgieter2011,
  NgobeniPotgieter2012}.

For informative reviews on several advanced aspects of the outer heliosphere 
see \citet{Jokipii2012} and \citet{Florinski-etal2011} and for a
review on the solar cycle from a solar physics point of view, see
\citet{Hathaway2010}.

The point to be made here is that the global features of the heliosphere, 
and in particular the variability and dynamics of the heliospheric boundary 
lagers, do influence the modulation and the longer-term variations of CRs, 
even at Earth. How important this all is, will unfold in the coming years.

\subsection{Solar wind and heliospheric magnetic field}

\subsubsection{Global magnetic field geometry}

\citet{Parker2001} reviewed how the expansion of the solar corona provides the 
solar wind with an embedded solar magnetic field that develops into the HMF. 
He had predicted a well-defined spiral structure for the HMF \citep{Parker1958}. 
Over the years modifications to this field have been proposed but the 
realization of \citet{Fisk1996} that the differential rotation of the Sun and the 
rigid rotation of polar coronal holes has a significant effect on the 
structure of the HMF, led to second generation global HMF models that are 
much more complex and controversial and as such not yet fully
appreciated of what it may imply for CR modulation \citep[e.g.,][and
  references therein]{Sternal-etal2011}. These specific features will
have to be studied with MHD modeling to resolve the dispute. The
Parker-type field and moderate modification thereof
\citep[e.g.,][]{SmithBieber1991} are still widely used in CR
modeling. See also the review by \citet{HeberPotgieter2006}.

A straightforward equation for the spiral HMF is
\begin{equation} \label{eq:b}
\mathbf{B}=B_o\left(\frac{r_0}{r}\right)^2\left(\mathbf{e}_r-\tan{\psi\mathbf{e}_{\phi}}\right),
\end{equation}
with unit vector components $\mathbf{e}_r$ and $\mathbf{e}_\phi$ in
the radial and azimuthal direction respectively; $r_0 =
1\mathrm{\ AU}$ for dimensional purposes and $\psi$ is the (spiral)
angle between the radial and the average HMF direction at a certain
position. It is given by
\begin{equation} \label{eq:tanpsi}
\tan{\psi}=\frac{\Omega\left(r-r_{\odot}\right)\sin{\theta}}{V},
\end{equation}
and indicates how tightly wound the HMF is, with $\Omega$ the angular
speed of the Sun, and with $r_{\odot} = 0.005\mathrm{\ AU}$ the radius
of the solar surface. A typical value at Earth is $\psi \approx
45^{\circ}$, increasing to $\sim 90^{\circ}$ with increasing radial
distance $r$ beyond 10~AU in the equatorial plane. The solar wind
speed is $V$ and $\theta$ is the polar angle so that the magnitude of
the HMF is 
\begin{equation} \label{eq:b_mag}
B=B_o\left(\frac{r_0}{r}\right)^2\sqrt{1+(\tan{\psi})^2},
\end{equation}
with an average value of 5~nT at Earth or as determined by $B_o$. 

The main attribute of the HMF is that it follows a 22-year cycle with a 
reversal about every $\sim$~11 years at the time of extreme solar activity 
\citep[e.g.,][]{PetrovayChristensen2010}. It exhibits many distinct shorter 
scale features \citep[e.g.,][]{BaloghJokipii2009}, also associated with the TS 
and the heliosheath region \citep{Burlaga-etal2005, BurlagaNess2011}. 
One of the largest of these magnetic structures was encountered by Voyager~1 
in 2009.7 when it already was $\sim$~17~AU beyond the TS when the shock had been 
observed at 94~AU. At this time the field direction suddenly changed 
indicating a sector crossing. This means that these features, which are 
well-known in the inner heliosphere, are also occurring in some evolved form 
in the inner heliosheath. For a review, see \citet{RichardsonBurlaga2011}. 
In this context, it is expected that more interesting and surprising 
observations and subsequent modeling will follow from the Voyager mission. 

A major corotating structure of the HMF of important to CR modulation is the 
heliospheric current sheet (HCS), which divides the solar magnetic field into 
hemispheres of opposite polarity. The $\sim$~11-year period when it is 
directed outwards in the northern hemisphere has become known as $A > 0$ 
epochs, such as during the 1970s and 1990s, while the 1980s and the period 
2002\,--\,2014 are known as $A < 0$ cycles. The HCS has a wavy 
structure, parameterized by using its tilt angle $\alpha$ \citep{Hoeksema1992} 
and is well correlated to solar activity. During high levels of activity 
$\alpha$~=~75\textdegree, but then becomes undetermined during times of extreme 
solar activity, while during minimum activity
$\alpha$~=~3\textdegree\,--\,10\textdegree. The waviness of the HCS
plays an important role in CR modulation \citep{Smith2001}. It is still
the best proxy for solar activity from this point of view. It is
widely used in numerical modeling and some aspects are discussed
below. A disadvantage is that it is not known how the waviness is
preserved as it moves into the outer heliosphere, and especially what
happens to it in the heliosheath. The waviness becomes compressed in
the inner heliosheath as the outward flow decreases across the TS. It
should also spread in latitudinal and azimuthal directions in the nose
of the heliosphere. A schematic presentation of how the waviness of
the HCS could differ from the nose to the tail regions of the
heliosphere is shown in Figure~\ref{fig3}. The dynamics of the HCS in the inner
heliosheath was investigated by \citet{Borovikov-etal2011} amongst
others. 

\epubtkImage{Figure3.png}{%
\begin{figure}[htb]
 \centerline{\includegraphics[width=0.5\textwidth]{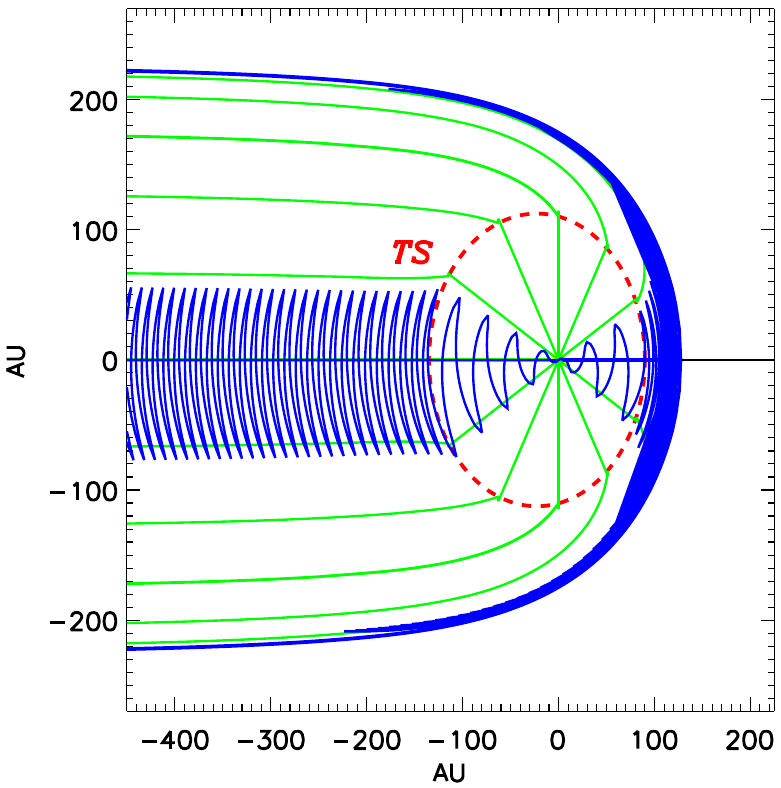}}
  \caption{A schematic presentation of how the waviness of the HCS
    could differ ideally from the nose to the tail regions of the
    heliosphere. The waviness depicted here corresponds to moderate
    solar activity. For an elaborate illustration of the dynamics of
    the HCS obtained with MHD models, see, e.g.,
    \citet{Borovikov-etal2011}. Image reproduced by permission from
    \citet{Kota2012}, copyright by Springer.}
  \label{fig3}
\end{figure}}

\citet{Drake-etal2010} suggested the compacted HCS could lead to magnetic 
reconnection, otherwise, it could become so densely wrapped that the 
distance between the wavy layers becomes less than the gyro-radius of the 
CRs, which may lead to different transport-effects. The behaviour of the HCS 
in the nose and tail directions of the heliosheath, and its role in 
transporting CRs, is a study in progress
\citep[e.g.,][]{Florinski2011}. It appears that the heliosheath may
even require additional transport physics to be included in the next
generation of transport models. What happens to the HMF closer to the
HP, in the sense of is it still embedded in the solar wind as the HP
is approached, is another study in progress.

\subsubsection{Global solar wind features}

A well reported feature of the global solar wind velocity is that it can be 
considered basically as radially directed from close to the Sun, across the 
TS and deep into the inner heliosheath. However, during periods of minimum 
solar activity this radial flow becomes distinctively latitude dependent, 
changing easily from an average of 450~\kms in the equatorial
plane to 800~\kms in the polar regions as observed by the
Ulysses mission \citep[see the reviews by][and reference
  therein]{HeberPotgieter2006, HeberPotgieter2008}. This aspect is also
illustrated in the bottom panel of Figure~\ref{fig1}. This significant
effect disappears with increasing solar activity. These features are
incorporated in most CR modulation models. 

The solar wind gradually evolves as it moves outward from the Sun. Its
speed is, on average, constant out to $\sim$~30~AU, then starts a slow
decrease caused by the pickup of interstellar neutrals, which reduces
its by $\sim$~20\% before the TS is reached. These pickup ions heat
the thermal plasma so that the solar wind temperature increases
outside $\sim$~25~AU. The solar wind pressure changes by a factor of 2
over a solar cycle and the structure of the solar wind is modified by
interplanetary coronal mass ejections (ICMEs) near solar maximum. The
first direct evidence of the TS was the observation of streaming
energetic particles by both Voyager~1 and 2 beginning $\sim$~2 years
before their respective TS crossings. The second evidence was a
decrease in the solar wind speed commencing 80~days before Voyager~2
crossed the TS. The TS seems to be a weak, quasi-perpendicular shock,
which transferred the solar wind flow energy mainly to the pickup ions
\citep{RichardsonStone2009}.

Across the TS, the radial solar wind speed slowed down from an average of 
$\sim$~400~\kms to $\sim$~130~\kms. It then had decreased essentially to 
zero as Voyager~1 approached the HP but based on non-plasma observations 
\citep{Krimigis-etal2011}. This appears to happen differently at Voyager~2 and 
will certainly be different in the tail direction of the heliosphere. 
Voyager~2, with a working plasma detector, has observed heliosheath plasma 
since August 2007, which indicates how it has evolved across the inner 
heliosheath. The radial speed slowly decreased as the plasma flow slowly 
turned tailward but remained above 100~\kms, which implies that
Voyager~2 was still a substantial distance from the HP in 2012
and that its approach towards the HP is also developing differently
from a solar wind point of view \citep{Richardson-etal2008,
RichardsonWang2011}. The inner heliosheath is clearly a highly
variable region. 

The magnitude of the radial component of the solar wind velocity in
terms of polar angle and radial distance in AU, as is typically used
in numerical models, up to just beyond the TS, is given by
\begin{eqnarray}
\label{eq:V}
V\left(r,\theta\right) = V_0\left[1-\exp\left(13.33\left(\frac{r_{\odot}-r}{r_0}\right)\right)\right]\nonumber\\
\left[1.475\mp0.4\tanh\left({6.8\left(\theta-\frac{\pi}{2}\pm\theta\right)}\right)\right]\nonumber\\
\left[\left(\frac{s+1}{2s}\right)-\left(\frac{s-1}{2s}\right)\tanh\left(\frac{r-r_{\mathrm{TS}}}{L}\right)\right],
\end{eqnarray}
where \textit{V}\sub{0}~=~400~\kms, \textit{s}~=~2.5 and
\textit{L}~=~1.2~AU. The TS is positioned at $r_{\mathrm{TS}}$,
\textit{L} is the TS scale length, and \textit{s} is the compression
ratio of the TS, which changes position over a solar cycle. The top
and bottom signs respectively correspond to the northern $(0 \leq
\theta \leq \pi/2)$ and southern hemisphere $(\pi/2 \leq \theta \leq
\pi)$ of the heliosphere, with $\theta_T = \alpha + 15\pi/180$. This
determines at which polar angle the solar wind speed changes from a
slow to a fast region during solar minimum activity conditions. This
equation is only for such conditions and must also be modified if the
solar wind velocity obtains a strong latitudinal and azimuthal
component when approaching the HP.

For reviews on observations and interpretations of the solar wind in
the outer heliosphere, see \citet{RichardsonStone2009} and
\citet{RichardsonBurlaga2011}.

\newpage 
\section{Cosmic Rays in the Heliosphere}

\subsection{Anomalous cosmic rays}

The anomalous component of cosmic rays (ACRs) was discovered in the early 
1970s \citep{Hovestadt-etal1973, Garcia-Munoz-etal1973}. This component, 
with kinetic energy $E$ between $\sim$~10 to 100~MeV/nuc, does not display the 
same spectral behaviour as galactic CRs but increases significantly with 
decreasing energy. Galactic CRs have harder spectra than ACRs. Their 
composition consists of hydrogen, helium, nitrogen, oxygen, neon, and argon 
and is primarily singly ionized \citep{CummingsStone2007}. They originate 
as interstellar neutrals that become ionized when flowing towards the Sun 
and then, as so-called pick-up ions, become accelerated in the solar wind 
\citep{Fisk-etal1974, Fisk1999}. \citet{Pesses-etal1981} suggested that ACRs 
were to be accelerated at the TS. Strictly speaking they are not CRs 
because they have a heliospheric origin, with the spectrum of ACRs 
determined by heliospheric processes. To become ACRs, these pick-up ions 
must be accelerated by four orders of magnitude. They are subjected to solar 
modulation and depict mostly, but not always, the same modulation features 
than CRs upstream of the TS \citep[e.g.,][]{McDonald-etal2000,
  McDonald-etal2003, McDonald-etal2010}. Only the ACRs with the
highest rigidity (oxygen) can reach Earth
\cite[e.g.,][]{Leske-etal2011, StraussPotgieter2010c}. See the
introductory review on this topic by \citet{Fichtner2001} and reviews
of recent developments by \citet{Giacalone-etal2012} and
\citet{Mewaldt2012}. 

The principal acceleration mechanism was
considered to be diffusive shock acceleration, a topic of considerable
debate since Voyager~1 crossed the TS \citep{Fisk2005}. At the
location of the TS there was no direct evidence of the effective local
acceleration of ACR protons but particles with lower energies were
effectively accelerated and have since become known as termination
shock particles (TSP). The higher energy ACRs thus seem
disappointingly unaffected by the TS but have increased gradually in
intensity away from the TS \citep{Stone-etal2005, Stone-etal2008,
  Decker-etal2005}. They clearly gain energy as they move inside the
inner heliosheath and seem to be trapped largely in this region. It is
expected that their intensity will drop sharply over the HP but some
should escape out of the inner heliosheath
\citep[e.g.,][]{Scherer-etal2008b}. Several very sophisticated
mechanisms have been proposed how these particles may gain their
energy beyond the TS and has become one of the most severely debated
issues in this field of research \citep[e.g.,][]{Gloeckler-etal2009,
  ZhangLee2011, ZhangSchlickeiser2012}. 

Typical observed proton, helium and oxygen spectra for TSP, ACRs and 
galactic CRs are shown in Figure~\ref{fig4} for early in 2005 when Voyager~1 already 
had crossed the TS (2004.96) when at 94.01~AU. Computed ACR spectra at the 
TS are shown for comparison. Figure~\ref{fig5} displays how the observed TSPs, ACRs, 
and galactic CRs for helium evolved and unfolded at Voyager~1 and Voyager~2 
from late in 2004 to early in 2008. 

\epubtkImage{Figure4.png}{%
\begin{figure}[htb]
 \centerline{\includegraphics[width=0.8\textwidth]{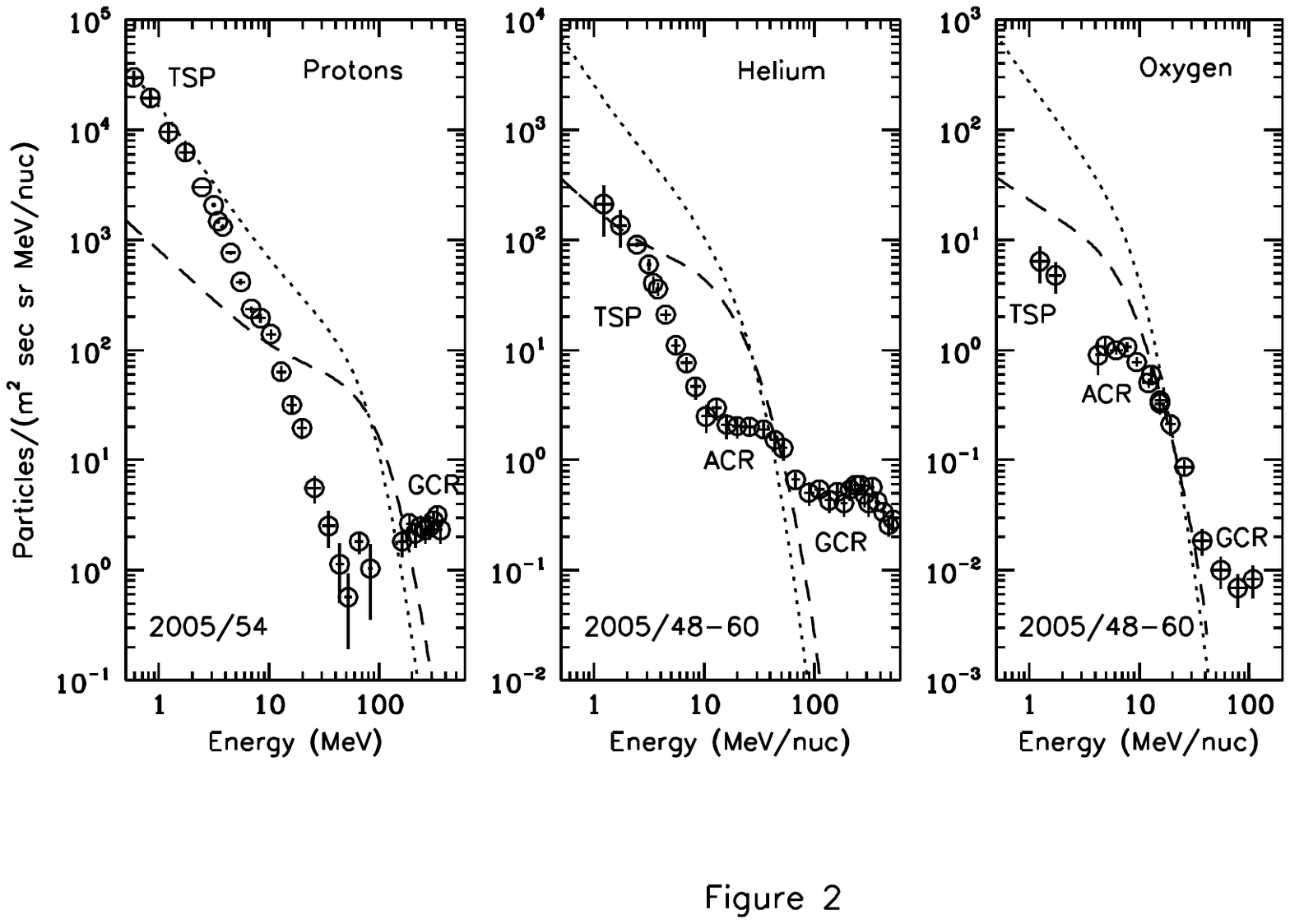}}
  \caption{Observed proton, helium, and oxygen spectra for TSP, ACRs
    and galactic CRs are shown for days in 2005 as indicated when
    Voyager~1 already had crossed the TS on 16 December 2004. Computed
    ACR spectra at the TS, assuming diffusion shock acceleration for
    two values of the TS compression ratio, are shown for
    comparison. Image reproduced by permission from
    \citet{Stone-etal2005}, copyright by AAAS.}
  \label{fig4}
\end{figure}}

\epubtkImage{Figure5.png}{%
\begin{figure}[htb]
 \centerline{\includegraphics[width=0.8\textwidth]{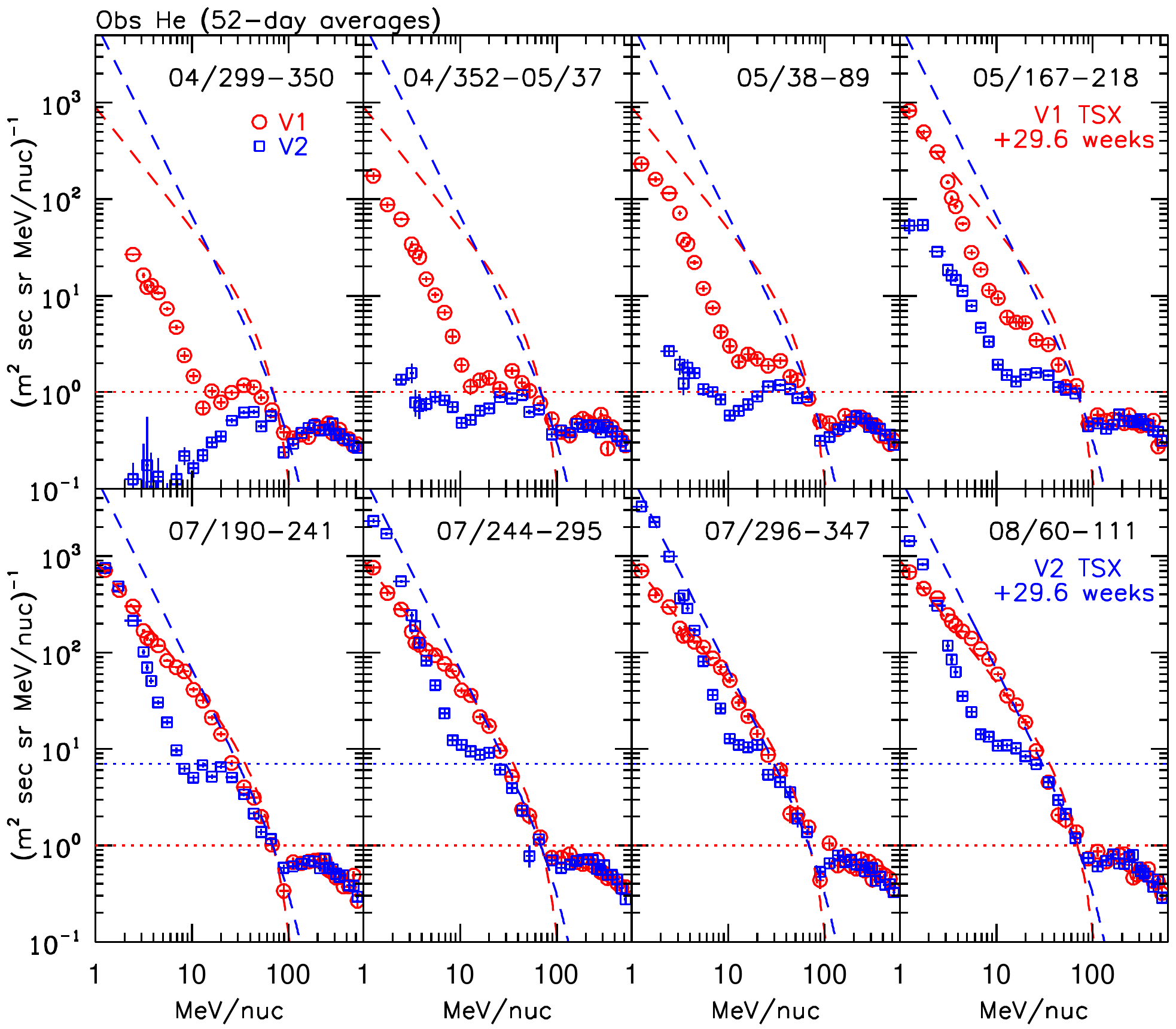}}
  \caption{A display of how the observed TSPs (exhibiting a power-law
    trent at low energies), ACRs (typically between 10~MeV and
    100~MeV) and galactic spectrum (typically above 100~MeV) for
    helium had evolved and unfolded at Voyager~1 (in red) and
    Voyager~2 (in blue) from late in 2004 to early in 2008. Voyager~1
    and Voyager~2 crossed the TS on 2004.96 and 2007.66
    respectively. Image reproduced by permission from
    \citet{Cummings-etal2008}, copyright by AIP.}
  \label{fig5}
\end{figure}}

Adding to the controversy, comparing computational results with spacecraft 
observations, it was found by \citet{Strauss-etal2010a} that the 
inclusion of multiply charged ACRs \citep{Mewaldt-etal1996a,
  Mewaldt-etal1996b, Jokipii1996} in a modulation model could explain
the observed strange spectrum of anomalous oxygen in the energy
range from 10\,--\,70~MeV per nucleon \citep{Webber-etal2007}. The
more effective acceleration of these multiply charged anomalous
particles at the TS causes a significant deviation from the usual
exponential cut-off spectrum to display instead of a power law decrease
up to 70~MeV per nucleon where galactic oxygen starts to
dominate. This can only happen if some acceleration takes place at the
TS. In addition, the model reproduces the features of multiply charged
oxygen at Earth so that a good comparison is obtained between
computations and observations. An extensive study on the intensity
gradients of anomalous oxygen was done by \citet{Cummings-etal2009} and
\citet{StraussPotgieter2010c}. For a comprehensive review on ACR
measurements at Earth and interesting conclusions, see
\citet{Leske-etal2011}.

Undoubtedly, the controversy indicates that we do not yet understand what is 
really happening to the ACRs in the inner heliosheath and only future 
observations with inquisitive modeling may enlighten us. On the other hand, 
the TSPs are accelerated at the TS. Surely, TSPs and ACRs are fascinating 
topics, from how they originate to their acceleration and modulation inside 
the heliosheath, and for the highest rigidity ACRs also up to Earth. For 
additional reviews of how these aspects have developed over time, see
\citet{HeberPotgieter2008}, \citet{Potgieter2008}, and
\citet{Florinski2009}.

\subsection{Galactic cosmic rays}

Cosmic rays are defined for the purpose of this overview as fully ionized 
nuclei as well as anti-protons, electrons, and positrons that are not 
produced on the Sun or somewhere in the heliosphere. As a rule they have 
kinetic energy $E \gtrsim 1\mathrm{\ MeV}$. 

\subsubsection{Local interstellar spectra}

A crucially important aspect of the modulation modeling of galactic CRs in 
the heliosphere is that the local interstellar spectra (LIS) need to be 
specified as input spectra at an assumed modulation boundary and then be 
modulated throughout the heliosphere as a function of position, energy, and 
time. A primary objective of the Voyager mission is to measure these LIS 
once the spacecraft enter the interstellar medium. Because of solar
modulation and the fact that the nature of the heliospheric diffusion
coefficients is not yet fully established, all cosmic ray LIS at
kinetic energies $E \lesssim 1\mathrm{\ GeV}$ remain
contentious. This is true from an astrophysical and heliospheric point
of view. 

Galactic spectra (GS), from a solar modulation point of view, are referred 
to as spectra that are produced from astrophysical sources, usually assumed 
to be evenly distributed through the Galaxy, typically very far from the 
heliosphere. Computed GS usually do not contain the contributions of any 
specific (local) sources within parsecs from the heliosphere so that an 
interstellar spectrum may be different from an average GS, which may again be 
different from a LIS (thousands of AU away) from the Sun, which might be 
different from a very LIS or what may be called a heliopause spectrum, right 
at the edge of the heliosphere, say 200~AU away from the Sun. Proper 
understanding of the extent of modulation cycles of galactic CRs in the 
heliosphere up to energies of $\sim$~30~GeV requires knowledge of these GS 
and LIS for the various species. More elaborate approaches to the 
distribution of sources have also been followed
\citep[e.g.,][]{BuschingPotgieter2008}, even considering contributions
of sources or regions relatively closer to the heliosphere
\cite[e.g.,][]{Busching-etal2008} with newer developments
\cite[e.g.,][]{Blasi-etal2012}. These spectra are calculated using
various approaches based on different assumptions but mostly using
numerical models, e.g., the well-known GALPROP propagation model
\citep{Moskalenko-etal2002, Strong-etal2007}. For energies below
$\sim$~10~GeV, which is of great interest to solar modulation studies,
the galactic propagation processes are acknowledged as less precise as
illustrated comprehensively by \citet{Ptuskin-etal2006} and
\citet{WebberHigbie2008, WebberHigbie2009}.
 
The situation for galactic electrons at low energies has always been 
considered somewhat better because electrons radiate synchrotron radiation 
so that radio data assist in estimating the electron GS at 
these low energies. For a discussion of this approach and some examples of 
consequent electron spectra, see \citet{Langner-etal2001},
\citet{WebberHigbie2008}, \citet{Strong-etal2011},
\citet{PotgieterNndanganeni2013b}, and references therein. 

\citet{PotgieterFerreira2002} and \citet{PotgieterLangner2004a} showed that 
the heliospheric TS could in principle re-accelerate low-energy galactic 
electrons to energies as high as $\sim$~1~GeV so that a heliopause spectrum 
could be different from a TS spectrum. Such a spectrum may even be higher 
than a LIS, depending on the energies considered. However, because the TS 
was observed as rather weak \citep{Richardson-etal2008}, 
obtaining such high energies now seems improbable. In fact, only a
factor of 2 increase was observed close to the TS for 6\,--\,14~MeV
electrons but since then Voyager~1 has observed an increase of a
factor of $\sim$~60 on its way to the HP \citep{Webber-etal2012,
  Nkosi-etal2011}. It seems that the influence of the heliospheric TS
on all LIS, in terms of the re-acceleration of these CRs, may
generally thus be neglected. 

The formation of a magnetic wall (barrier) at the HP, if significant, may 
cause a drop in low energy CRs, surely in the flux of TSPs and the ACRs 
while high energy CRs are not expected to change much. If low energy 
particles are partially trapped inside the inner heliosheath, the LIS of low 
energy CRs will not be known until well beyond the HP.

For a compilation of computed galactic spectra based on the GALPROP code for 
CR protons, anti-protons, electrons, positrons, helium, boron and 
carbon and many more, see \citet{Moskalenko-etal2002}. Peculiarly, the solar 
modulation of many of these species and their isotopes has not been properly 
modeled, probably because of uncertainties in their LIS.

\subsubsection{Main cosmic ray modulation cycles}

The dominant and the most important time scale in CRs related to solar 
activity is the 11-year cycle. This quasi-periodicity is convincingly 
reflected in the records of sunspots since the early 1600s and also in the 
galactic CR intensity observed at ground and sea level since the 1950s. 
This was the period when neutron monitors (NMs) were widely deployed as 
Earth bound CR detectors, especially during the International 
Geophysical Year (IGY). The year 2007 was celebrated as the 50th anniversary 
if the IGY and was called the International Heliophysical Year (IHY). These 
NMs have been remarkably reliable with good statistics over five full 
11-year cycles. 

The discovery of another important cycle, the 22-year cycle, 
was a milestone in the exploration and modeling of CRs in the heliosphere. 
It is directly related to the reversal of the HMF during each period of 
extreme solar activity. The causes of these cycles will be discussed in more 
detail in later sections. Figure~\ref{fig6} displays the 11-year and
22-year cycles in galactic CRs as observed by the Hermanus NM in South
Africa at a cut-off rigidity of 4.6~GV and a mean response energy of
$\sim$~18~GeV. 

\epubtkImage{Figure6.png}{%
\begin{figure}[htb]
 \centerline{\includegraphics[width=0.8\textwidth]{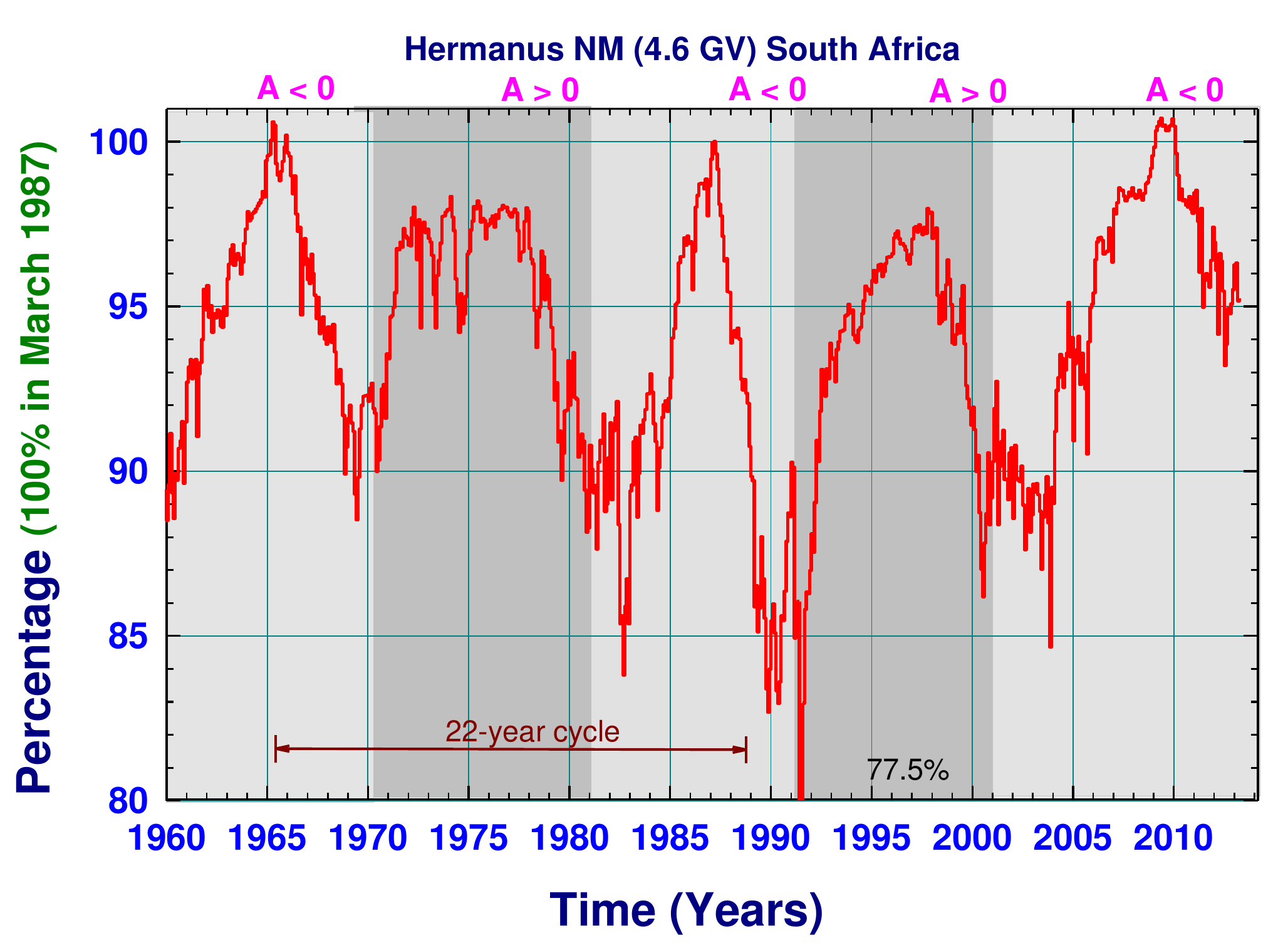}}
  \caption{An illustration of the 11-year and 22-year cycles in the
    solar modulation of CRs as observed by the Hermanus NM in South
    Africa at a cut-off rigidity of 4.6~GV in terms of percentage
    with March 1987 at 100\%.}
  \label{fig6}
\end{figure}}

Additional short periodicities are evident in NM and other CR data,
e.g., the 25\,--\,27-day variation owing to the rotational Sun, and
the daily variation owing to the Earth's rotation \citep[e.g.,][and
references therein]{Alania-etal2011}. These variations seldom have
magnitudes of more than 1\% with respect to the previous quite time
fluxes. Corotating interaction regions (CIRs), caused when a fast
solar wind region catches up with a lower region, usually merge as
they propagate outwards to form various types of larger interaction
regions. The largest ones are known as global merged interaction
regions (GMIRs), discussed in some detail later. They are related to
coronal mass ejections (CMEs) that are prominent with increased solar
activity but dissipating during solar minimum. Although CIRs may be
spread over a large region in azimuthal angle, they do not cause
long-term CR periodicities on the scale (amplitude) of the 11-year
cycle. An isolated GMIR may cause a decrease similar in magnitude than
the 11-year cycle but it usually lasts only several months (as
happened in 1991). A series of GMIRs, on the other hand, may
contribute significantly to long-term CR modulation during periods of
increased solar activity, in the form of large discrete steps,
increasing the overall amplitude of the 11-year cycle
\citep[e.g.,][]{PotgieterleRoux1992, leRouxPotgieter1995}. 

The galactic CR flux is not expected to be constant along the trajectory of 
the solar system in the galaxy. Interstellar conditions should differ 
significantly over very long time-scales, for example, when the Sun moves in 
and out of the galactic spiral arms \citep{BuschingPotgieter2008}. It is 
accepted that the concentration of Be\super{10} nuclei in polar ice exhibits 
temporal variations on a very long time scale in response to changes in the 
flux of the primary CRs. Exploring CR modulation over time scales of 
hundreds of years and longer and during times when the heliosphere was
significantly different from the present epoch is a very interesting
topic and a work in progress. See the reviews by, e.g.,
\citet{Scherer-etal2006}, \citet{McCrackenBeer2007}, and \citet{Usoskin2013}.

There are indications of CR periods of 50\,--\,65 years and
90\,--\,130 years, also for a periodicity of about 220 and 600
years. Quasibiennial oscillations have also been detected as a
prominent scale of variability in CR data \citep{Laurenza-etal2012}. It is
not yet clear whether these variabilities should be considered
`perturbations', stochastic in nature or truly time-structured to be
figured as superposition of several periodic processes. Cases of
strong `perturbations' of the consecutive 11-year cycles are the
`grand minima' in solar activity, with the prime example the Maunder
minimum (1640\,--\,1710) when sunspots almost completely
disappeared. Assuming the HMF to have vanished as well or without any
reversals during the Maunder minimum would be an
oversimplification. The heliospheric modulation of CRs could have
continued during this period but much less pronounced (with a small
amplitude). It is reasonable to infer that less CMEs occurred so that
the total flux of CRs at Earth then should have been higher than
afterwards. In this context, see the reviews by, e.g.,
\citet{Beer-etal2011} and \citet{McCracken-etal2011}.

An interesting reoccurring phenomenon, called the Gnevyshev Gap has been 
observed in all solar-terrestrial parameters and consists of a relatively 
short period of decreased solar activity during the extreme maximum phase of 
each 11-year cycle, yielding structured maxima with a first peak at the end 
of the increasing activity phase and a second one at the start of the 
declining phase. For a review, see \citet{Storini-etal2003}.

\clearpage 
\section{Solar Modulation Theory}

The paradigm of CR transport in the heliosphere has developed soundly over 
the last $\sim$~50 years. The basic processes are considered to be known. 
However, it is still quite demanding to relate the modulation of galactic 
CRs and of the ACRs to their true causes and to connect these causes over a 
period of a solar cycle or more, from both a global and microphysics level. 
The latter tends to be of a fundamental nature, attempting to understand the 
physics from first principles (ab initio) whereas global descriptions 
generally tend to be phenomenological, mostly driven by observations and/or 
the application of new numerical methods and models. In this context, global 
numerical models with relevant transport parameters are essential to make 
progress. Obviously, major attempts are made to have these models based on 
good assumptions, which then have to agree with all major heliospheric 
observations.

Understanding the basics of solar modulation of CRs followed only in the 
1950s when \citet{Parker1965} formulated a constructive transport theory. At 
that stage NMs already played a major role in observing solar activity 
related phenomena in CRs. Although Parker's equation contained an 
anti-symmetric term in the embedded tensor accounting for regular gyrating 
particle motion, it was only until the development of numerical models in 
the mid-1970s \citep{Fisk1971, Fisk1976, Fisk1979} that progress in
appreciation the full meaning of transport theory advanced
significantly.

\subsection{Basic transport equation and theory}

A basic transport equation (TPE) was derived by
\citet{Parker1965}. \citet{GleesonAxford1967} came to the same
equation more rigorously. They also derived an approximate solution to
this TPE, the so called force-field solution, which had been widely
used and was surpassed only when numerical models became available
\citep{GleesonAxford1968}. For a formal overview of these
theoretical aspects and developments, see \citet{Schlickeiser2002}. See also
\citet{Quenby1984}, \citet{Fisk1999}, and \citet{Moraal2011} for overviews of the TPEs
relevant to CR modulation. The basic TPE follows from the equations of
motion of charged particles in fluctuating magnetic fields (on both
large and small scales) and averages over the pitch and phase angles
of propagation particles. It is based on the reasonable assumption
that CRs are approximately isotropic. This equation is remarkably
general and is widely used to model CR transport in the
heliosphere. The heliospheric TPE according to \citet{Parker1965}, but
in a rewritten form, is
\begin{equation} \label{eq:parker_tpe}
\underbrace{\frac{\partial{f}}{\partial{t}}}_{\mathrm{a}} = -(\underbrace{\mathbf{V}}_{\mathrm{b}}+
\underbrace{\left\langle\mathbf{v}_{d}\right\rangle}_{c})\cdot\nabla{f}+\underbrace{\nabla\cdot(\mathbf{K}_s\cdot\nabla{f})}_{d}+
\underbrace{\frac{1}{3}\left(\nabla\cdot\mathbf{V}\right)\frac{\partial{f}}{\partial\ln{P}}}_{e} \,,
\end{equation}
where $f (\boldsymbol{r},P,t)$ is the CR distribution
function, $P$ is rigidity, $t$ is time, $r$ is the position in 3D,
with the usual three coordinates $r$, $\theta$, and $\phi$ specified
in a heliocentric spherical coordinate system where the equatorial
plane is at a polar angle of $\theta$~=~90\textdegree. A steady-state
solution has $\partial f / \partial t = 0$ (part~a), which means that all
short-term modulation effects (such as periods shorter than one solar
rotation) are neglected, which is a reasonable assumption for solar
minimum conditions. Terms on the right hand side respectively
represent convection (part b), with $\mathbf{V}$ the solar wind velocity; averaged
particle drift velocity $\langle \mathbf{v}_d \rangle$ caused by
gradients and curvatures in the global HMF (part~c); diffusion (part~d), with
$\mathbf{K}_{s}$ the symmetrical diffusion tensor and then the term
describing adiabatic energy changes (part~e). It is one of the four major
modulation processes and is crucially important for galactic CR
modulation in the inner heliosphere. If $(\nabla \cdot \mathbf{V}) >
0$, adiabatic energy losses are described, which become quite large in
the inner heliosphere \citep[see the comprehensive review
  by][]{Fisk1979}. If $(\nabla \cdot \mathbf{V}) < 0$, energy gains
are described, which may be the case for ACRs in the heliosheath
\citep[illustrated, e.g., by][]{Langner-etal2006b,
  Strauss-etal2010b}. If $(\nabla \cdot \mathbf{V}) = 0$, no adiabatic
energy changes occur for CRs, perhaps the case beyond the TS. This is
probably an over simplification but it was shown that the effect is
insignificant for galactic CRs but crucially important for ACRs
\citep[e.g.,][]{Langner-etal2006a}. For recent elaborate illustrations
of these effects, see \citet{Strauss-etal2010a, Strauss-etal2010b}.

In cases when anisotropies are large, other types of transport
equations must be used \citep[see, e.g., ][]{Schlickeiser2002}. Close
to the Sun and Jupiter, and even close to the TS, where large observed
anisotropies in particle flux sometimes occur
\citep[e.g.,][]{Kota2012}, Equation~(\ref{eq:parker_tpe}) thus needs to
be modified or even replaced to describe particle propagation based on
the Fokker--Planck equation. The `standard' TPE as given by
Equation~(\ref{eq:parker_tpe}) also needs modification by inserting
additional terms relevant to the conditions beyond the TS
\citep[e.g.,][]{Strauss-etal2010a, Strauss-etal2010b}.

For clarity on the role of diffusion, particle drifts, convection, and 
adiabatic energy loss, the TPE can be written in terms of a heliocentric 
spherical coordinate system as follows:
\begin{eqnarray}\label{eq:spherical_compact_tpe}
  &&\left[\frac{1}{r^2}\frac{\partial}{\partial{r}}\left(r^2K_{rr}\right)+
	\frac{1}{r\sin\theta}\frac{\partial}{\partial\theta}\left(K_{\theta{r}}\sin\theta\right)+\frac{1}{r\sin\theta}\frac{\partial{K}_{\phi{r}}}{\partial\phi}-V\right]\frac{\partial{f}}{\partial{r}}\quad\\
  &+&\left[\frac{1}{r^2}\frac{\partial}{\partial{r}}\left(rK_{r\theta}\right) + \frac{1}{r^2\sin\theta}\frac{\partial}{\partial\theta}\left(K_{\theta\theta}\sin\theta\right) + \frac{1}{r^2\sin\theta}\frac{\partial{K}_{\phi\theta}}{\partial{\phi}}\right]\frac{\partial{f}}{\partial\theta}\nonumber\\
  &+&\left[\frac{1}{r^2\sin\theta}\frac{\partial}{\partial{r}}\left(rK_{r\phi}\right) + \frac{1}{r^2\sin\theta}\frac{\partial{K}_{\theta\phi}}{\partial\theta} + \frac{1}{r^2\sin^2\theta}\frac{\partial{K}_{\phi\phi}}{\partial\phi} - \Omega\right]\frac{\partial{f}}{\partial\phi}\nonumber\\
  &+&K_{rr}\frac{\partial^2f}{\partial{r}^2} + \frac{K_{\theta\theta}}{r^2}\frac{\partial^2f}{\partial\theta^2} + \frac{K_{\phi\phi}}{r^2\sin^2\theta}\frac{\partial^2f}{\partial\phi^2}\nonumber\\
  &+&\frac{2K_{r\phi}}{r\sin\theta}\frac{\partial^2f}{\partial{r}\partial\phi}
  +
  \frac{1}{3r^2}\frac{\partial}{\partial{r}}\left(r^2V\right)\frac{\partial{f}}{\partial\ln{p}}=0 \,,\nonumber
\end{eqnarray}
where it is assumed that the solar wind is axis-symmetrical and
directed radially outward, i.e., $\mathbf{V}=V\mathbf{e}_r$.  This
version of the TPE is rearranged in order to emphasize the various
terms that contribute to diffusion, drift, convection, and adiabatic
energy losses, so that Equation~(\ref{eq:spherical_compact_tpe}) becomes
\begin{eqnarray}\label{eq:arranged_spherical_tpe}
  && \overbrace{\left[\frac{1}{r^2}\frac{\partial}{\partial{r}}\left(r^2K_{rr}\right)+\frac{1}{r\sin\theta}\frac{\partial{K}_{\phi{r}}}{\partial\phi}\right]\frac{\partial{f}}{\partial{r}} + \left[\frac{1}{r^2\sin\theta}\frac{\partial}{\partial\theta}\left(K_{\theta\theta}\sin\theta\right)\right]\frac{\partial{f}}{\partial\theta}}^{\text{diffusion}}\\
  &+& \overbrace{\left[\frac{1}{r^2\sin\theta}\frac{\partial}{\partial{r}}(rK_{r\phi})+\frac{1}{r^2\sin^2\theta}\frac{\partial{K}_{\phi\phi}}{\partial\phi}-\Omega\right]\frac{\partial{f}}{\partial\phi}}^{\text{diffusion}}\nonumber\\
  &+& \overbrace{K_{rr}\frac{\partial^2f}{\partial{r}^2}+\frac{K_{\theta\theta}}{r^2}\frac{\partial^2f}{\partial\theta^2}+\frac{K_{\phi\phi}}{r^2\sin^2\theta}\frac{\partial^2f}{\partial\phi^2}+\frac{2K_{r\phi}}{r\sin\theta}\frac{\partial^2f}{\partial{r}\partial\phi}}^{\text{diffusion}}\nonumber\\
  &+& \overbrace{\left[-\langle\mathbf{v}_d\rangle_{r}\right]\frac{\partial{f}}{\partial{r}}+\left[-\frac{1}{r}\langle\mathbf{v}_d\rangle_{\theta}\right]\frac{\partial{f}}{\partial\theta}+\left[-\frac{1}{r\sin\theta}\langle\mathbf{v}_d\rangle_{\phi}\right]\frac{\partial{f}}{\partial\phi}}^{\text{drift}}\nonumber\\
  &-& \overbrace{V\frac{\partial{f}}{\partial{r}}}^{\text{convection}}\nonumber\\
  &+& \overbrace{\frac{1}{3r^2}\frac{\partial}{\partial{r}}\left(r^2V\right)\frac{\partial{f}}{\partial\ln{p}}}^{\text{adiabatic energy losses}}=0.\nonumber
\end{eqnarray}

Here $K_{rr}$, $K_{r \theta}$, $K_{r \phi}$, $K_{\theta r}$,
$K_{\theta \theta}$, $K_{\theta \phi}$, $K_{\phi r}$, $K_{\phi
  \theta}$, and $K_{\phi \phi}$,  are the nine elements of the 3D
diffusion tensor, based on a Parkerian type HMF. Note that 
$K_{rr}$, $K_{r \phi}$, $K_{\theta \theta}$, $K_{\phi r}$, $K_{\phi
\phi}$
describe the diffusion processes and that
$K_{r \theta}$, $K_{\theta r}$, $K_{\theta \phi}$, $K_{\phi
\theta}$, describe particle drifts, in most cases consider to be
gradient, curvature, and current sheet drifts. For this equation it is
assumed that the solar wind velocity has only a radial component, which
is not the case in the outer parts of the heliosheath, close to the
HP. 

A source function may also be added to this equation, e.g., 
if one wants to study the modulation of Jovian electrons
\citep[e.g.,][]{Ferreira-etal2001}. 

The components of the drift velocity are given in the next
Section~\ref{sec:diffusion-coefficients}. It is important to note that
the drift velocity in Equation~(\ref{eq:parker_tpe}) is multiplied with
the gradient in $f$ so when the CR intensity gradients are reduced by
changing the contribution from diffusion, drift effects on CR
modulation become implicitly reduced. This reduction will unfold
differently from changing the drift coefficient explicitly.

It is more useful to calculate the differential intensity of CRs,
e.g., as a function of kinetic energy to obtain spectra, which can be
compared to observations. In this context, a few useful definitions
and relations are as follows:

Particle rigidity is defined as
\begin{equation}
  P = \frac{pc}{q} = \frac{mvc}{Ze} \,,
\end{equation}
where $p$ is the magnitude of the particle's relativistic momentum,
$q=Ze$ its charge, $v$ its speed, $m$ its relativistic mass, and $c$
the speed of light in outer space.

For a relativistic particle, the total energy in terms of momentum is
given by
\begin{equation}
  E_t^2 = \left(E+E_{0}\right)^2  = p^2c^2 + m_0^2c^4 \,,
\end{equation}
with $E$ the kinetic energy and $E_{0}$ the rest-mass energy of the
particle (e.g., $E_0=938\mathrm{\ MeV}$ for protons and
$E_0=5.11\times10^{-1}\mathrm{\ MeV}$ for electrons) and $m_0$ the
particle's rest mass.

The kinetic energy per nucleon in terms of particle rigidity is then
\begin{equation}
  E = \sqrt{P^2\left(\frac{Ze}{A}\right)^2+E_0^2}-E_0 \,,
\end{equation}
so that rigidity can be written in terms of kinetic energy per nucleon as
\begin{equation}
  P = \left(\frac{A}{Ze}\right)\sqrt{E\left(E+2E_0\right)},
\end{equation}
with $Z$ the atomic number and $A$ the mass number of the CR particle.
The ratio of a particle's speed to the speed of light, $\beta$, in
terms of rigidity is given by
\begin{equation}
  \beta = \frac{v}{c} = \frac{P}{\sqrt{P^2+\left(\frac{A}{Ze}\right)^2E_0^2}}
\end{equation}
and in terms of kinetic energy it is
\begin{equation}
  \beta = \frac{v}{c} = \frac{\sqrt{E(E+2E_0)}}{E+E_0}.
\end{equation}
The relation between rigidity, kinetic energy, and $\beta$ is
therefore 
\begin{equation}
  P = \frac{A}{Z}\sqrt{E\left(E+2E_0\right)} = \left(\frac{A}{Z}\right)\beta\left(E+E_0\right).
\end{equation}

Particle density within a region $d^3r$, for particles with momenta
between $\mathbf{p}$ and $\mathbf{p}+d\mathbf{p}$, is related to the
full CR distribution function (which includes a pitch angle
distribution) by
\begin{equation}
  n = \int{{F}(\mathbf{r},\mathbf{p},t)d^3p} =
  \int\limits_p{p^2\left[\int\limits_{\Omega}{F(\mathbf{r},\mathbf{p},t)
        \, d\Omega}\right]dp} \,,
\end{equation}
where $d^3p=p^2 \, dp \, d\Omega$. The differential particle density, $U_p$,
is related to $n$ by
\begin{equation}
  n=\int{U_p(\mathbf{r},p,t) \, dp} \,,
\end{equation}
which gives
\begin{equation}
  U_p(\mathbf{r},p,t) =
  \int\limits_{\Omega}{p^2F(\mathbf{r},\mathbf{p},t) \, d\Omega}.
\end{equation}
The omni-directional (i.e., pitch angle) average of
$F(\mathbf{r},\mathbf{p},t)$ is calculated as
\begin{equation}
  f(\mathbf{r},p,t) =
  \frac{\int\limits_{\Omega}{F(\mathbf{r},\mathbf{p},t) \, d\Omega}}{\int\limits_{\Omega} \, d\Omega} =
  \frac{1}{4\pi}\int\limits_{\Omega}{F(\mathbf{r},\mathbf{p},t) \, d\Omega}
  \,,
\end{equation}
which leads to
\begin{equation}
  U_p(\mathbf{r},p,t) = 4\pi{p}^2f(\mathbf{r},p,t).
\end{equation}

The differential intensity, in units of particles/unit area/unit
time/unit solid angle/unit momentum, is
\begin{equation}
  j_p = \frac{vU_p(\mathbf{r},p,t)}{\int\limits_{\Omega}d\Omega} =
  \frac{vU_p(\mathbf{r},p,t)}{4\pi} = vp^2f(\mathbf{r},p,t) \,,
\end{equation}
so that 
\begin{equation}
  j(\mathbf{r},p,t) = \frac{v}{4\pi}U_p\frac{dp}{dE_t} =
  \frac{1}{4\pi}U_p = p^2f(\mathbf{r},p,t) \,,
\end{equation}
where $j(\mathbf{r},p,t)$ is the differential intensity in units of
particles/area/time/solid angle/energy. The relation between $j$ and
$f$ is simply $(P)^2$.

\subsection{Basic diffusion coefficients}
\label{sec:diffusion-coefficients}

The diffusion coefficients of special interest in a 3D heliocentric
spherical coordinate system are
\begin{eqnarray}\label{eq: kapparr}
K_{rr} &=& K_{\parallel }\cos ^{2}\psi +K _{\perp r}\sin ^{2}\psi,\nonumber\\
K _{\perp \theta } &=& K _{\theta \theta}, \nonumber\\
K_{\phi\phi} &=& K_{\perp r}\cos ^{2}\psi+K _{\parallel }\sin ^{2}\psi,\nonumber\\
K_{\phi{r}}&=& \left( K_{\perp r}-K_{\parallel }\right) \cos\psi \sin \psi = K_{r\phi} \,,
\end{eqnarray}
where $K_{rr}$ is the effective radial diffusion coefficient, a combination of 
the parallel diffusion coefficient $K_{\vert \vert}$ and the radial 
perpendicular diffusion coefficient $K_{\bot r}$, with $\psi $ the
spiral angle of the average HMF; 
$K_{\theta\theta} = K_{\bot \theta}$ is the effective perpendicular diffusion
coefficient in the polar direction. Here, $K_{\phi \phi}$ describes
the effective diffusion in the azimuthal direction and $K_{\phi r}$ is
the diffusion coefficient in the $\phi r$-plane, etc. Both are
determined by the choices for $K_{\vert \vert}$ and $K_{\bot}$. Beyond
$\sim$~20~AU in the equatorial plane $\psi \to 90^{\circ}$, so that
$K_{rr}$ is dominated by $K_{\bot r}$ whereas $K_{\phi \phi}$ is
dominated by $K_{\vert \vert}$, but only if the HMF is Parkerian in
its geometry. These differences are important for the modulation of
galactic CRs in the inner heliosphere
\citep[e.g.,][]{Ferreira-etal2001}. The important role of
perpendicular diffusion (radial and polar) in the inner heliosphere
has become increasingly better understood over the past decade since
it was realized that it should be anisotropic \citep{Jokipii1973},
with reasonable consensus that $K_{\bot \theta} > K_{\bot r}$ away
from the equatorial regions. The expressions for these diffusion
coefficients become significantly more complicated when advanced
geometries are used for the HMF \citep[e.g.,][]{Burger-etal2008,
  Effenberger-etal2012}.

A typical empirical expression as used in numerical models for the diffusion 
coefficient parallel to the average background HMF is given by
\begin{equation}\label{eq:kappa_parallel}
  K_{\parallel} = K_{\parallel,0}\beta\frac{B_n}{B_m}\left(\frac{P}{P_0}\right)^{a}
	\left[\frac{\left(\frac{P}{P_0}\right)^c+\left(\frac{P_k}{P_0}\right)^c}{1+
	\left(\frac{P_k}{P_0}\right)^c}\right]^{\left(\frac{b-a}{c}\right)} \,,
\end{equation}
where $(K_{\vert \vert})_0$ is a constant in units of
10\super{22}~cm\super{2}~s\super{-1}, with the rest of the equation
written to be dimensionless with $P_0$~=~1~GV and $B_m$ the modified
HMF magnitude with $B_n$~=~1~nT (so that the units remain
cm\super{2}~s\super{-1}). Here, $a$ is a power index that can change
with time (e.g., from 2006 to 2009); $b = 1.95$ and together with $a$
determine the slope of the rigidity dependence respectively above and
below a rigidity with the value $P_k$, whereas $c = 3.0$ determines the
smoothness of the transition. This means that the rigidity dependence
of $K_{\vert \vert}$ is basically a combination of two power laws. The
value $P_k$ determines the rigidity where the break in the power law
occurs and the value of $a$ specifically determines the slope of the
power law at rigidities below $P_k$ \citep[e.g.,][]{Potgieter-etal2013}.

Perpendicular diffusion in the radial direction is usually assumed to be 
given by 
\begin{equation} \label{eq:kappa_per_r}
K_{\perp{r}} = 0.02 K_{\parallel}.
\end{equation}

This is quite straightforward but still a very reasonable and widely used 
assumption in numerical models. It is to be expected that this ratio could not 
just be a constant but should at least be energy-dependent.
Advanced and complicated fundamental approaches had been 
followed, e.g., by \citet{Burger-etal2000} with subsequent CR modulation results that are 
not qualitatively different from a global point of view. Advances in 
diffusion theory and subsequent predictions for the heliospheric diffusion coefficients 
\citep[e.g.,][]{TeufelSchlickeiser2002} make it possible to narrow down the 
parameter space used in typical modulation models, e.g., the 
rigidity dependence at Earth. This is a work in progress.

The role of polar perpendicular diffusion, $K_{\bot \theta}$, in the inner 
heliosphere has become increasingly better understood over the past decade 
since it was realized that perpendicular diffusion should be anisotropic, with reasonable 
consensus that $K_{\bot \theta} > K_{\bot r}$ away from the equatorial 
regions \citep{Potgieter2000}. Numerical modeling shows explicitly that in 
order to explain the small latitudinal gradients observed for protons by 
Ulysses during solar minimum modulation in 1994, an enhancement of 
latitudinal transport with respect to radial transport is required
\citep[e.g.,][and reference therein]{HeberPotgieter2006, HeberPotgieter2008}. 
The perpendicular diffusion coefficient in the polar direction is thus
assumed to be given by 
\begin{equation}\label{eq:kappa_per_theta}
K_{\perp\theta} = 0.02 K_{\parallel} f_{\perp\theta},
\end{equation}
with
\begin{equation}
f_{\perp\theta} = {A^+}\mp+{A^-}\tanh\left[8\left({\theta_A}-{90^\circ}\pm{\theta_F}\right)\right].
\end{equation}

Here $A^\pm = (d \pm 1)/2$, $\theta_F =35^{\circ}$, $\theta_A = \theta$ for 
$\theta \le 90^{\circ}$ but $\theta_A = 180^{\circ} - \theta$ with
$\theta \ge 90^{\circ}$, and $d = 3.0$. This means that $K_{\theta
  \theta} = K_{\bot \theta}$ is enhanced towards the poles by a factor
$d$ with respect to the value of $K_{\vert \vert}$ in the equatorial
regions of the heliosphere. This enhancement is an implicit way of
reducing drift effects by changing the CR intensity gradients
significantly. For motivations, applications and discussions, see
\citet{Potgieter2000}, \citet{Ferreira-etal2003a, Ferreira-etal2003b},
\citet{Moeketsi-etal2005}, and
\citet{NgobeniPotgieter2008, NgobeniPotgieter2011}. The procedure
usually followed to solve Equation~(\ref{eq:spherical_compact_tpe}) is
described by, e.g., \citet{Ferreira-etal2001} and \citet{Nkosi-etal2008}.

\subsection{The drift coefficient}

The pitch angle averaged guiding center drift velocity for a near isotropic 
CR distribution is given by $\langle \mathbf{v}_d \rangle =
\nabla \times (K_{d}\mathbf{e}_{B})$, with $\mathbf{e}_{B} =
\boldsymbol{B}/B$ where $B$ is the magnitude of the background HMF
usually assumed to have a basic Parkerian geometry in the equatorial
plane. This geometry gives very large drifts over the polar regions of
the heliosphere so that it is standard practice to modify it in the
polar regions, e.g., \citet{SmithBieber1991} and \citet{Potgieter1996,
  Potgieter2000}. 

Under the assumption of weak
scattering, the drift coefficient is straightforwardly given as
\begin{equation}\label{eq:kappa_d_weak_scattering}
(K_d)_{ws} = (K_d)_0 \frac{\beta{P}}{3B_{m}},
\end{equation}
where $(K_d)_0$ is dimensionless; if $(K_d)_0 = 1.0$, it describes
what \citet{Potgieter-etal1989} called 100\% drifts (i.e., full `weak
scattering' gradient and curvature drifts). Drift velocity components
in terms of $K_{r \theta}$, $K_{\theta r}$, $K_{\phi \theta}$,
$K_{\theta \phi}$ are
\begin{eqnarray} \label{drift1}
\left\langle \mathbf{v}_{d}\right\rangle _{r} &=&-\frac{\textrm{A}}{r\sin
\theta }\frac{\partial }{\partial \theta }(\sin \theta K_{\theta r}),  
\nonumber \\
\left\langle \mathbf{v}_{d}\right\rangle _{\theta } &=&-\frac{\textrm{A}}{r}%
\left[ \frac{1}{\sin \theta }\frac{\partial }{\partial \phi }(K_{\phi \theta
})+\frac{\partial }{\partial r}(rK_{r\theta })\right] ,  \nonumber \\
\left\langle \mathbf{v}_{d}\right\rangle _{\phi } &=&-\frac{\textrm{A}}{r}%
\frac{\partial }{\partial \theta }(K_{\theta \phi }), 
\end{eqnarray}
with $A = \pm 1$; when this value is positive (negative) an $A > 0$
($A < 0$) polarity cycle is described. The polarity cycle around 2009
is indicated  by $A < 0$, as was also the case for the years around
1965 and 1987. During  such a cycle, positively charged CRs are
drifting into the inner heliosphere  mostly through the equatorial
regions, thus having a high probability of encountering the wavy HCS. 

Idealistic global drift patterns of galactic CRs in the heliosphere are 
illustrated in Figure~\ref{fig7} for positively charged particles in
an $A > 0$ and $A < 0$ magnetic polarity cycle respectively, together
with a wavy HCS as expected during solar minimum conditions. Different
elements of the diffusion tensor are also shown in the left panel with
respect to the HMF spiral for illustrative purposes. 

\epubtkImage{Figure7.png}{%
\begin{figure}[htb]
 \centerline{\includegraphics[width=0.9\textwidth]{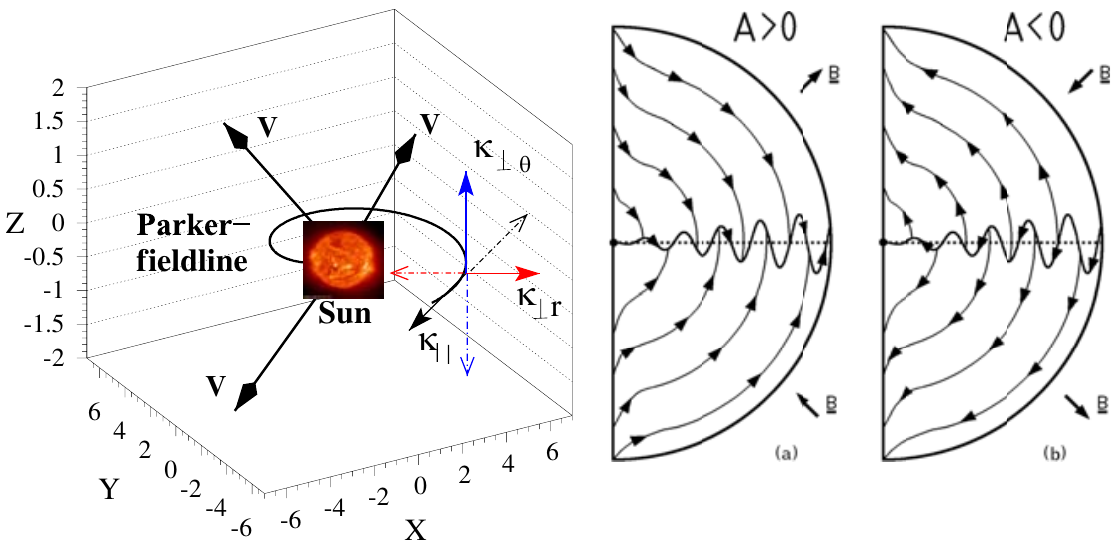}}
  \caption{The parallel and two perpendicular diffusion orientations,
    indicated by the corresponding elements of the diffusion tensor,
    are shown with respect to the HMF spiral direction (left) for
    illustrative purposes. The arrows with V indicate the radially
    expanding solar wind (convection). Idealistic global
    drift patterns of positively charged particles in an $A > 0$ and
    $A < 0$ magnetic polarity cycle are schematically shown in the
    right panel, together with a wavy HCS as expected during solar
    minimum  conditions. Image reproduced by permission from
    \citet{HeberPotgieter2006}, copyright by Springer.}
  \label{fig7}
\end{figure}}

A formal and fundamental description of global curvature, gradient and 
current sheet drifts in the heliosphere is still unsettled. The spatial and 
rigidity dependence of $K_{d}$ is entirely based on the assumption of 
weak-scattering. A deviation from this weak scattering form is given by 
\begin{equation}
\label{eq:kappa_drift}
K_d = (K_{d})_{ws}\frac{\left(\frac{P}{P_{d0}}
\right)^2}{1+\left(\frac{P}{P_{d0}}\right)^2}.
\end{equation}
This means that below $P_{d0}$ (in GV) particle drifts are
progressively reduced with respect to the weak scattering case. This
is required to explain the small latitudinal gradients at low
rigidities observed by Ulysses \citep{HeberPotgieter2006,
  HeberPotgieter2008, DeSimone-etal2011}. This reduction is in line
with what \citet{Potgieter-etal1989}, \citet{Webber-etal1990},
\citet{FerreiraPotgieter2004}, and \citet{Ndiitwani-etal2005} found
when describing modulation in terms of drifts with a HCS tilt angle
dependence in numerical models. Theoretical arguments and numerical
simulations have been presented requiring the reduction of particle
drifts, in particular with increasing solar activity. For a summary of
the essence of the problem, see \citet{TautzShalchi2012}. This aspect
must fit into the picture where Ulysses observations of CR latitudinal
gradients especially at lower energies, require particle drifts to be
reduced. It thus remains a theoretical challenge to explain why
reduced drifts in the heliosphere is needed to explain some of these
major CR observations, and what type of particle drifts apart from
gradient and curvature drifts may occur in the heliosphere.

\subsection{Gradient, curvature, and current sheet drifts}

The realization that particle drifts could not be neglected in the
solar modulation of CRs was elevated by the development of numerical
models, which reached sophisticated levels already in the late 1980s
and early 1990s including a full tensor. The importance of particle
drifts in the heliosphere was at first discussed only theoretically
but it was soon discovered compellingly in various existing CR
observations. It should be noted that it was rather fortuitous that
the sharp peak in the 1965 intensity-time profiles of CRs was followed
by a really flat intensity-time profile in the 1970s, not to repeat
again so evidently. The mini-modulation-cycle around 1974 had little
to do with drifts \citep[e.g.,][]{Wibberenz-etal2001}. The recent $A <
0$ polarity cycle also did not produce such a sharp peak as during
previous $A < 0$ cycles as shown in Figure~\ref{fig6}. See also the
review by \citet{Potgieter2013}.

Convincing theoretical arguments for the importance of particle drifts were 
presented by \citet{Jokipii-etal1977} and later followed by persuasive 
numerical modeling \citep[e.g.,][]{JokipiiKopriva1979,
JokipiiThomas1981, KotaJokipii1983, PotgieterMoraal1985}, which
illustrated that gradient and curvature drifts could cause charge-sign
dependent modulation and a 22-year cycle. The main reason for
this to occur is that the solar magnetic field reverses polarity every
$\sim$~11~years so that galactic CRs of opposite charge will reach
Earth from different heliospheric directions. This also makes the wavy
HCS of the HMF a very important modulation feature with its tilt angle
\citep{Hoeksema1992} being a very useful modulation parameter. When
protons drift inwards mainly through the equatorial regions of the
heliosphere ($A < 0$ polarity cycles) they encounter the dynamic HCS
and get progressively reduced by its increasing waviness as solar
activity surges. This produces the sharp peaks in the galactic CR
intensity-time profiles whereas during the $A > 0$ cycles the profiles
are generally flatter \citep[for recent updates see,
  e.g.,][]{Potgieter2011,  Krymsky-etal2012, Potgieter2013}. The
effect reverses for negatively charged galactic CRs causing
charge-sign dependent effects as reviewed by, e.g.,
\citet{HeberPotgieter2006, HeberPotgieter2008} and \citet{Kota2012}.

Figure~\ref{fig8} is an illustration of the computed intensity
distribution caused by drifts for the two HMF polarity cycles, in this
case for 1~MeV/nuc anomalous oxygen in the meridional plane of the
heliosphere. The position of the $r_{\mathrm{TS}}$~=~90~AU is
indicated by the white dashed line. The contours are significantly
different in the inner heliosphere, with the intensity reducing
rapidly towards Earth. The distribution is quite different beyond the
TS, where the region of preferred acceleration of these ACRs is
assumed to be away from the TS, closer to the HP near the equatorial
plane, positioned at 140~AU, for illustrative purposes
\citep{Strauss-etal2010a, Strauss-etal2010b}. 

\epubtkImage{Figure8.png}{%
\begin{figure}[htb]
 \centerline{\includegraphics[width=0.9\textwidth]{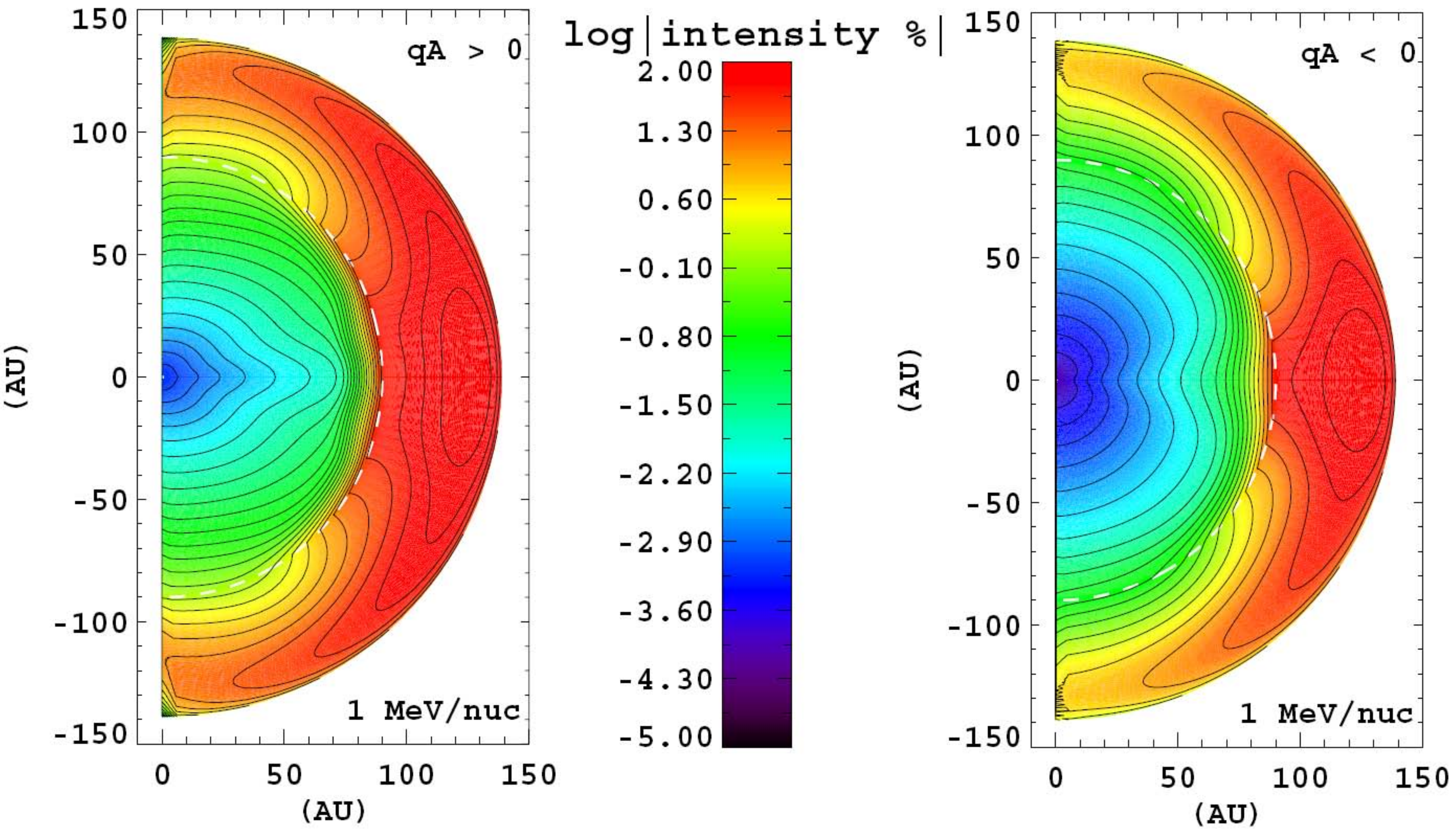}}
  \caption{An illustration of the computed intensity distribution
    caused by drifts for the two HMF polarity cycles, in this case for
    1~MeV/nuc ACR oxygen in the meridional plane of the
    heliosphere. The position of the TS at 90~AU is indicated by the
    white dashed line. Note how the coloured contours differ in the
    inner heliosphere for the two cycles and how the intensity
    decreases towards Earth, and how the distribution is quite
    different beyond the TS. In this case, the region of preferred
    acceleration for these ACRs is assumed near the equatorial plane
    and close to the HP at 140~AU. Image reproduced by permission
    from~\citet{Strauss-etal2011b}, copyright by COSPAR.}
  \label{fig8}
\end{figure}}

Another indication of the role of gradient and curvature drifts came in the 
form of a 22-year variation in the direction of the daily anisotropy vector 
in the galactic CR intensity as measured by NMs from one 
polarity cycle to another \citep{Levy1976, PotgieterMoraal1985}. 
\citet{Potgieter-etal1980} also discovered a 22-year cycle in the differential 
response function of NMs used for geomagnetic latitude surveys 
at sea-level in 1965 and 1976 \citep[see also][]{Moraal-etal1989}. 

Some of the key outcomes of gradient, curvature and 
current sheet drifts as applied to the solar modulation of CRs, are:

\begin{enumerate}

\item Particles of opposite charge will experience solar modulation
  differently because they sample different regions of the heliosphere
  during the same polarity epoch before arriving at Earth or at
  another observation point. Particle drift effects inside the
  heliosheath, on the other hand, may be different from upstream
  (towards the Sun) of the TS.

\item A well-established 22-year cycle occurs in the solar modulation
  of galactic CRs, which is not evident in other standard proxies for
  solar activity (see Figure~\ref{fig6}). This is also evident in the
  directional changes of the diurnal anisotropy vector with every HMF
  polarity reversal \citep[e.g.,][]{PotgieterMoraal1985,
    Nkosi-etal2008, NgobeniPotgieter2010} and from the changes in
  differential response functions of NMs obtained during
  geomagnetic latitude surveys. 
	
\item The wavy HCS plays a significant role in establishing the
  features of this 22-year cycle in the solar modulation of CRs.

\item Cosmic ray latitudinal and radial intensity gradients in the
  heliosphere are significantly different during the two HMF polarity
  cycles and repeated ideal modulation conditions will display a
  22-year cycle \citep[e.g.,][]{HeberPotgieter2006,
    HeberPotgieter2008, Potgieter-etal2001, DeSimone-etal2011}. An
  illustration is shown in Figure~\ref{fig9} of how the computed radial
  gradients change with kinetic energy for the two polarity cycles at
  different positions in the heliosphere. In Figure~\ref{fig10} the difference
  caused by drifts in the computed latitudinal proton gradients
  between the two polarity cycles is shown as a function of rigidity
  in comparison with the observed gradient from PAMELA and Ulysses for
  the period 2007 \citep{DeSimone-etal2011}.

\item Cosmic ray proton spectra are softer during $A > 0$ cycles so
  that below 500~MeV the $A > 0$ solar minima spectra are always
  higher than the corresponding $A < 0$ spectra
  \citep{Beatty-etal1985, PotgieterMoraal1985}. This means that the
  adiabatic energy losses that CRs experience in $A < 0$ cycles are
  somewhat different than during the $A > 0$ cycles
  \citep{Strauss-etal2011a, Strauss-etal2011c}. This also causes the
  proton spectra for two consecutive solar minima to cross at a few
  GeV \citep{ReineckePotgieter1994}.

\item Drift effects are not necessarily the same during every 11-year
  cycle, not even at solar minimum because the recent solar minimum
  was different than other $A < 0$ cycles. The intriguing interplay
  among the major modulation mechanisms changes with the solar cycles
  \citep{Potgieter-etal2013}.

\item Drifts also influence the effectiveness by which
  galactic CRs are re-accelerated at the solar wind TS
  \citep[e.g.,][]{Jokipii1986, PotgieterLangner2004a}. However, it is unclear
  whether particle drifts play a significant role in the heliosheath
  and to what extent drift patterns are different than in the inner
  heliosphere.

\end{enumerate}

Particle drifts, as a modulation process, has made a major impact on solar 
modulation theory and was eclipsed only in the early 2000s when major 
efforts were made to understand diffusion theory better together with the 
underlying heliospheric turbulence theory (as reviewed by
\citealp{Bieber2003} and \citealp{McKibben2005}). It was also then
finally realized that particle drifts do not dominate solar modulation
over a complete solar cycle but that it is part of an intriguing
interplay among basically four mechanisms and that this play-off
changes over the solar cycle and from one cycle to another. The latest
prolonged solar minimum brought additional insight in how this
interplay can change as the Sun keeps on surprising us
\citep{Potgieter-etal2013}.

\epubtkImage{Figure9.png}{%
\begin{figure}[htbp]
 \centerline{\includegraphics{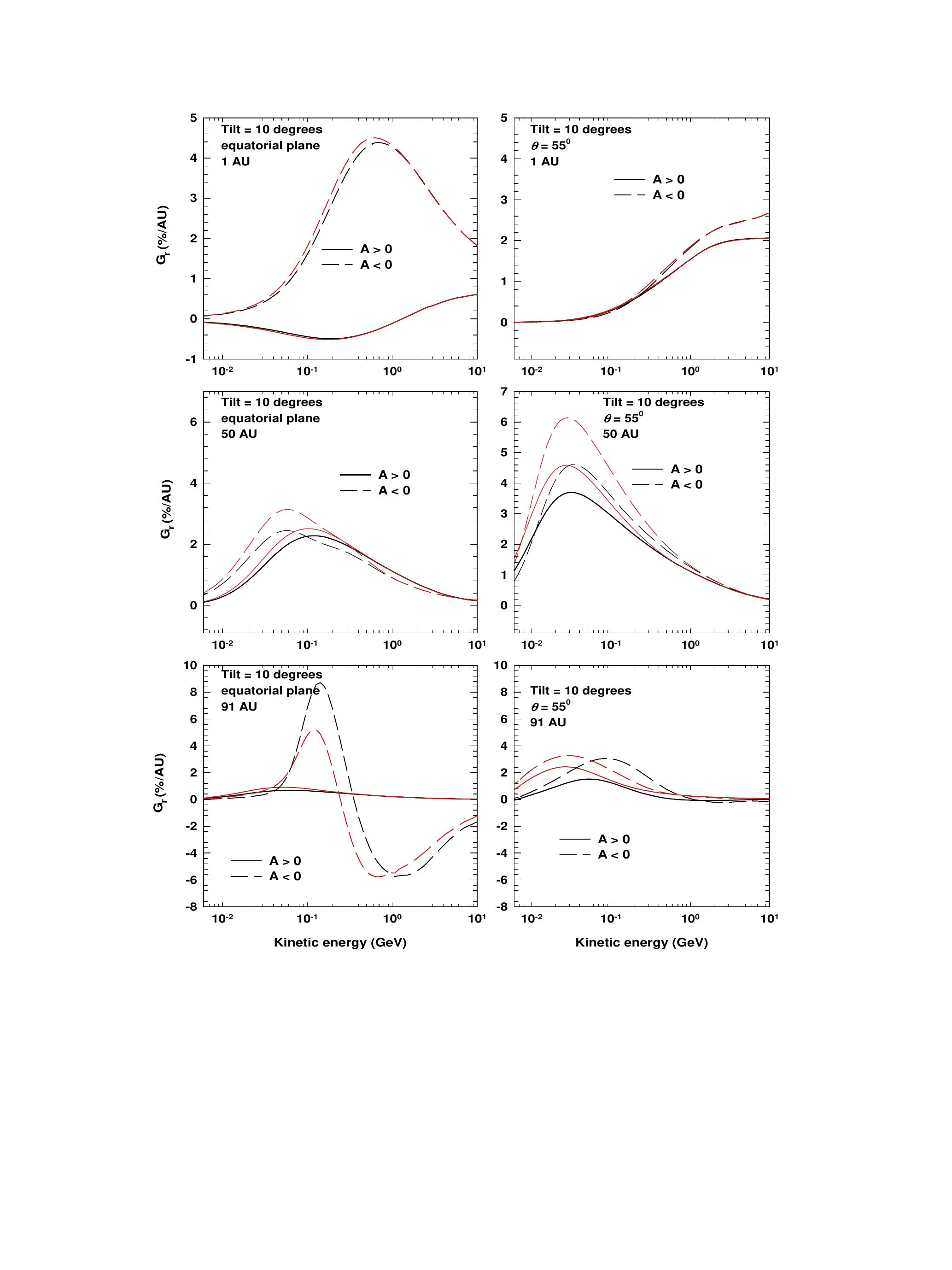}}
  \caption{\emph{Left panels:} Computed radial gradients for galactic
    protons, in \% AU\super{-1}, as a function of kinetic energy for
    both polarity cycles and for solar minimum conditions in the
    equatorial plane at 1, 50, and 91~AU, respectively (top to bottom
    panels). \emph{Right panels:} Similar but at a polar angle
    $\theta$~=~55\textdegree. Two sets of solutions are shown in all
    panels, first without a latitude dependence (black lines) and
    second with a latitude-dependent compression ratio for the TS
    (red lines). In this case, the TS is at 90~AU and the HP is at
    120~AU. Image reproduced by permission from
    \citet{NgobeniPotgieter2010}, copyright by COSPAR.}
  \label{fig9}
\end{figure}}

\epubtkImage{Figure10.png}{%
\begin{figure}[htb]
 \centerline{\includegraphics[width=0.75\textwidth]{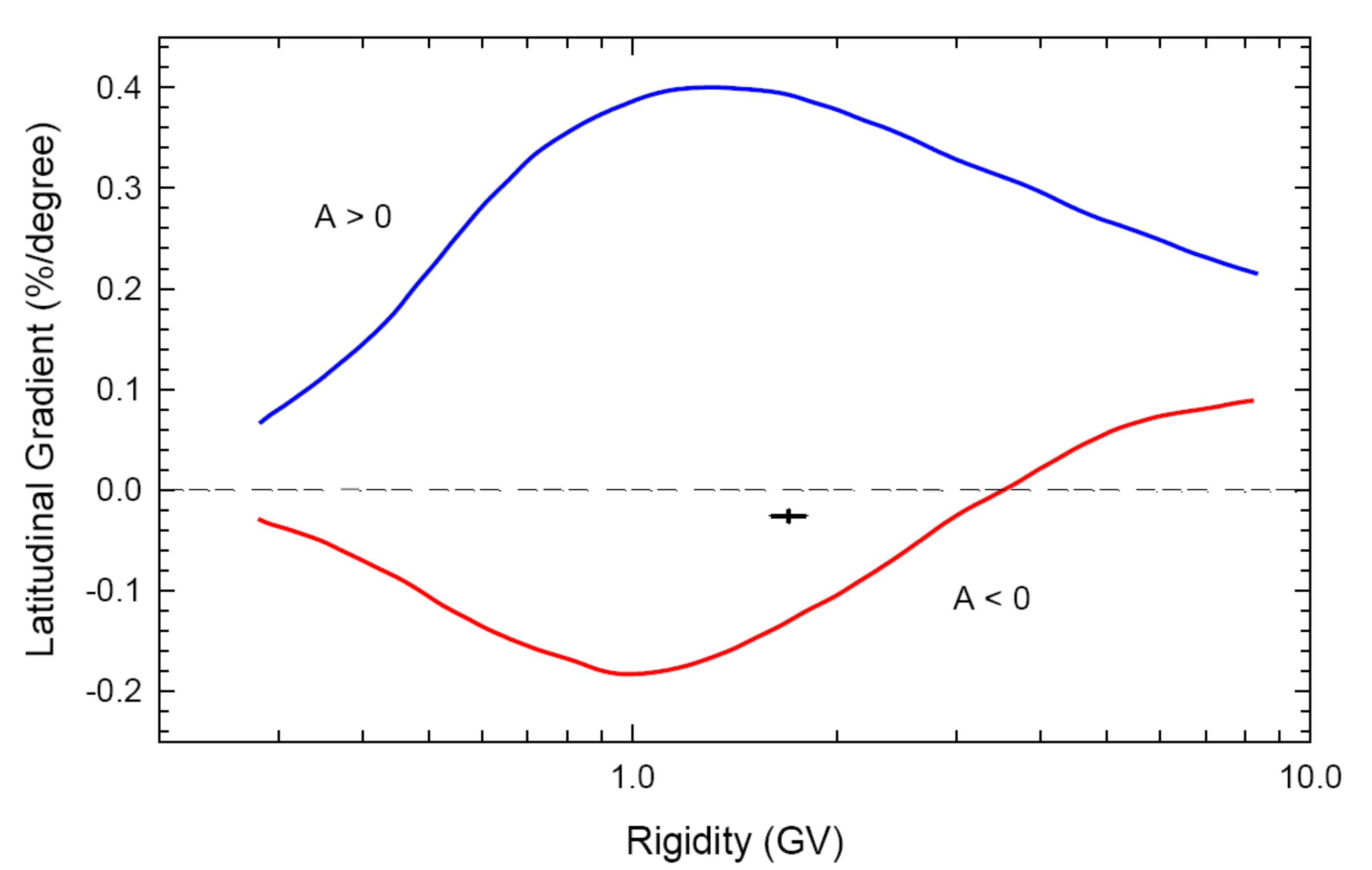}}
  \caption{The difference caused by drifts in the computed latitudinal
    gradients (\% degree\super{-1}) for protons in the inner
    heliosphere for the two HMF polarity cycles as a function of rigidity
    \cite{Potgieter-etal2001}. Marked by the black line-point is the
    latitudinal gradient calculated from a comparison between Ulysses
    and PAMELA observations for 2007. Image reproduced by permission
    from \citet{DeSimone-etal2011}.}
  \label{fig10}
\end{figure}}

\subsection{Aspects of diffusion and turbulence theory relevant to
  solar modulation}

Progress has been made over the last decade to improve the understanding of 
heliospheric turbulence, in particular to overcome some of the deficiencies 
of standard scattering theories such as quasi-linear theory (QLT). New 
approaches have taken into account the dynamical 
character and the three-dimensional geometry of magnetic field fluctuations. 
Historically, it is important to note that in order to reconcile 
observations with the theoretical mean free paths, \citet{Bieber-etal1994} 
advocated a composite model for the turbulence, which consists of
$\sim$~20\% slab and $\sim$~80\% 2D fluctuations. At high rigidities
the simpler theory can be used to describe particle transport parallel
to the mean field, but at low rigidities the dynamical theory
predicts a much more efficient scattering, which reduces the parallel
mean path compared to standard QLT for ions, but gives large values
for electrons. \citet{Droge2005} showed that standard QLT underestimated
mean free paths of low energy electrons by almost two orders of
magnitude even with a corrected slab fraction for magnetic
turbulence. A general conclusion seems to be that the parallel mean
free path of CRs is a key input parameter for CR transport but not
nearly the only one. \citet{TeufelSchlickeiser2002}, amongst others,
produced analytical formulae for the parallel mean free path as a
function of rigidity at Earth for CR protons and electrons, which
differ significantly at lower energies for these two species. 

The diffusion of particles across the mean HMF is an important area of 
study. It is very nearly radially directed towards the Sun for distances 
beyond 10~AU. For CRs to reach Earth, they must cross the mean HMF, 
otherwise it would require large parallel mean-free paths and subsequently 
very large anisotropies, which are not observed. Cross-field diffusion 
remains puzzling, with the dimensionality of the underlying turbulence of 
critical importance. Conceptually, it is understood that the local transport 
across individual magnetic lines and the motion of particles along spatially 
meandering magnetic field lines can both occur. Simulations of particle 
transport in irregular magnetic fields were performed by, e.g.,
\citet{GiacaloneJokipii1994, GiacaloneJokipii1999} and
\citet{QinShalchi2012} with insightful results. Later, a new theory
with different assumption, was developed by \citet{Matthaeus-etal2003}
and \citet{Bieber-etal2004}. This nonlinear guiding-center (NLGC)
theory is promising for understanding perpendicular transport. This
topic is evidently a very specialized field of theoretical research
and for these details the reader is referred to the reviews by
\citet{Bieber2003} and \citet{Giacalone2011}. Developing a full ab
initio theory of CR transport and modulation, especially for
perpendicular diffusion, by integrating turbulence quantities with
diffusion coefficients throughout the heliosphere is a work in
progress \citep[e.g.,][]{Burger-etal2000,
  Pei-etal2010}. Unfortunately, the \textit{ab initio} approach cannot
as yet produce elements of the diffusion tensor than can explain all
observations consistently when used in global modulation models so
that phenomenological approaches remain very useful. For comprehensive
monographs on transport theory, see \citet{Schlickeiser2002} and
\citet{Shalchi2009}. 

For elaborate reviews on turbulence effects in the heliosphere and on the 
fundamental process of reconnection and acceleration of particles, see
\citet{FiskGloeckler2009, FiskGloeckler2012},
\citet{Lazarian-etal2012}, \citet{Lee-etal2012},
\citet{MatthaeusVelli2011}, and several other contributions in the
same issue of \textit{Space Science Reviews}.

\subsection{Development of numerical modulation models}

The wide-ranging availability of fast computers has brought significant 
advances in numerical modeling of solar modulation. In-situ observations 
have always been limited so that numerical modeling plays an important role 
to broaden our understanding of solar modulation. For comprehensive global 
modeling it is essential to have a sound transport theory, reliable
numerical schemes with appropriate boundary conditions, local interstellar spectra 
as initial input spectra and properly considered  transport parameters. Furthermore, a 
basic knowledge of the solar wind and HMF and how they change throughout the 
whole heliosphere is required. This is quite a task and takes major efforts 
to accomplish.

\citet{Fisk1971} developed the first numerical solution of the TPE by
assuming a steady-state and spherical symmetry, i.e., a
one-dimensional (1D) model with radial distance as the only spatial
variable, and of course an energy dependence. Later, a polar angle
dependence was included to form an axisymmetric (2D) steady-state
model without drifts \citep{Fisk1976}. In 1979,
\citet{JokipiiKopriva1979} and \citet{Moraal-etal1979} presented their
separately developed 2D steady-state models including gradient and
curvature drifts for a flat HCS. The first 2D models to emulate the
waviness of the HCS were developed by \citet{PotgieterMoraal1985} and
\citet{BurgerPotgieter1989}. Three-dimensional (3D)
steady-state models including drifts with a 3D wavy HCS were developed
by \citet{JokipiiThomas1981} and \citet{KotaJokipii1983} with similar models developed later by
\citet{Hattingh-etal1997} and
\citet{Gil-etal2005}. \citet{HaasbroekPotgieter1998}, \citet{Fichtner-etal2000}
and \citet{Ferreira-etal2001} independently developed steady-state
models including the Jovian magnetosphere as a source of low-energy electrons. 

The first 1D time-dependent model (numerically thus three dimensions:
radial distance, energy, and time) was developed by
\citet{PerkoFisk1983}. Extension to two spatial dimensions was done by
\citet{leRouxPotgieter1990} including drifts and the effect of
outwards propagating GMIRs at large radial distance thus enabling the
study of long-term CR modulation effects
\citep{Potgieter-etal1993}. \citet{Fichtner-etal2000} developed a 3D
time-dependent model for electrons, but approximated adiabatic cooling
of electrons at lower energies by doing a momentum averaging of the
Parker TPE. 

The inclusion of the effects of a heliospheric TS was done by
\citet{Jokipii1986} who developed the first 2D time-dependent,
diffusion shock acceleration model. \citet{PotgieterMoraal1988}
demonstrated that it was possible to include shock acceleration in a
steady-state spherically symmetric model by specifying the appropriate
boundary conditions with regard to the CR streaming and spectra at the
TS. This model was expanded to 2D by \citet{Potgieter1989}. Later on
2D shock acceleration models with discontinuous and continuous
transitions of the solar wind velocity across the TS were developed to
study ACRs \citep[e.g.,][]{SteenbergMoraal1996, leRoux-etal1996,
  LangnerPotgieter2004a,
  LangnerPotgieter2004b}. \citet{HaasbroekPotgieter1998} developed a
model that could handle all possible geometrical elongation of the
heliosphere by assuming a non-spherical heliospheric boundary
geometry. All the above mentioned numerical models were developed
using the Crank--Nicholson and Alternating Direction Implicit (ADI)
schemes, with some deviations depending on the complexity of the
studied physics \citep{Fichtner2005}. 

The intricacy of the TPEs applicable to CR modulation may cause numerical 
models to have notorious problems with instability when solving in higher 
(five numerical) dimensions. Solving the relevant TPEs by means of 
stochastic differential equations (SDEs) has become therefore quite popular 
after earlier noteworthy attempts were not truly appreciated 
\citep[e.g.,][]{Fichtner-etal1996, Gervasi-etal1999, Yamada-etal1999, Zhang1999}. This 
method has several advantages, most notably unconditional numerical 
stability and an independence of a spatial grid size. The method is highly 
suitable for parallel processing. Recently, models developed around SDEs 
have become rather sophisticated \citep[e.g.,][]{Kopp-etal2012} but not always 
focussed on additional insight into CR modulation. It has been illustrated 
that additional physical insights can be extracted from this approach, e.g., 
\citet{FlorinskiPogorelov2009}, \citet{Strauss-etal2011a,
  Strauss-etal2011c, Strauss-etal2012c, Strauss-etal2012d}, and
\citet{Bobik-etal2012}. Some aspects are discussed next.

\subsubsection{Illustrations of SDE based modeling}

The following figures are illustrations of CR modulation based on the SDE 
approach to solar modulation as examples of what can be done additionally to 
the `standard' numerical approaches. Figure~\ref{fig11} shows
pseudo-particle traces for 100~MeV protons in the $A < 0$ cycle and
varying values of the tilt angle $\alpha$, projected onto the
meridional plane in comparison with a projection of the wavy HCS
onto the same plane. These particles propagate mainly in latitudes
covered by the HCS, and their transport is greatly affected by the
waviness of the HCS but mainly for small values of $\alpha$, as
diffusion disrupts the drift pattern. Diffusion can in principle
almost wipe out these drift patterns as illustrated by
\citet{Strauss-etal2012c}. This realistic picture of the combination
of global and HCS drifts is in sharp contrast to the idealist picture
shown in Figure~\ref{fig7}. Studies that are focusing on drifts as the
sole modulation process give scenarios that are unrealistic
\citep[e.g.,][]{Roberts2011}. Perfect drift dominated CR transport does
not exist because CR modulation essentially is a convection-diffusion
process. As mentioned, inspection of  Equation~(\ref{eq:parker_tpe})
shows that when changing diffusion the CR intensity gradients are
changed directly, which subsequently changes the effects of drifts
implicitly \citep{Potgieter-etal2013} while explicit changes in drifts
and consequently of drift effects are done by changing the drift
coefficient directly (Equation~(\ref{eq:kappa_d_weak_scattering})).

In Figure~\ref{fig12} binned propagation times for 100~MeV galactic electrons between 
Earth and the HP (at 140~AU) for three scenarios: the $A < 0$ polarity cycle 
(left panel), the $A > 0$ cycle (middle panel), and the non-drift case 
(right panel). For the latter, the propagation time follows a normal 
distribution, peaking at $\sim$~400~days, while for the different drift 
cycles, the distribution tends to be more Poisson like (CR cannot reach 
Earth infinitely fast) with lower propagation times. The reason for the 
shorter propagation times, $\sim$~240~days for the $A > 0$ cycle and
$\sim$~110~days for the $A < 0$ cycle, is that drifts cause a
preferred direction of transport for these CR electrons, thereby
allowing them to propagate faster to Earth. The propagation times for
the $A < 0$ cycle is shorter than for the $A > 0$ cycle because these
electrons can easily escape through the heliospheric poles than
drifting along the HCS in the $A > 0$ cycle \citep{Strauss-etal2011a,
  Strauss-etal2011c}.

The propagation times and energy loss of 100~MeV protons propagating
from the HP (at 100~AU) to Earth as a function of the HCS tilt angle
$(\alpha)$ for $A < 0$ polarity cycles of the HMF are shown in
Figure~\ref{fig13}. Note that the increase in propagation time
significantly slows down above $\alpha$~=~40\textdegree. The energy
loss levels off above $\alpha$~=~40\textdegree. See
\citet{Strauss-etal2011a, Strauss-etal2011c} for additional
illustrations.

\epubtkImage{Figure11.png}{%
\begin{figure}[htbp]
 \centerline{\includegraphics[width=0.8\textwidth]{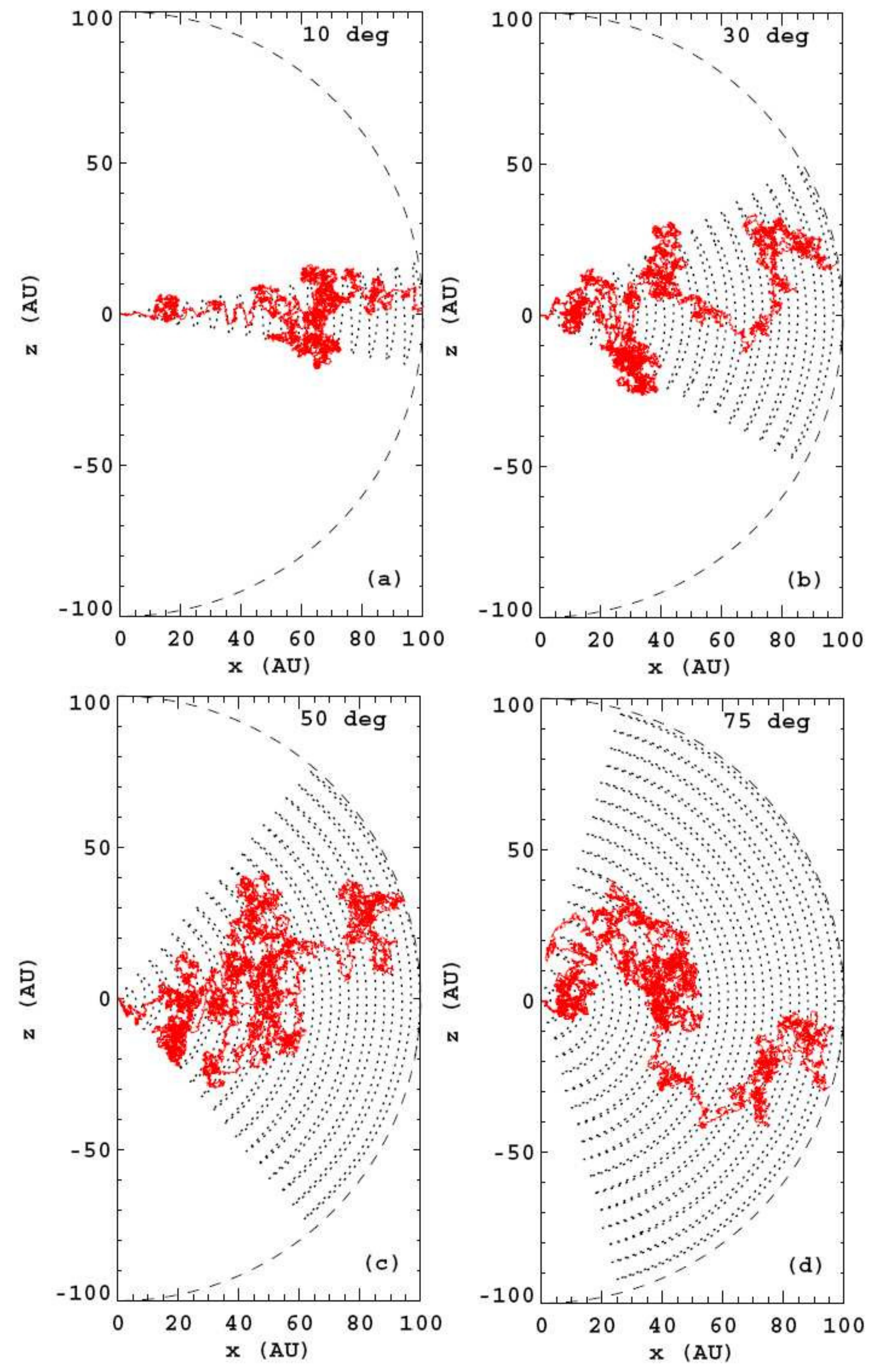}}
  \caption{Pseudo-particle traces (trajectories) for galactic protons
    in the $A < 0$ HMF cycle projected onto the meridional plane for
    four values of the HCS tilt angle, shown as red lines. The HP
    position is indicated by the dashed lines, while the dotted lines
    show a projection of the waviness of HCS onto the same plane. The
    simulation is done for 100~MeV protons. Image reproduced by
    permission from \citet{Strauss-etal2012c}, copyright by Springer.}
  \label{fig11}
\end{figure}}

\epubtkImage{Figure12.png}{%
\begin{figure}[htb]
 \centerline{\includegraphics[width=0.9\textwidth]{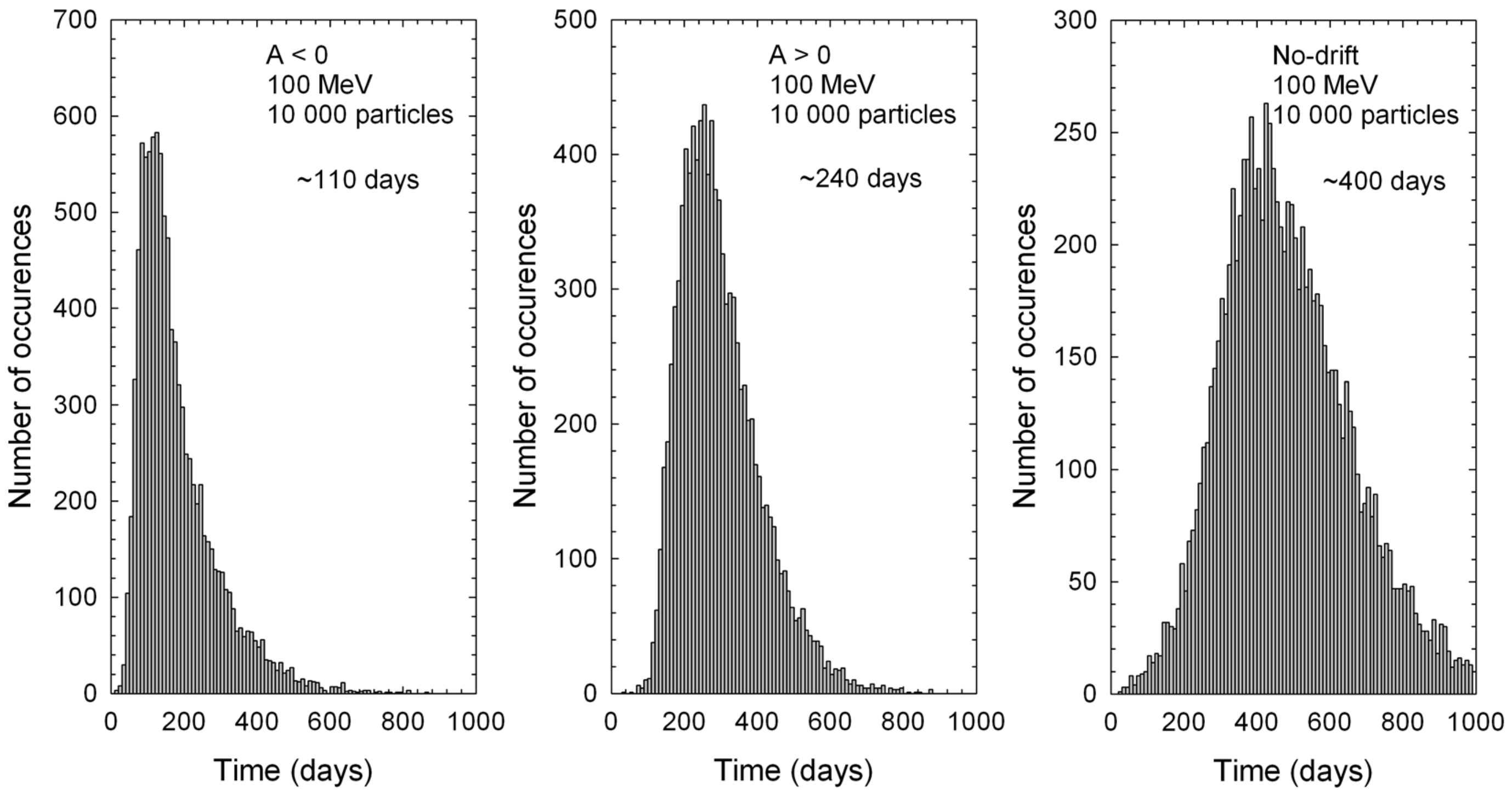}}
  \caption{Binned propagation times for galactic electrons released at
    Earth at 100~MeV for the $A < 0$ (left panel), $A > 0$ (middle panel),
    and for the no-drift scenarios (right panel). For each computation
    10000 particle trajectories were integrated using the SDE approach
    to modulation modeling. Image reproduced by permission from
    \citet{Strauss-etal2011a}, copyright AAS.}
  \label{fig12}
\end{figure}}

\epubtkImage{Figure13.png}{%
\begin{figure}[htb]
 \centerline{\includegraphics[width=0.6\textwidth]{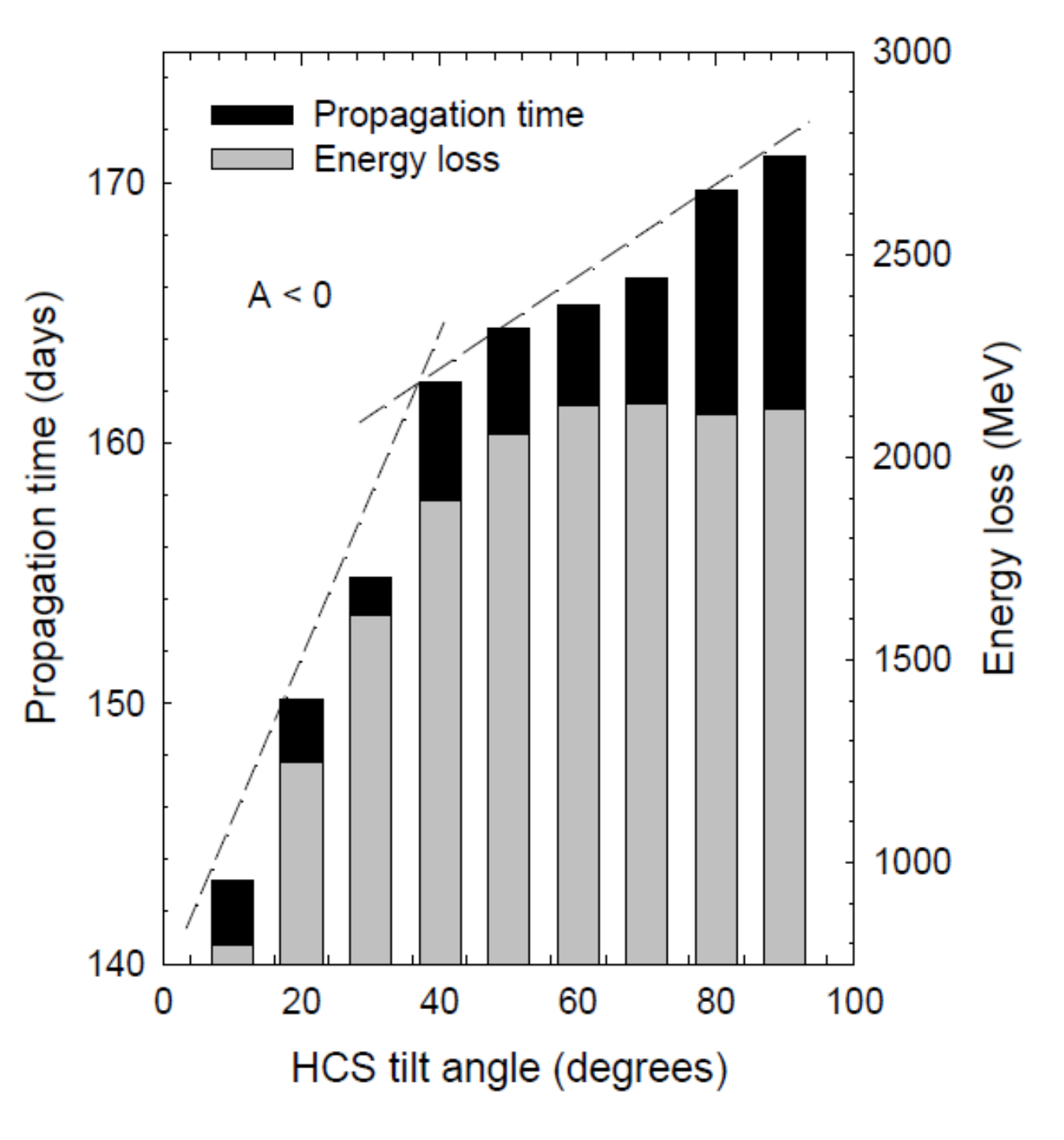}}
  \caption{The propagation times and energy loss of 100~MeV protons
    propagation from the HP to Earth as a function of the HCS tilt
    angle ($\alpha$) for $A < 0$ polarity cycles of the HMF. Note the
    change in propagation time at $\alpha$~=~40\textdegree\ and how
    the energy loss levels off above $\alpha$~=~50\textdegree. Image
    reproduced by permission from~\citet{Strauss-etal2012c}, copyright
    by Springer.}
  \label{fig13}
\end{figure}}

\clearpage
\subsection{Charge-sign dependent modulation}

It was not until around 1976\,--\,1978 that particle drifts were considered 
seriously as a competitive modulation process, in addition to the 
conventional convective, diffusive, and adiabatic energy loss processes. It 
was experienced as controversial when introduced and it took 10~years to 
become widely accepted. Even today the full extent of its relevance and 
importance over a complete solar activity cycle is debated. See, as an 
example, the critical review by \citet{Cliver-etal2011}. The early stages of 
this theoretical development, and the status of the research round that 
time, were reviewed by \citet{Quenby1984}. The realization also came that the 
only way to understand the full scope of particle drifts on galactic CR 
modulation was with numerical modeling. Therefore, since the early 1980s 
increasingly more sophisticated numerical models had been introduced that 
kept on improving as discussed above. See \citet{Potgieter2013} for a
full review on this topic, parts of which is also repeated here.

Simultaneous measurements of CR electrons and positrons (protons and 
anti-protons) serve as a crucial test of our present understanding of how 
large charge-sign dependent modulation in the heliosphere is, as a function 
of energy and position over a complete solar activity cycle. It is expected 
that the effects of drifts on CRs should become more evident closer to 
minimum solar activity. Observations of CR particles and their 
anti-particles have been done over the years and are presently been made 
simultaneously by PAMELA \citep[e.g.,][]{Boezio-etal2009, Sparvoli-etal2012, 
BoezioMocchiutti2012} and the AMS-02 mission
\citep[e.g.,][]{Battiston-etal2010}. PAMELA is a satellite-borne
experiment designed for cosmic-ray antimatter studies. The instrument
is flying on board the Russian Resurs-DK1 satellite since June 2006,
following a semi-polar near-Earth orbit. For an overview of
balloon-based observations, see \citet{Seo2012}.

Before these simultaneous measurements, charge-sign dependent solar
modulation was mostly studied using `electrons', which was actually
the sum of electrons and positrons, together with CR protons and
helium of the same rigidity. The first sturdy observational evidence
of charge-sign dependent solar modulation was reported by
\citet{Webber-etal1983} and modelled by \citet{PotgieterMoraal1985}
using a first generation drift model. This is shown in
Figure~\ref{fig14} as electron spectra during two consecutive solar
minimum modulation periods in 1965 and 1977. The corresponding
charge-sign dependent effect is illustrated in Figure~\ref{fig15},
comparing proton and electron measurements made during two consecutive
solar minimum periods (1965 as $A < 0$ and 1977 as $A > 0$). The ratio of
differential intensities, DI(1977)/DI(1965\,--\,66), is shown for both
electrons and protons as a function of $E$. Evidently,
electrons behaved different from protons during these consecutive
solar minimum epochs, again convincingly revealing a 22-year
modulation cycle and charge-sign dependence.

\epubtkImage{Figure14.png}{%
\begin{figure}[htbp]
 \centerline{\includegraphics[width=0.5\textwidth]{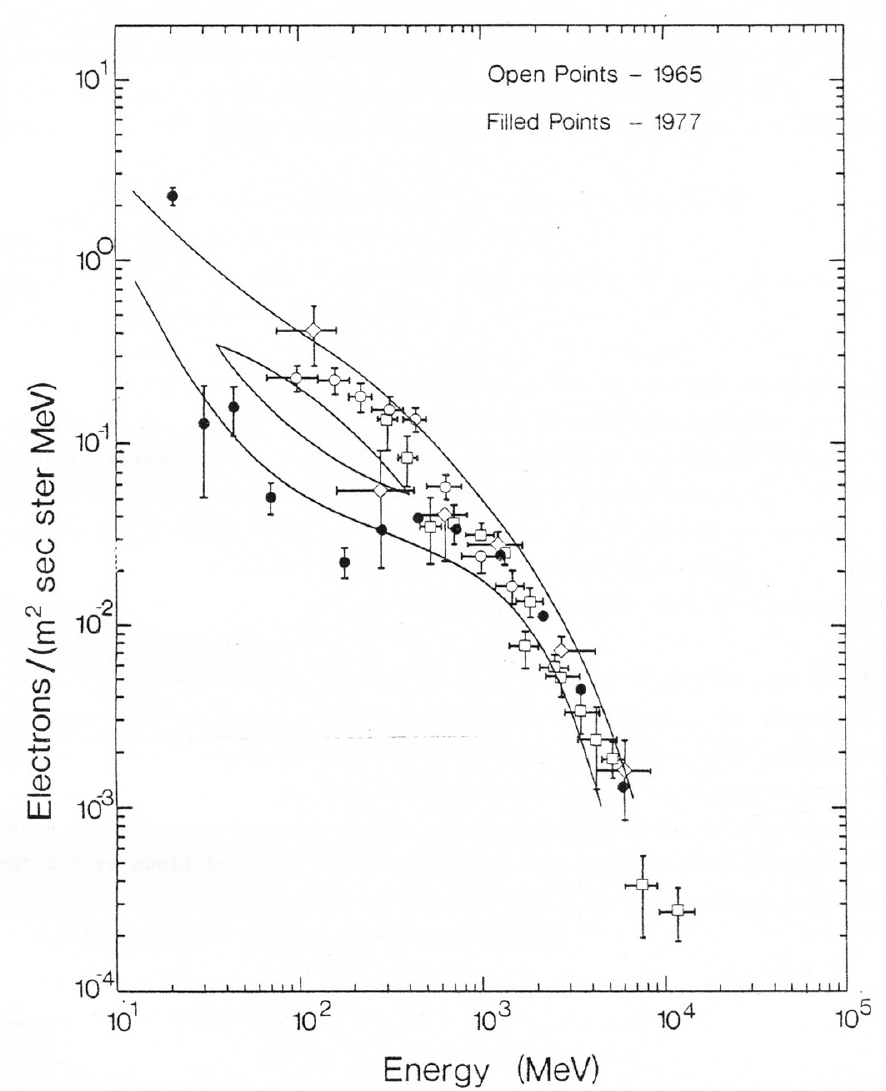}}
  \caption{Galactic CR electron observations for two consecutive solar
    minimum modulation periods in 1965 (open circles) and 1977 (filled
    circles) compared to the predictions of a first generation
    drift-modulation model (band between solid lines) containing
    gradient, curvature, and current sheet drifts. Clearly, a 22-year
    modulation cycle is portrayed \citep[][and
    references therein]{PotgieterMoraal1985}.}
  \label{fig14}
\end{figure}}

\epubtkImage{Figure15.png}{%
\begin{figure}[htbp]
 \centerline{\includegraphics[width=0.6\textwidth]{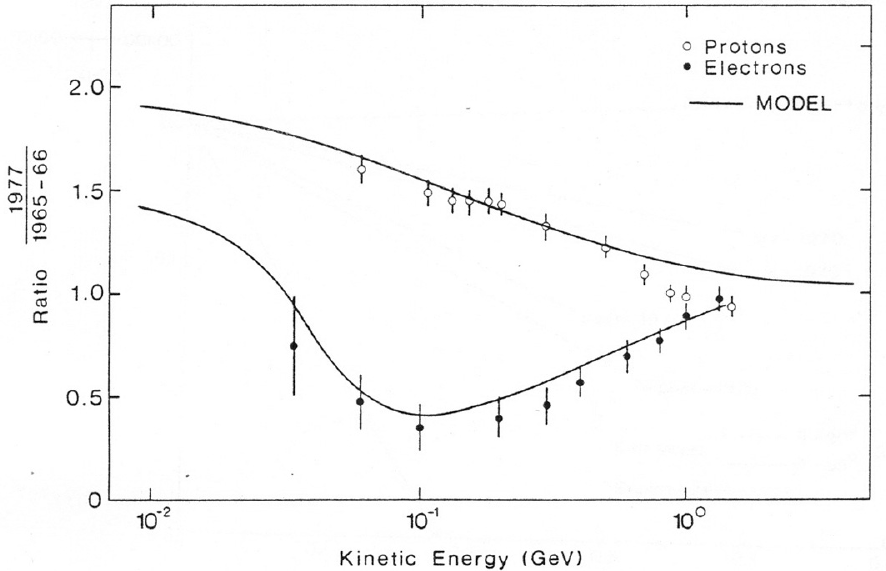}}
  \caption{Ratios of proton and electron measurements for 1977 ($A >
    0$ polarity cycle) to 1965\,--\,66 ($A < 0$ polarity cycle) as a
    function of kinetic energy compared to the predictions made with a
    drift-modulation model illustrating how differently protons behave
    to electrons during two solar minimum periods with opposite solar
    magnetic field polarity \citep[][and references
      therein]{Webber-etal1983, PotgieterMoraal1985}.}
  \label{fig15}
\end{figure}}

A newer generation drift-modulation model was used by \citet{Ferreira2005} to 
illustrate how the modulation of galactic electrons differ form one polarity 
cycle to another as shown in Figure~\ref{fig16} (right panel). For
electrons the influence of particle drifts is evident over an energy
range from 50~MeV to 5~GeV with a maximum effect around 200~MeV to
500~MeV as shown in the left panel. It also illustrates how the drift
effect changes with increasing distance from the Sun.

\epubtkImage{Figure16.png}{%
\begin{figure}[htb]
 \centerline{\includegraphics[width=0.9\textwidth]{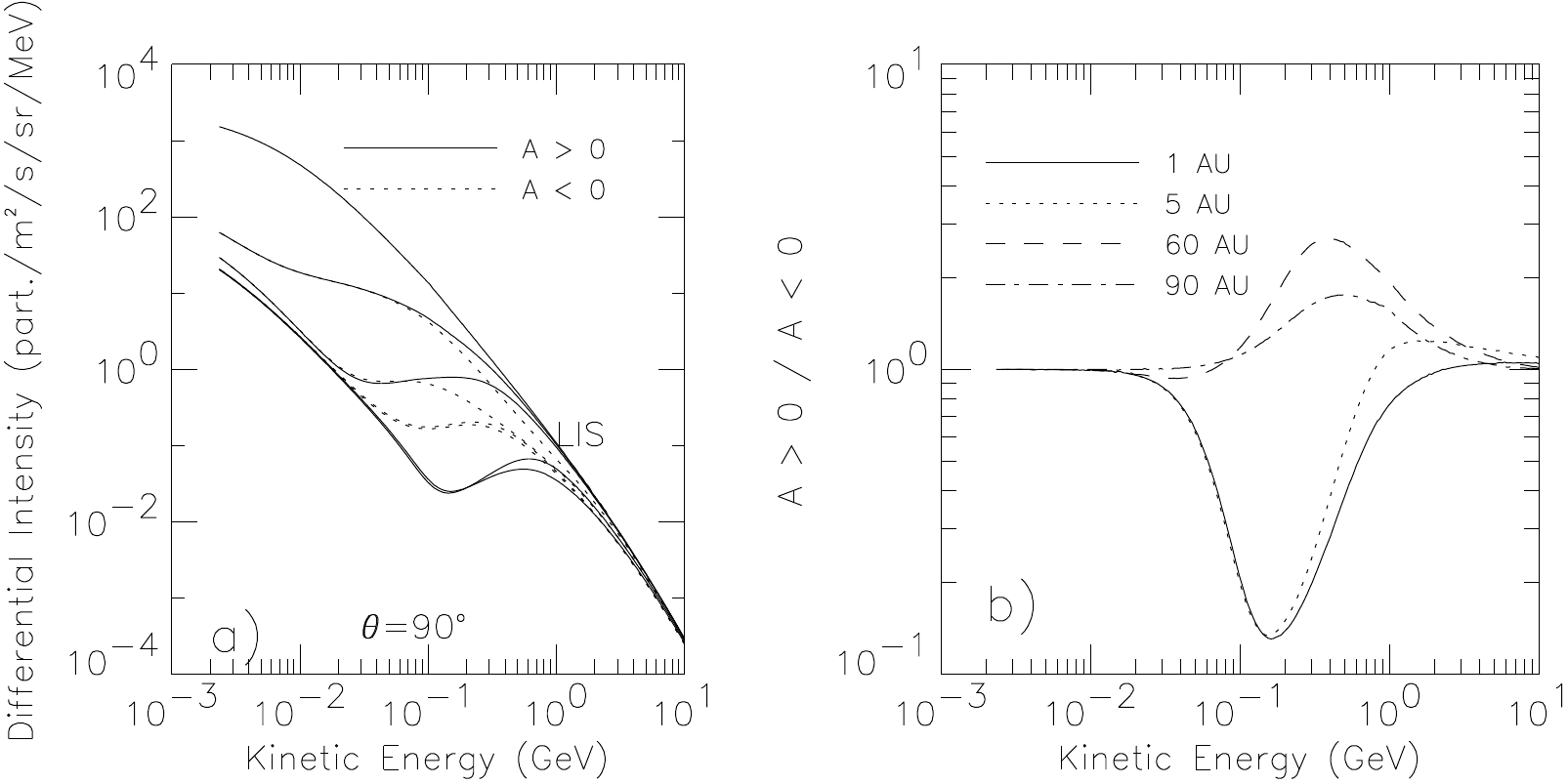}}
  \caption{\emph{Panel~(a):} Computed differential intensities of galactic
    electron at 1, 5, 60 and 90~AU (from bottom to top) in the
    heliospheric equatorial plane for the $A > 0$ and $A < 0$ polarity
    cycles. \emph{Panel~(b):} Ratio of the computed intensities for the $A >
    0$ and $A < 0$ cycles as a function of kinetic energy and for the
    radial distance as in (a). Image reproduced by permission from
    \citet{Ferreira2005}, copyright by COSPAR.}
  \label{fig16}
\end{figure}}

Drift-modulation models predicted that during $A > 0$ polarity cycles the 
ratio of electron to proton ($\e^{-}/\p$) intensities as a function of time (with 
solar activity as described by the wavy HCS) should exhibit an inverted V 
shape while during $A < 0$ cycles it should exhibit an upright V around 
minimum modulation periods \citep[see the reviews
  by][]{Potgieter-etal2001, HeberPotgieter2006, HeberPotgieter2008,
  Strauss-etal2012b}. This means that as a function of time electrons
would exhibit a sharper intensity time profile than protons or helium
during $A > 0$ solar epochs. This was displayed eloquently by Ulysses
observations of electrons, helium and protons at 1.3~GV and 2.5~GV for
the period 1990 to 2004 as reproduced in Figure~\ref{fig17}. The $\e^{-}/\p$
ratio indeed formed an inverted V around the 1997 solar minimum
shown in the bottom panel. This effect was also shown by
\citet{Ferreira-etal2003a} and \citet{Ferreira-etal2003b} for the 1987
solar minimum ($A < 0$ cycle) when a V-shape was displayed in the
$\e^{-}/\mathrm{He}$ ratio.

\epubtkImage{Figure17.png}{%
\begin{figure}[htbp]
 \centerline{\includegraphics[width=0.9\textwidth]{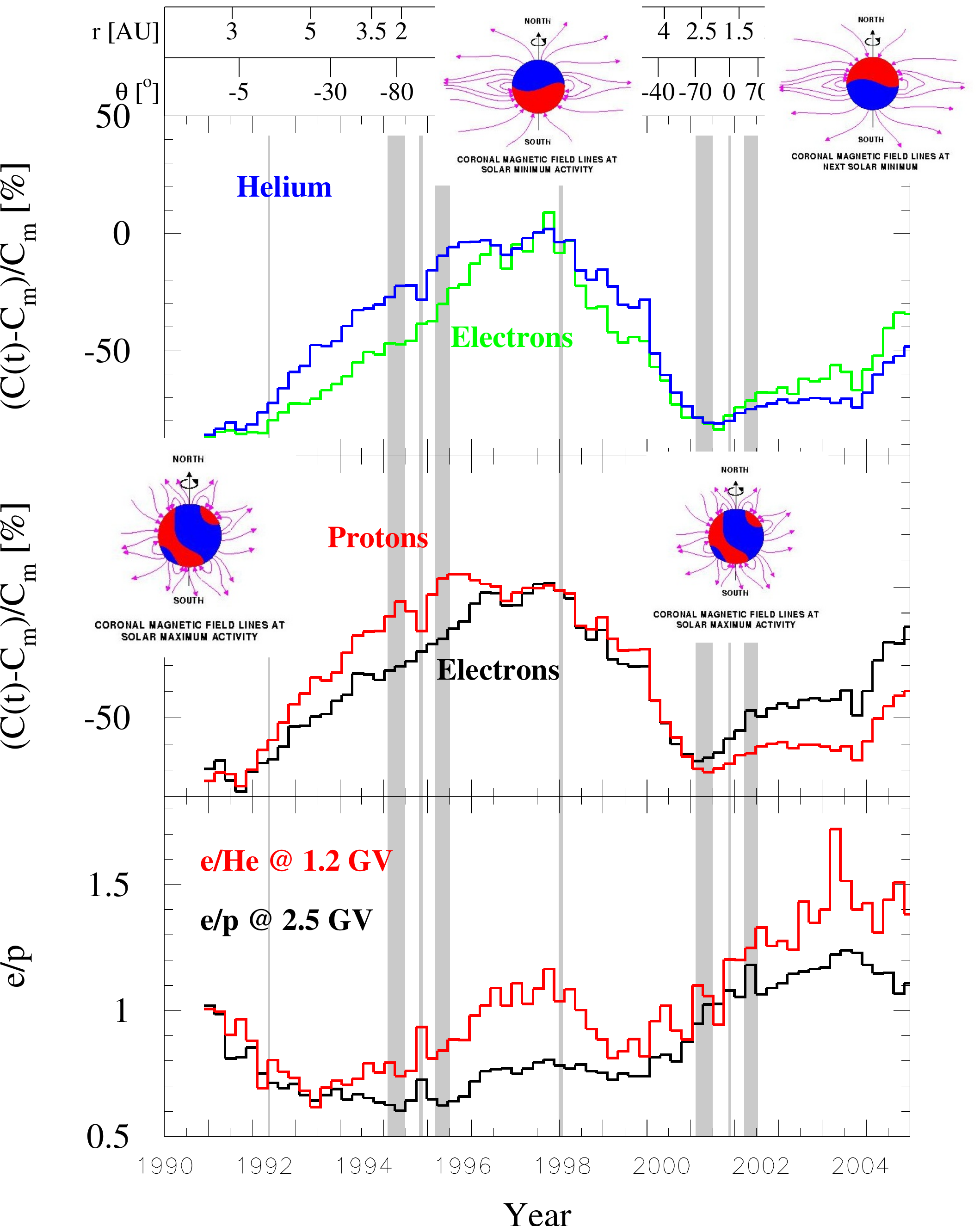}}
  \caption{Observed \% changes respectively of helium (1.2~GV),
    electrons (1.2~GV and 2.5~GV), and protons (2.5~GV), as a function
    of time (solar activity) for the Ulysses mission from 1990 to
    2005. The period from 1990 to 2000 was an $A > 0$ polarity epoch but
    changed to an $A < 0$ epoch around 2000\,--\,2001. Clearly the electrons
    exhibited a sharper profile over this $A > 0$ cycle than protons and
    helium in accord with predictions of drift-modulation
    models. Adapted by Heber from \citet{Heber-etal2002, Heber-etal2003,
      Heber-etal2009}.}
  \label{fig17}
\end{figure}}

\epubtkImage{Figure18.png}{%
\begin{figure}[htb]
 \centerline{\includegraphics{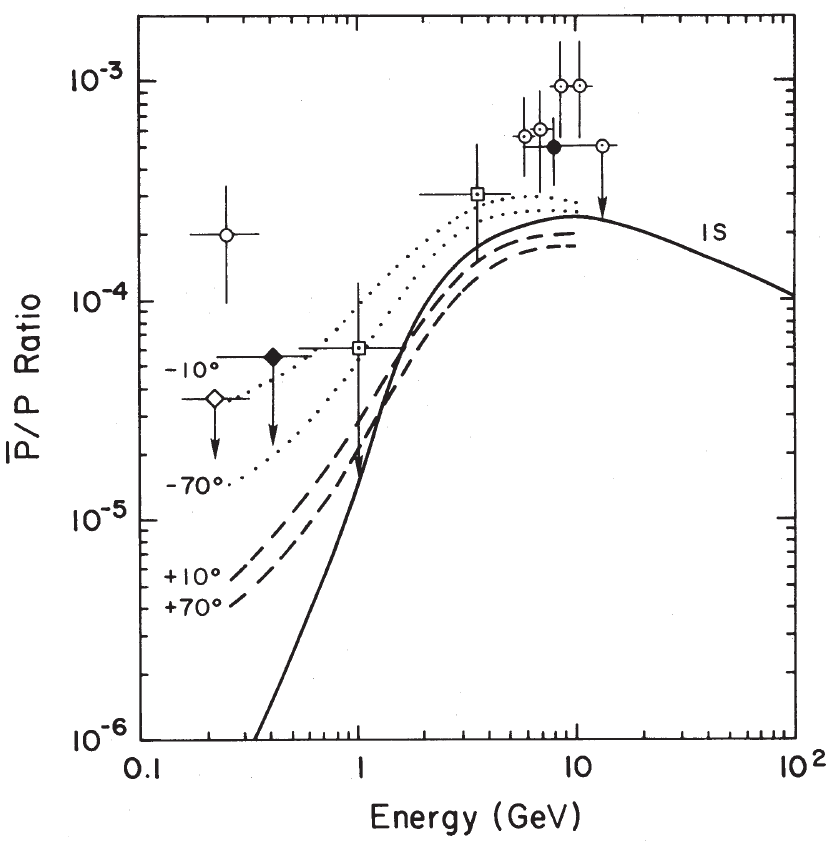}}
  \caption{A comparison of the observed anti-proton to proton ratio
    (below 10~GeV with first generation drift-model computations for
    solar minimum conditions with the HCS tilt angle
    $\alpha$~=~10\textdegree\ and for solar maximum conditions with
    $\alpha$~=~70\textdegree. The -- and + signs indicate $A < 0$ and
    $A > 0$ polarity cycles, respectively. The corresponding ratio for
    the galactic spectra is indicated as IS. Image reproduced by
    permission from \citet{WebberPotgieter1989}, copyright by AAS.}
  \label{fig18}
\end{figure}}

A first illustration of the charge-sign dependent effect for the modulation 
of protons and anti-protons was made by \citet{WebberPotgieter1989}. This is 
shown in Figure~\ref{fig18}. These first generation drift models predicted a change 
of a factor of 10 in the $\bar{\p}/\p$ ratio at 200~MeV from $A > 0$ to
$A < 0$ solar minima but less for increased solar activity
periods. This was done in an attempt to reproduce the data displayed
in this figure with the GS as assumed in those
days. Comprehensive modeling has later been done for electron-positron
($\e^{-}/\e^{+}$) and proton-antiproton modulation
\citep{LangnerPotgieter2004a, LangnerPotgieter2004b,
PotgieterLangner2004a} and heavier CR species such as boron and
carbon \citep{PotgieterLangner2004b}. They relied heavily on the
computations of galactic spectra with the GALPROP code
\citep[e.g.,][]{Moskalenko-etal2002, Strong-etal2007}. Apart from
drifts, their model also included the effects of the solar wind
TS. These results are shown in Figure~\ref{fig19} 
respectively for $\e^{-}/\e^{+}$ and $\bar{\p}/\p$ as a
function of $E$ for the two solar polarity cycles during typical solar
minimum modulation conditions. This is compared to the corresponding
ratio of the local interstellar spectra (LIS). See also
\citet{DellaTorre-etal2012}.

\epubtkImage{Figure19-20.png}{%
\begin{figure}[htb]
 \centerline{
   \includegraphics[height=5cm]{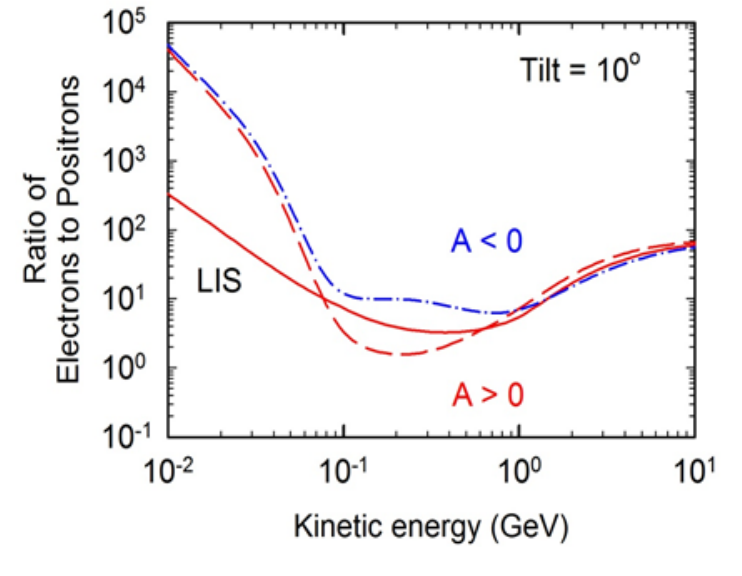}\qquad
   \includegraphics[height=5cm]{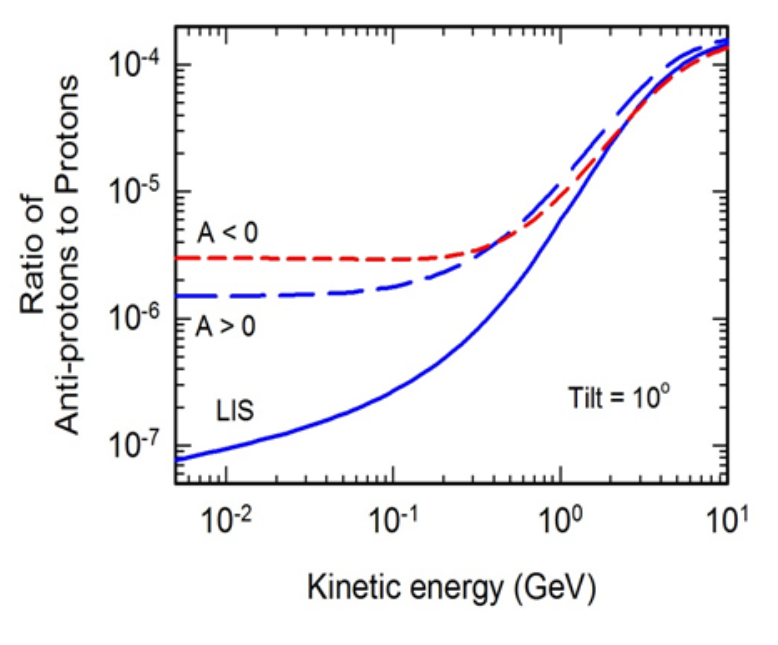}
 }
  \caption{\emph{Left:} Computed electron to positron ratios at Earth for two
    polarity cycles ($A > 0$, e.g., 1997, 2020 and $A < 0$, e.g., 1985,
    2009) of the HMF compared to the ratio of the LIS at the
    heliopause (120~AU). Differences above about 80~MeV are caused by
    gradient, curvature, and current sheet drifts in the heliosphere
    during solar minimum activity. \emph{Right:} Similar to left
    panel, but for computed ratios of galactic protons to
    anti-protons \citep{LangnerPotgieter2004a}.}
  \label{fig19}
\end{figure}}


Evidently, the charge-sign effect is significant but seems less than what 
was predicted by the first generation models. \citet{Strauss-etal2012a} 
applied a newly developed 3D modulation model based on the SDE approach 
to study the modulation of galactic protons and anti-protons inside the heliosphere. 
They also found a maximum drift effect of a factor of 10
below $\sim$~100~MeV for the $\bar{\p}/\p$, which gradually subsides
with increasing $E$ to become negligible above 10~GeV. They
predict that for the next solar minimum period ($A > 0$ cycle) this
ratio will be lower than in the preceding $A < 0$ cycle and noted that
if modulation conditions during the next solar minimum would be the
same as in the previous $A < 0$ minimum of 2009, CR intensities will
be even higher at Earth because of drifts. They illustrated
convincingly with pseudo-particle traces how protons drift differently
than anti-protons through the heliosphere to Earth during the same
modulation cycle. See also \citet{Bobik-etal2011, Bobik-etal2012}. For
an illustration of charge-sign-dependent effects in the outer
heliosphere and heliosheath see, e.g., \citet{LangnerPotgieter2004a}.

In order to establish how large the effect is over an energy range 
considered relevant to solar modulation, precise and simultaneous 
measurements of CRs and their antiparticles by the same instrumentation are 
necessary. The PAMELA mission has introduced an era of such precise 
measurements of protons (anti-protons) and electrons (positron) done 
simultaneously to energies down to $\sim$~100~MeV so that solar modulation 
can also be studied and particle drifts thoroughly tested. The preliminary 
proton and electron data as monthly averages, from mid-2006 to the end of 
December 2009, were reported in the PhD theses of
\citet{DiFelice2010} and \citet{DeSimone2011}, and the Master's theses of
\citet{Vos2011} and \citet{Munini2011}. The finalized proton spectra for
this period were reported by \citet{Adriani-etal2013} with the
electron and positron data for this period to be finalized in
2013. Yearly averages were reported by \citet{Sparvoli-etal2012} and
\citet{BoezioMocchiutti2012}. The PAMELA data shown here as Figure~\ref{fig21}
illustrate how protons between 0.5\,--\,1.0~GV responded differently
over the mentioned period than electrons at the same rigidity. Protons
had increased by an average of a factor of $\sim$~2.5 over 4.5~years
whereas electrons of the same rigidity had increased by only a factor
of $\sim$~1.4 over the same period. This trend once again reveals the
inverted V shape as discussed above.

\epubtkImage{Figure21.png}{%
\begin{figure}[htb]
 \centerline{\includegraphics[width=0.8\textwidth]{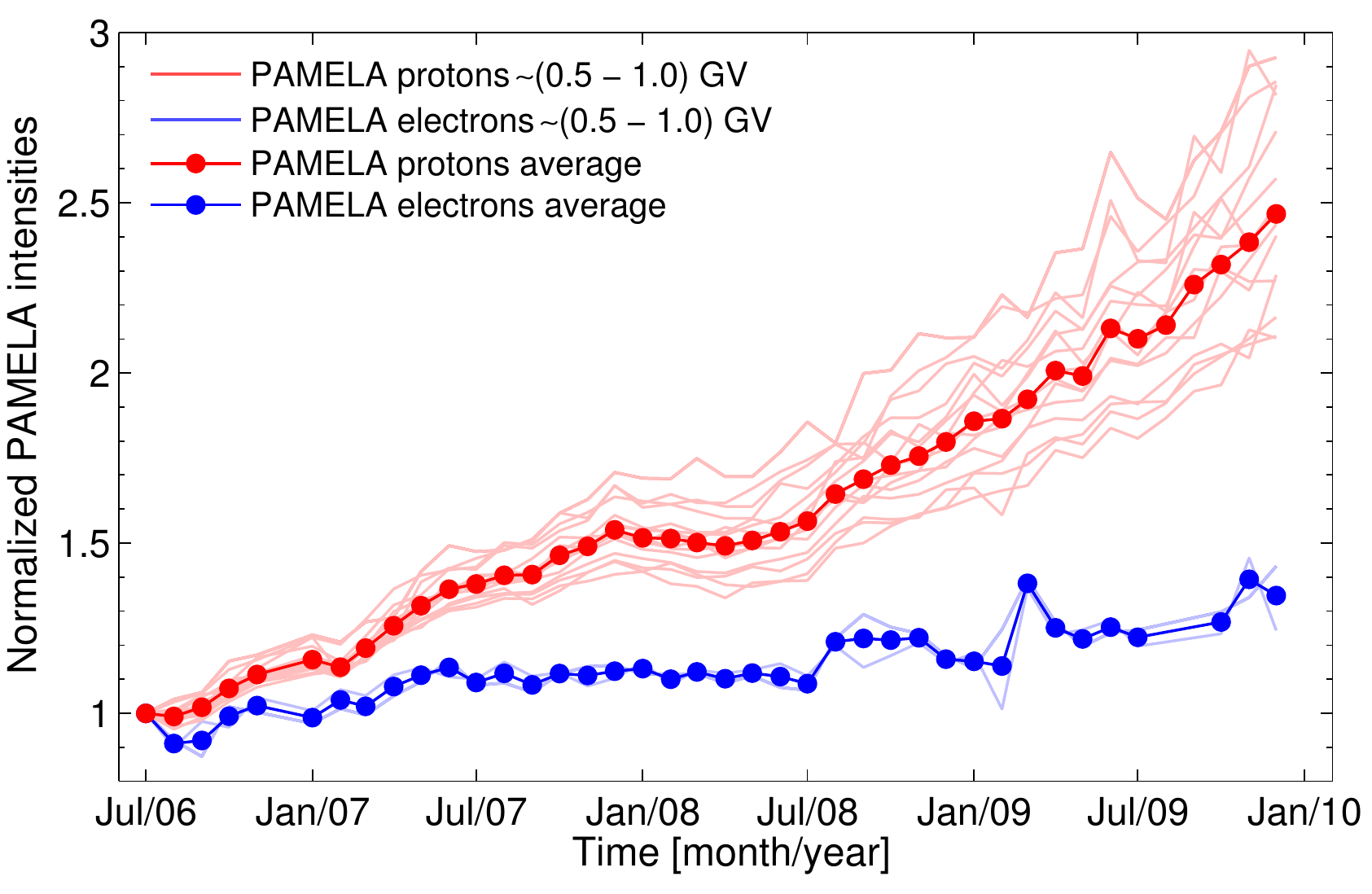}}
  \caption{Normalized proton (light-red lines) and electron
    (light-blue lines) differential intensities at (0.75~\textpm~0.2)
    GV as a function of time, from July 2006 to December 2009. The red
    and blue symbols represent the average intensities of protons and
    electrons, respectively. These intensity-time profiles are
    normalized to the intensity measured in July 2006. Image
    reproduced by permission from \citet{Vos2011}. See also
    \citet{DiFelice2010} and \citet{DeSimone2011}.}
  \label{fig21}
\end{figure}}

\citet{DeSimone2011} reported for the year 2009 that
$\e^{-}/\e^{+} = 6$ at 200~MeV changed to
$\e^{-}/\e^{+} = 15$ at 8~GeV, which demonstrates a similar trend
as the prediction in Figure~\ref{fig6} but seemly larger. This provides ample
motivation to apply a comprehensive drift model to the PAMELA data for
CRs with opposite charge to establish the exact extent of drifts
during the recent prolonged solar minimum. 

What happens to the $\e^{-}/\p$ or $\e^{-}/\mathrm{He}$ ratios at times of
solar maximum activity was demonstratively discussed by, e.g.,
\citet{Ferreira2005}, \citet{HeberPotgieter2006, HeberPotgieter2008},
and \citet{Potgieter2011}. PAMELA observations after 2009 could be
most helpful in this regard.

It follows from Figure~\ref{fig19} that the optimum energy range for
observations to test the extent of drift effects for electrons and
positrons is between 50~MeV and 5~GeV. Below 50~MeV, electrons observed
at Earth get `contaminated' by Jovian electrons
\citep{PotgieterNndanganeni2013a}.

\clearpage
\subsection{Main causes of the complete 11-year and 22-year solar
  modulation cycles}

Apart from the 11-year and 22-year cycles, regular steps are superposed on 
the intensity-time profiles of CR modulation. A departure point for these 
time-dependent steps (both increases and decreases) from a global point of 
view is that `propagating barriers' are formed and later dissipate in the outer 
heliosphere during the 11-year activity cycle. These `barriers' are 
basically formed by solar wind and magnetic field co-rotating structures 
which are inhibiting the easy access of CRs to a relative degree. This is 
especially applicable to the phase of the solar activity cycle before and 
after solar maximum conditions when large steps in the particle intensities 
had been observed. A wide range of interaction regions occur in the 
heliosphere, with GMIRs the largest, as introduced by \citet{Burlaga-etal1993}. 
They observed that a clear relation exists between CR decreases (recoveries) 
and the time-dependent decrease (recovery) of the HMF magnitude and extent 
local to the observation point. The paradigm on which these modulation 
barriers is based is that interaction and rarefaction regions form with 
increasing radial distance from the Sun \citep[see also][and
  references therein]{PotgieterleRoux1989}. These relatively narrow
interaction regions can grow latitudinally and especially azimuthally
and as they propagate outwards they spread, merge and interact to form
eventually GMIRs that can become large in extent and capable of
causing the large step-like changes in CRs. \citet{Potgieter-etal1993}
illustrated that their affects on long-term modulation depend on their
rate of occurrence, the radius of the heliosphere (i.e., how long they
stay inside the modulation volume), the speed with which they
propagate, their spatial extent and amplitude, especially their
latitudinal extent (to disturb drifts), and the background modulation
conditions they encounter. Drifts on the other hand, normally dominate
periods of solar minimum modulation up to four years so that during an
11-year cycle a transition must occur (depending how solar activity
develops) from a period dominated by drifts to a period dominated by
these propagating structures. The largest of the step decreases and
recoveries shown in Figure~\ref{fig6} are caused by these GMIRs. 

The 11-year and 22-year cycles together with the step-like modulation are 
good examples of the interplay among the main modulation mechanisms as 
illustrated by \citet{leRouxPotgieter1995}. They showed that it was 
possible to simulate, to the first order, a complete 22-year modulation 
cycle by including a combination of drifts with time-dependent tilt angles 
and GMIRs in a time-dependent modulation model. See also
\citet{Potgieter1995, Potgieter1997} and references
therein. \Citet{leRouxFichtner1999} confirmed that a series of GMIRs
cannot on their own reproduce the fully observed 11-year modulation
cycle. This is achieved by adding a well-defined time variation in the
propagation process such as for the diffusion coefficients, or using
the time-dependent wavy HCS. A major issue with time-dependent
modeling, apart from the global dynamic features such as the wavy HCS,
is what to assume for the time dependence of the diffusion
coefficients mentioned above. 

A subsequent step in understanding long-term modulation came when
\citet{Cane-etal1999} pointed out that the step decreases observed at
Earth could not be primarily caused by GMIRs because they occurred
well before any GMIRs could form beyond 10\,--\,20~AU. Instead, they
suggested that time-dependent global changes in the HMF over an
11-year cycle are responsible for long-term
modulation. \citet{PotgieterFerreira2001} and
\citet{FerreiraPotgieter2004} combined these changes with
time-dependent drifts, naming it the compound modeling approach. It
was assumed that all the diffusion coefficients change with time
proportional to $B(t)^{-n(P,t)}$, with $B(t)$ the observed,
time-dependent HMF magnitude close to Earth, and $n(P,t)$ a function
of rigidity and the HCS tilt angle, which introduces an additional time
dependence related to solar activity. These changes are then
propagated outwards at the solar wind speed to form propagating
modulation barriers throughout the heliosphere, changing with the
solar cycle. With simply $n = 1$, and with $B(t)$ changing by an
observed factor of 2 over a solar cycle, this approach resulted in a
variation of the diffusion coefficients by a factor of 2 only, which
is perfect for simulating the 11-year modulation at NM energies at
Earth, as seen in Figure~\ref{fig6}, but not at all for lower
rigidities. In order to reproduce spacecraft observations at energies
below a few GeV, $n(P,t)$ must depend on time (solar activity) and on
rigidity. For a more advanced treatment and recent application of this
approach, see \citet{Manuel-etal2011a, Manuel-etal2011b}.

\citet{FerreiraPotgieter2004} confirmed that using the HCS tilt angles as 
the only time-dependent modulation parameter resulted in compatibility only 
with solar minimum observations. Using the compound approach resolved this 
problem. Applied at Earth and along the Ulysses and Voyager~1 and 2 
trajectories, this approach is remarkably successful over a period of
22~years, e.g., when compared with 1.2~GV electron and helium
observations at Earth, it produces the correct modulation amplitude
and most of the modulation steps. Some of the simulated steps did not
have the correct magnitude and phase, indicating that refinement of
this approach is needed by allowing for merging of the propagating
structures. However, solar maximum modulation could be largely
reproduced for different CR species using this relatively simple
concept, while maintaining all the other major modulation features
during solar minimum, such as charge-sign dependence discussed above. 

\epubtkImage{Figure22.png}{%
\begin{figure}[htb]
 \centerline{\includegraphics[width=\textwidth]{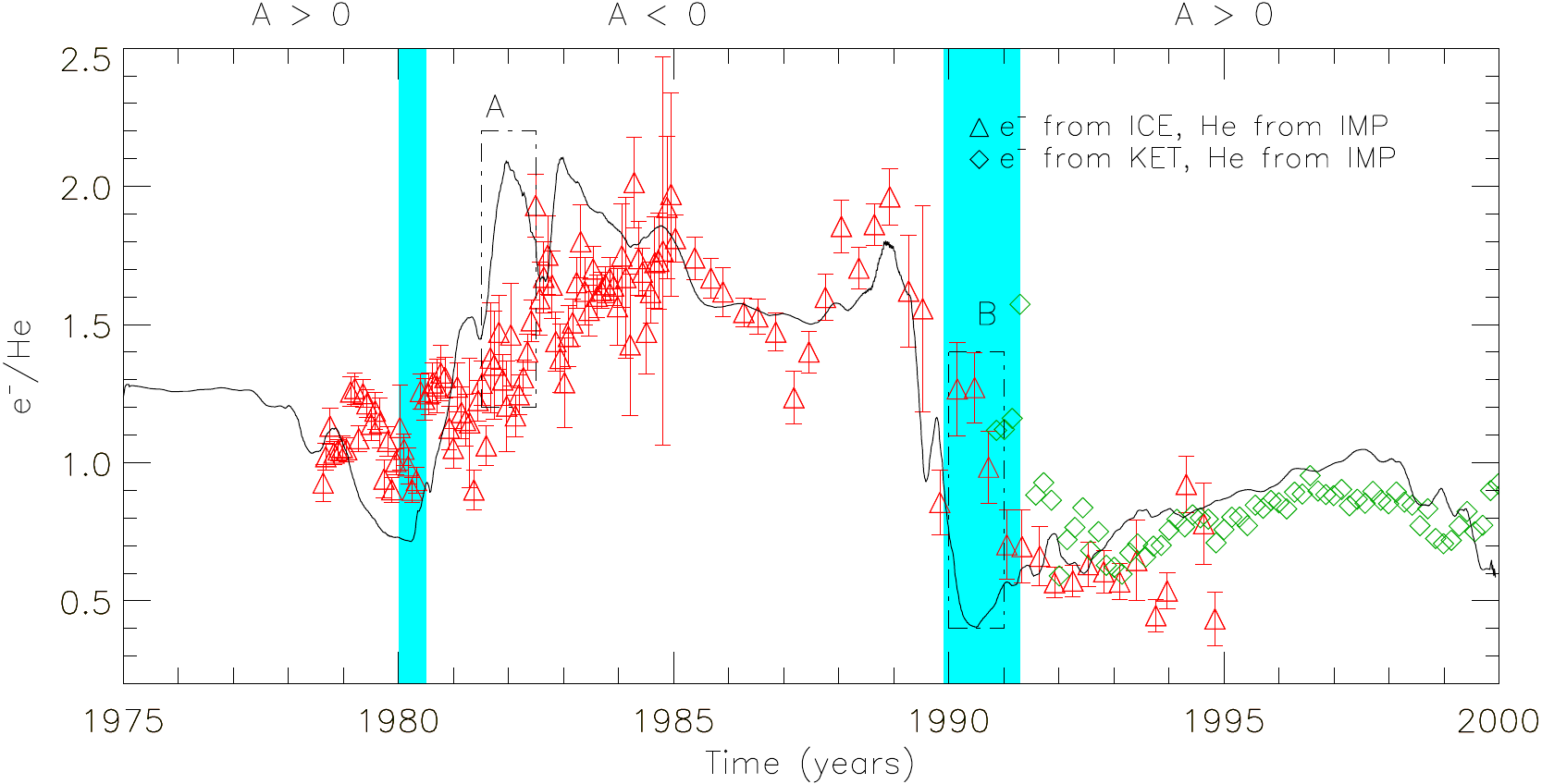}}
  \caption{Computed 1.2~GV $\e^{-}/\mathrm{He}$ ratio at Earth for
    1976\,--\,2000 in comparison with the observed $\e^{-}/\mathrm{He}$
    obtained from electron measurements of ISEE3/ICE, He measurements
    from IMP and electron measurements from KET
    \citep{Heber-etal2003}. The shaded areas correspond to the period
    with no well-defined HMF polarity. Two periods indicated by labels
    A and B with relatively large differences between the computed
    ratios and the observations require further investigation. Image
    reproduced by permission from \citet{Ferreira-etal2003a},
    copyright by COSPAR.}
  \label{fig22}
\end{figure}}

An important accomplishment of this compound approach is that it also 
produces the observed charge-sign dependent CR modulation from solar minimum 
to maximum activity. Figure~\ref{fig22} depicts the computed 1.2~GV
electron ($\e^{-}$) to helium ratio at Earth for 1976\,--\,2000 in
comparison with the observed $\e^{-}/\mathrm{He}$ obtained from electron
measurements of ISEE3/ICE, helium measurements from IMP and electron
measurements from KET \citep{Heber-etal2002, Heber-etal2003}. The
shaded areas correspond to the period when there was not a well
defined HMF polarity. Two periods, labelled A and B, were found with
relatively large differences between the computed ratios and the
observations that require further refinement. 

The compound approach also involves two other important modifications. 
First, a significant increased polar perpendicular diffusion is required to 
account for the observed latitude dependence of CR protons and the lack 
thereof for electrons along the Ulysses trajectory over the 22 year cycle. 
This is mainly to reduce the large latitudinal gradients caused by 
unmodified drifts and in addition to the time dependence of the diffusion 
coefficients. This effect is illustrated in Figure~\ref{fig23} using
the computed 2.5~GV electron to proton ratio ($\e^{-}/\p$) along the Ulysses
trajectory and at Earth in comparison with the 2.5~GV $\e^{-}/\p$
observations from KET \citep{Heber-etal2002, Heber-etal2003}. The top
panels show the position of Ulysses during this period. In order to
model the observed e-/p as a function of time, the latitude dependence
of both electrons and protons must be correctly modeled
\citep{Ferreira-etal2003b}. Second, during periods of large solar
activity, drifts must be reduced additionally to improve explaining
the observed electron to He intensity ratio at Earth and the electron
to proton ratio along the Ulysses trajectory during the period when
the HMF polarity reverses. For example, drifts had to be reduced from
a 50\% level at the beginning of 1999 to a 10\% level by the end of
1999, to vanish during 2000, but to quickly recover after the polarity
reversal during 2001 to levels above 10\%. For the period after 2001,
the model predicts a steady increase in drifts from 10\% to 20\% by
the end of 2002. This indicates that in order to produce realistic
charge-sign dependent modulation during extreme solar maximum
conditions, the heliosphere must become diffusion (non-drift)
dominated. \citet{Ndiitwani-etal2005} calculated the percentage drifts
required over a full modulation cycle, especially during extreme solar
maximum, to find compatibility between the compound model and the
observed $\e^{-}/\p$. This is shown in Figure~\ref{fig24} in comparison to
the observed tilt angles as proxy for solar activity. Obviously,
little drifts are required during solar maximum in contrast to
$\sim$~90\% at solar minimum activity. For charge-sign dependent
effects in the outer heliosphere and heliosheath, see
\citet{Langner-etal2003} and \citet{LangnerPotgieter2004a}.

\epubtkImage{Figure23.png}{%
\begin{figure}[htb]
 \centerline{\includegraphics[width=0.6\textwidth]{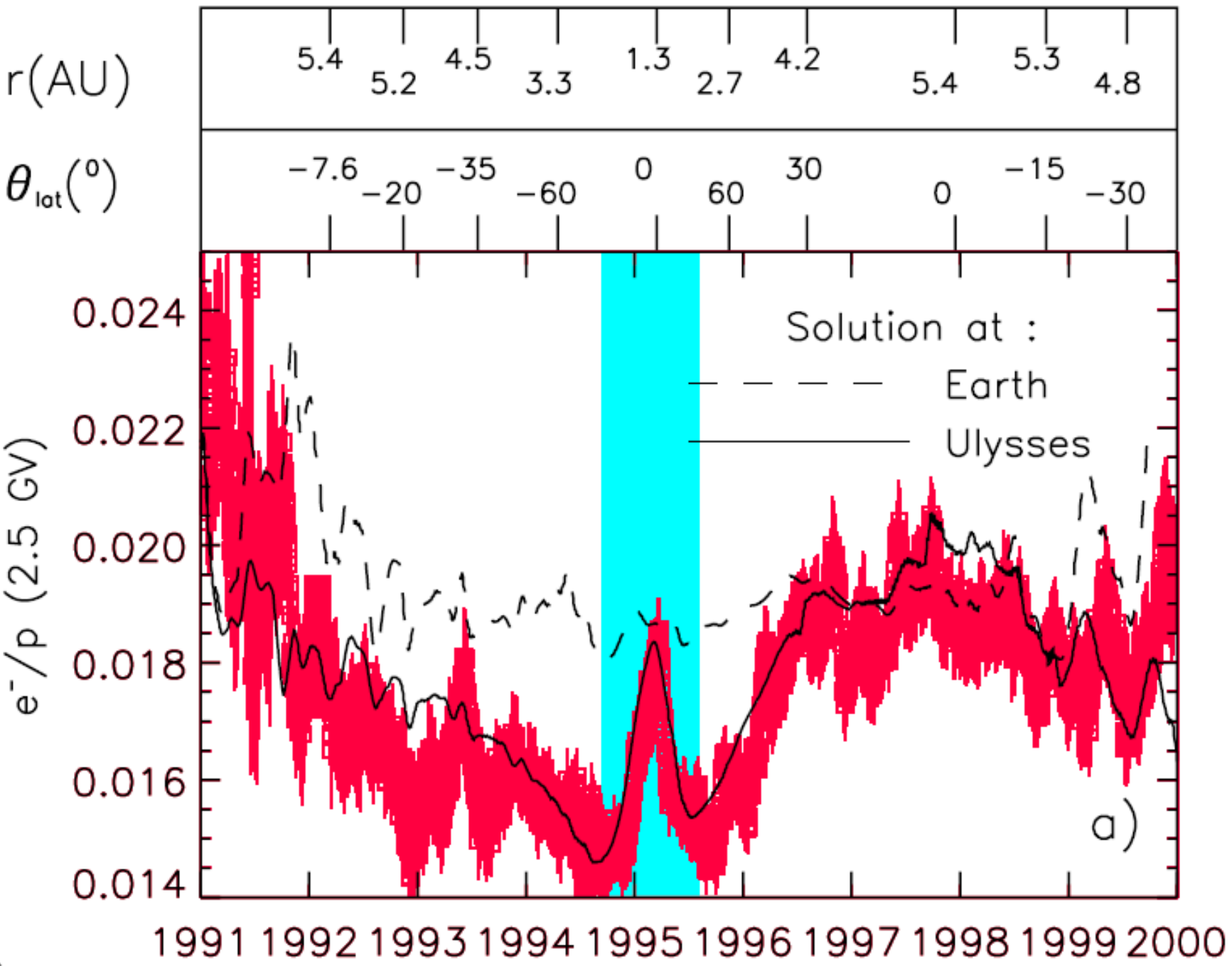}}
  \caption{Computed 2.5~GV electron to proton ratio ($\e^{-}/\p$) as a
    function of time along the Ulysses trajectory (solid line) and at
    Earth (dotted line) in comparison with the 2.5~GV ratio observed
    with KET \citep{Heber-etal2002}. Top panels show the position of
    Ulysses. In order to simulate this observed ratio as a function of
    time, the latitude dependence of both electrons and protons must
    be correctly modeled \citep{Ferreira-etal2003a,
      Ferreira2005}. Image reproduced by permission
    from \citet{Ferreira-etal2003b}, copyright by EGU.}
  \label{fig23}
\end{figure}}

\epubtkImage{Figure24.png}{%
\begin{figure}[htb]
 \centerline{\includegraphics[width=0.9\textwidth]{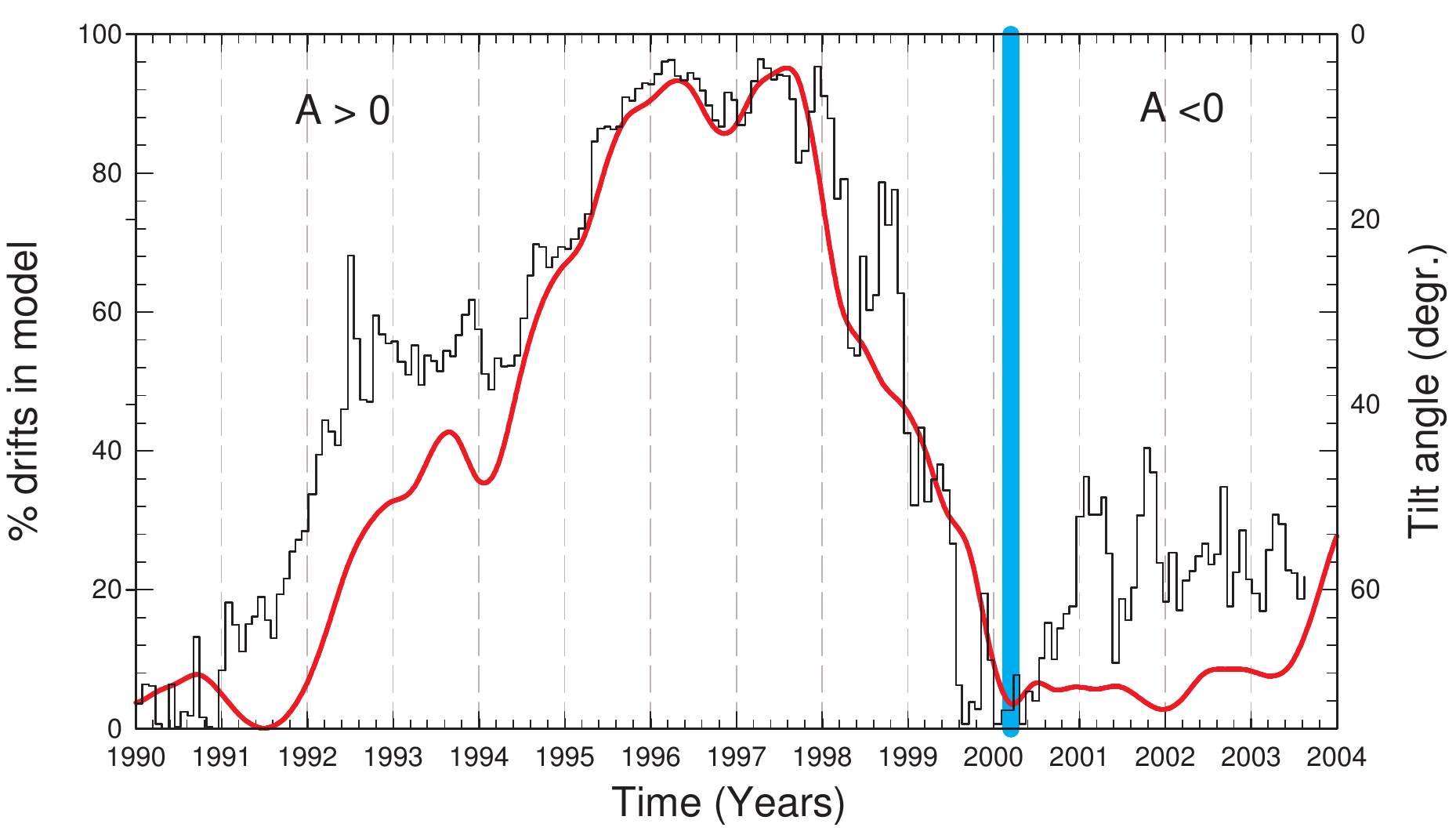}}
  \caption{Percentage of drifts (continuous line) in the compound
    model that gives realistic modulation for various stages of the
    solar cycle for both the 2.5~GV electron and protons. As a proxy
    for solar activity the tilt angles, as used in the model are shown
    for illustrative purposes. Image reproduced by permission from
    \citet{Ndiitwani-etal2005}, copyright by EGU.}
  \label{fig24}
\end{figure}}

\newpage
A question that is relevant within the context of Voyager~1 and 2 
observations is how much modulation occurs inside the heliosheath. The 
process is of course highly energy-dependent. An illustrative example of the 
amount of modulation that CR protons may experience in the heliosheath in 
the nose direction is shown in Figure~\ref{fig25}. The percentage of
modulation in the equatorial plane in the heliosheath is given with
respect to the total modulation (between 120~AU and 1~AU) as a
function of kinetic energy for both polarity cycles, for solar minimum
and moderate maximum conditions \citep{Langner-etal2003}. See also
\citet{Strauss-etal2013} for a recent illustration of modulation in
the outer heliosphere.

\epubtkImage{Figure25.png}{%
\begin{figure}[htb]
 \centerline{\includegraphics[width=0.6\textwidth]{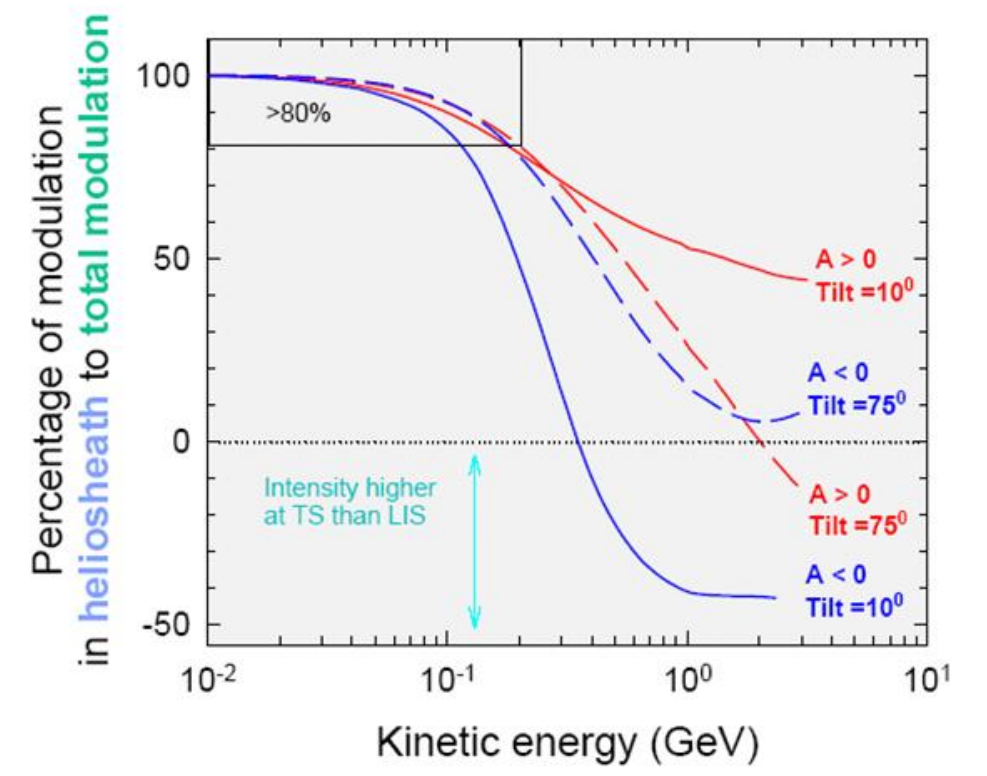}}
  \caption{Computed percentage of galactic CR modulation in the
    heliosheath with respect to the total modulation (between 120~AU
    and 1~AU) for the two magnetic polarity cycles ($A > 0$ and $A <
    0$), for solar minimum ($\alpha$~=~10\textdegree) and for moderate
    maximum ($\alpha$~=~75\textdegree) conditions, in the equatorial
    plane in the nose direction of the heliosphere. Negative
    percentages mean that the galactic CRs are reaccelerated at the TS
    under these assumed conditions \citep{Langner-etal2003}.}
  \label{fig25}
\end{figure}}

Evidently, at $E < \sim 0.02\mathrm{\ GeV}$ modulation $>$~80\% may
occur in the heliosheath for both polarity cycles. For all four the
conditions, the heliosheath modulation will eventually reach 0\% (not
shown) but at different energies, indicating that it differs
significantly with energy as well as with drift cycles. How much
gradient and curvature drifts actually occur in the heliosheath is
still unanswered. The negative percentages indicate that the intensity
is actually increasing in the heliosheath as one moves inward from the
outer boundary toward the TS because of the re-acceleration of CRs at
the TS. This depends on many aspects, in particular the TS compression
ratio as discussed by \citet{Langner-etal2003, Langner-etal2006a,
  Langner-etal2006b}.

In the study of the long-term CR modulation in the heliosphere, several 
issues need further investigation and research. Despite the apparent success 
of the compound numerical model described above the amount of merging taking 
place beyond 20~AU needs to be studied with MHD models, especially the 
relation between CMEs and GMIRs, and how these large barriers will modify 
the TS and the heliosheath. The full rigidity dependence of the compound 
model is as yet not well described. A major issue with time-dependent 
modeling, apart from global dynamic features such as the wavy HCS, is what 
to use for the time dependence of the diffusion coefficients in
Equation~(\ref{eq:spherical_compact_tpe}), on top of the already complex
issue of what their steady-state energy (rigidity) and spatial
dependence are in the inner heliosphere. It has now also become
pressing to understand the diffusion tensor in the outer heliosphere
and beyond the TS. Equation~(\ref{eq:parker_tpe}) is probably more
complex and has been extended to include mechanisms otherwise
considered to be negligible. Fundamentally, from first principles, it
is not yet well understood how gradient and curvature drifts reduce
with solar activity. This aspect needs now also to be investigated for
the region beyond the TS. For example, what happens to the wavy HCS in
the heliosheath and how will the strong non-radial components of the
solar wind and the associated HMF affect drifts and CR modulation in
the heliosheath? To answer this, more study is needed to gain a better
understanding of the finer details of these complex magnetic fields and
the relation to long-term CR modulation, especially how this field
changes with solar activity. An aspect that should be kept in mind is
that the study of more realistic fields requires an ever more complex
diffusion tensor and description of drifts. Other interesting aspects
of CRs beyond the TS are discussed by \citet{Potgieter2008}. Another
important question is how the heliospheric modulation volume varies
with time over scales of hundreds to thousands of years, which are
most relevant for the study of space climate. The Sun encounters
different interstellar environments during its passage through the
galaxy, and hence the outer heliospheric structure should change
\citep[e.g.,][]{Scherer-etal2008a}. 

The paleo-cosmic ray records can be used to study the properties of the 
heliosphere and solar processes over 10000 and more
years. \citet{McCracken-etal2011} reported that both varied greatly
over such as period, ranging from $\sim$~26 Grand Minima of duration
50\,--\,100 years when the Sun was inactive, to periods similar to the
recent 50 years of strong solar activity. The 11 and 22 year cycles of
solar activity continued through the Sp\"{o}rer and Maunder Grand
Minima. They speculated that the solar dynamo exhibits a 2300 year
periodicity, wherein it alternates between two different states of
activity and argued that paleo-cosmic ray evidence suggests that the
Sun has now entered a more uniform period of activity, following the
sequence of Grand Minima (Wolf, Sp\"{o}rer, Maunder, and Dalton) that
occurred between 1000 and 1800~AD. See also the discussion by
\citet{Usoskin-etal2001} and the review by \citet{Usoskin2013}.

\clearpage 
\section{A Few Observational Highlights}

\subsection{The unusual solar minimum of 2007\,--\,2009}

The past solar minimum activity period and the consequent minimum modulation 
conditions for galactic CRs was unusual. It was expected that the new 
activity cycle would begin early in 2008, assuming a 10.5~year periodicity. 
Instead, solar minimum modulation conditions had continued until the end of 
2009, characterized by a much weaker HMF compared to previous cycles. The 
tilt angle of the wavy HCS, on the other hand, had not decreased as rapidly 
as the magnitude of the HMF at Earth during this period, but eventually also 
reached a minimum value at the end of 2009. \citet{AhluwaliaYgbuhay2011}, 
\citet{Stozhkov-etal2009}, \citet{Mewaldt-etal2010} and
\citet{Krymsky-etal2012} all reported that CRs with high rigidity
reached record setting intensities during this time \citep[see
  also][]{Heber-etal2009, McDonald-etal2010}. It followed from
observations for this period that the delay between the time for
minimum sunspot numbers and maximum CR intensities was at
least three times longer than during previous even numbered solar
cycles \citep[e.g.,][]{Kane2011}. The decay phase of the sunspot cycle
23 exhibited two unusual features, it lasted quite long while the HMF
at Earth reached the lowest value since in situ measurements in space
began in 1963. See also \citet{Cliver-etal2011}, \citet{ZhaoFisk2011},
\citet{AhluwaliaJackiewicz2012}, and the reviews by
\citet{AhmalBadrudin2012} and \citet{Mewaldt2012}.

Since the beginning of the space age, the highest CR proton spectrum 
was observed by PAMELA in December 2009. This was unexpected because during 
previous $A < 0$ polarity cycles, proton spectra were always lower than for 
$A > 0$ cycles at kinetic energies less than a few-GeV, in full accord with 
drift models. The PAMELA experiment has provided important results on
the antiproton \citep{Adriani-etal2009a} and positron galactic abundances
\citep{Adriani-etal2009b, Boezio-etal2009, BoezioMocchiutti2012}.

The high-resolution PAMELA spectrometer also allows to perform hydrogen and 
helium spectral measurements up to 1.2~TV \citep{Sparvoli-etal2012}, which is 
the highest limit achieved by this kind of
experiments. \citet{Adriani-etal2013} presented observations down to
400~MV of the absolute flux of protons from July 2006 until the end of
2009. Large proton statistics collected by the instrument allowed the
measurement of the proton flux for each Carrington rotation. In Figure~\ref{fig26}
these spectra are shown from July 2006 to the very beginning of
2010. The spectrum at the end of December 2009 was the highest
recorded. In January 2010, solar activity picked up significantly so
that the proton intensity had started to decrease. In order to
emphasize how decreasing solar modulation conditions allowed galactic
protons to increase at Earth, especially at low kinetic energies, the
spectra in Figure~\ref{fig26} are used to calculate and plot the intensity
ratios as a function of kinetic energy with respect to July 2006. This
is shown in Figure~\ref{fig27}.

\epubtkImage{Figure26.png}{%
\begin{figure}[htb]
 \centerline{\includegraphics[width=0.75\textwidth]{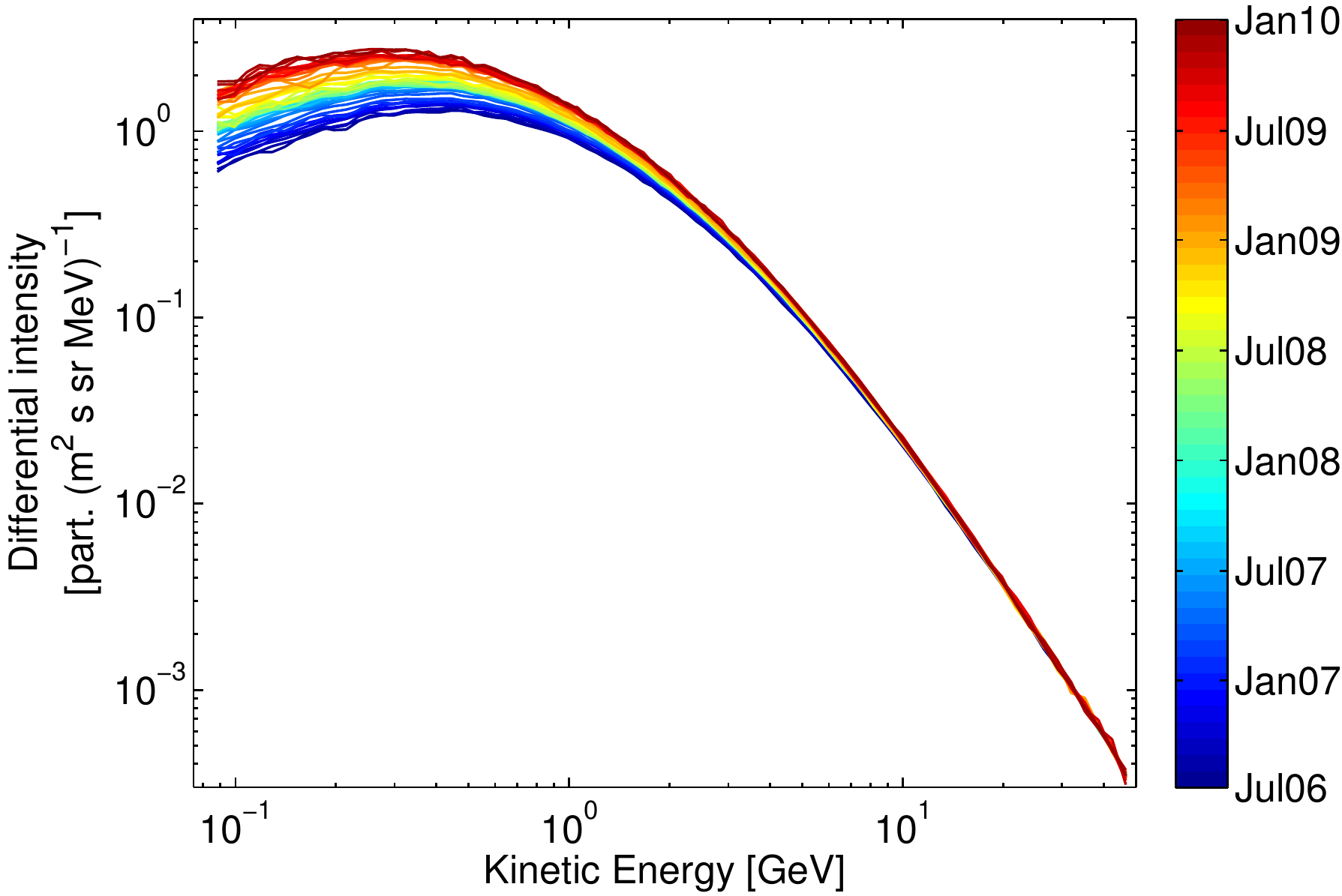}}
  \caption{Proton spectra, averaged over one Carrington rotation, as
    observed by the PAMELA space instrument from July 2006 to the
    beginning of 2010 (see the colour coding on the left). The
    spectrum at the end of December 2009 was the highest recorded. See
    \citet{Adriani-etal2013} and also \citet{Potgieter-etal2013}.}
  \label{fig26}
\end{figure}}

\epubtkImage{Figure27.png}{%
\begin{figure}[htb]
 \centerline{\includegraphics[width=0.75\textwidth]{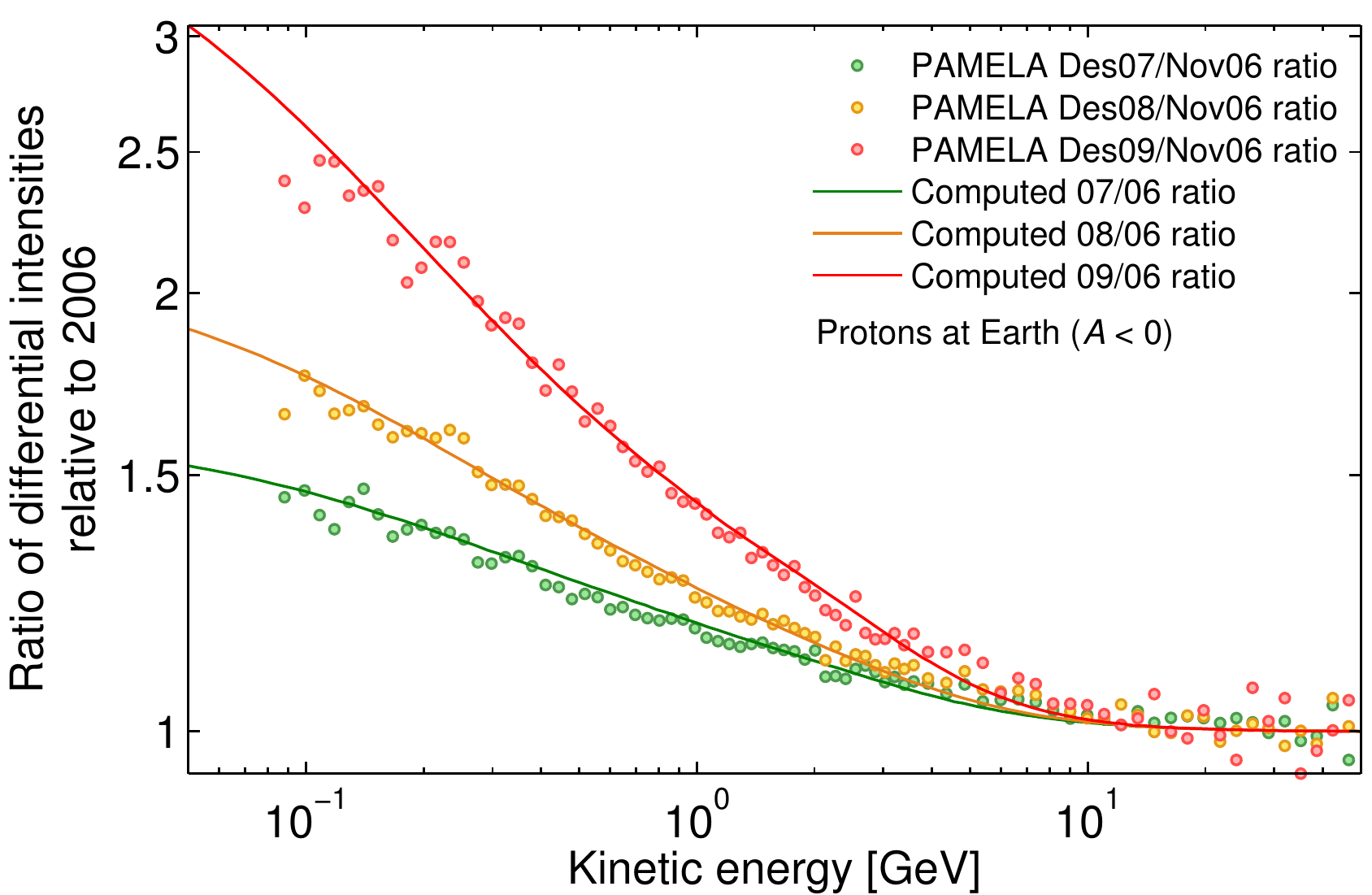}}
  \caption{Computed ratios of differential intensities for selected
    periods in 2007, 2008, 2009 with respect to Nov.~2006 as a
    function of kinetic energy in comparison with PAMELA proton
    observations \citep{Potgieter-etal2013}.}
  \label{fig27}
\end{figure}}

\clearpage
\subsection{Samples from the inner heliosphere}

The Ulysses mission produced solar modulation observations from 1990 to 2009 
as reviewed by \citet{HeberPotgieter2006, HeberPotgieter2008} and
\citet{Heber2011}. These and other CR observations had yielded several
new and surprising insights and are summarized as follows:

\begin{itemize}

\item It was observed by Ulysses that the galactic CR flux was not
  symmetric to the heliographic equator implying a
  North-South-asymmetry in CR modulation
  \citep{McKibben-etal1996}. Surprisingly, neither the solar wind
  experiments nor the magnetic field investigations reported this
  asymmetry. Later magnetic field investigations were interpreted to
  give a deficit of the magnetic flux in the southern hemisphere
  suggesting a relatively large latitudinal offset of the location of
  the HCS by 10\textdegree\ \citep{Smith-etal2000}. Recently,
  \citet{ErdosBalogh2010} disputed this number, arguing that the
  Ulysses magnetic field measurements did not give evidence for such a
  large displacement of the HCS, only a southward displacement of
  2\textdegree\,--\,3\textdegree\ could be possible. It remains thus
  an open question whether this observation was an occurrence of
  events that pertained during this rapid pole to pole passage of
  Ulysses or was related to an asymmetrical magnetic flux.

\item Small latitudinal galactic CR gradients were observed at solar
  minimum, confirming that the LIS cannot be observed in the inner
  polar regions of the heliosphere. A particular motivation of the
  Ulysses mission was to explore this possibility. Drift dominated
  models of that time actually allowed for this
  \citep{JokipiiKopriva1979}. In contrast, the observed proton
  spectrum was highly modulated to large heliolatitudes.

\item The latitudinal gradients for CR protons as a function of
  rigidity was observed to reach a maximum around 2~GV, to decrease
  significantly below these values, in sharp contrast to what drift
  dominated models predicted at that time, a factor of 10 at 200~MeV.

\item A renowned observation was that recurrent particle events
  occurred at high heliolatitudes without corresponding features in
  the solar wind and magnetic field data. It appears that similar
  effects did not occur during the last polar excursion by
  Ulysses. \citet{Dunzlaff-etal2008} reported that measurements in the
  fast solar wind show differences: In cycle~22 the recurrent cosmic
  ray decreases showed a clear maximum near
  25\textdegree\ heliolatitude and were still present beyond
  40\textdegree, whereas in cycle~23 neither such a pronounced maximum
  nor significant decreases were observed above 40\textdegree. The
  periodicity in the CR intensity that could be clearly seen
  in the slow solar wind appeared to have vanished in the fast solar
  wind.

\item Essentially no latitudinal gradients were observed for any
  galactic CRs at solar maximum, indicating that drifts, mainly
  responsible for setting-up these gradients during solar minimum
  conditions, were almost absent at solar maximum. 
	
\item The observed electron to proton ratios (implicitly also
  containing the radial and latitudinal gradients) indicated that
  large particle drifts were occurring during solar minimum but
  diminished significantly toward solar maximum when rather diminished
  drifts were required in models to explain the observed values (see
  Figure~\ref{fig24}).

\item Jovian electrons were observed at high heliolatitudes, implying
  effective latitudinal transport of electrons at these energies
  ($\sim$~10~MeV). This still has to be explained from basic
  theoretical considerations, in particular if this could be the cause
  of extraordinary effective transport for these low-energy electrons
  including large polar perpendicular diffusion \citep[see
    also][]{Ferreira2005}.

\item Anomalous fluxes of oxygen, nitrogen, and neon were observed
  along the Ulysses trajectory and at Earth and with spatial gradients
  different from galactic CRs \citep[e.g.,][]{HeberMarsden2001,
    Cummings-etal2009}. 

\end{itemize}

Several other space missions, at or near Earth, and several balloon
experiments also made numerous and valuable observations of CR
modulation \citep[see the review by][]{Mewaldt2012}. Some of the
recent PAMELA results are discussed above within a given context.

\subsection{Samples from the outer heliosphere}

A major surprise came from the outer heliosphere when Voyager~1 crossed 
the TS in 2004 and found that the intensity of the ACRs did not reach a peak 
intensity at the shock for energies more than a few MeV/nuc. As 
mentioned above, the ACR intensity kept increasing far into the heliosheath 
implying that they are accelerated somewhere else and probably by 
different processes other than diffusive shock acceleration. Continued 
observations from the Voyagers will hopefully offer an explanation. The TS 
was observed to be disappointingly weak so that it is unlikely that it will 
reaccelerate galactic CRs to the extent that it influences the spectral 
shape of the CR spectra in the heliosphere.

With Voyager~1 very close to the HP it is relevant to ask what
signatures of the HP in CRs can be expected? A magnetic barrier or
`wall', if present, should cause a significant increase of the
magnetic field magnitude at the HP. It is unlikely that high energy
galactic CRs will change abruptly but the effect should be strongly
energy-dependent so that the low energy part of the CR spectrum should
rise steeply at the HP while the ACR intensities drop sharply. 

Apart from the ACRs, the galactic electrons at low energies ($E < 100$~MeV) 
showed extraordinary behaviour in the heliosheath. This is shown in
Figure~\ref{fig28}. Note the jumps in intensity and accompanying
changes in the radial gradients. After crossing the TS in late 2004,
the 6\,--\,14~MeV galactic electron intensity measured at Voyager~1
increased rapidly and irregularly. By about 2008.5 this intensity was
5 times that measured when Voyager~1 crossed the shock. The local
radial intensity gradient was as high as
(18.5~\textpm~1.5)~\%/AU at times, to drop to as low
(8~\textpm~1)~\%/AU as shown in the figure \citep[see
  also][]{Nkosi-etal2011}. \citet{Webber-etal2012} argued that these
sudden changes in intensity and in the gradients of electrons and
protons are evidence that Voyager~1 crossed into regions of
significantly different propagation conditions.

Even more spectacular is what happened in August 2012 to the intensity of 
0.5~MeV protons (mainly ACRs) and 6\,--\,14~MeV galactic electrons,
and $E > 100$~MeV proton count rates when Voyager~1 was at 121.7~AU,
as displayed in Figure~\ref{fig29}. Within a few days the intensity of
the dominant energetic component above 1\,--\,2~MeV decreased by more
than 90\%. At the same time a sudden increase of a factor of $\sim$~2
occurred in lower energy (6\,--\,100~MeV) electrons and
$\sim$~30\,--\,50\% for the higher energy nuclei above 100~MeV. The
magnitude of this intensity change for ACRs has not been observed in
the 35~years of the Voyager mission except close to Jupiter. This
simultaneous abrupt reduction of ACR (and TSPs) intensities at lower
energies and increase in galactic CR intensities at somewhat higher
energies was interpreted by \citet{WebberMcDonald2013} as evidence of
the crossing of the HP, or at least the crossing of a
`heliocliff'. This is another spectacular milestone for the Voyager
mission. It should be noted that this apparent crossing has not yet
been confirmed by magnetic field observations. The crossing of the HP
by Voyager~2 is expected to happen soon.

\epubtkImage{Figure28.png}{%
\begin{figure}[htb]
 \centerline{\includegraphics[width=0.75\textwidth]{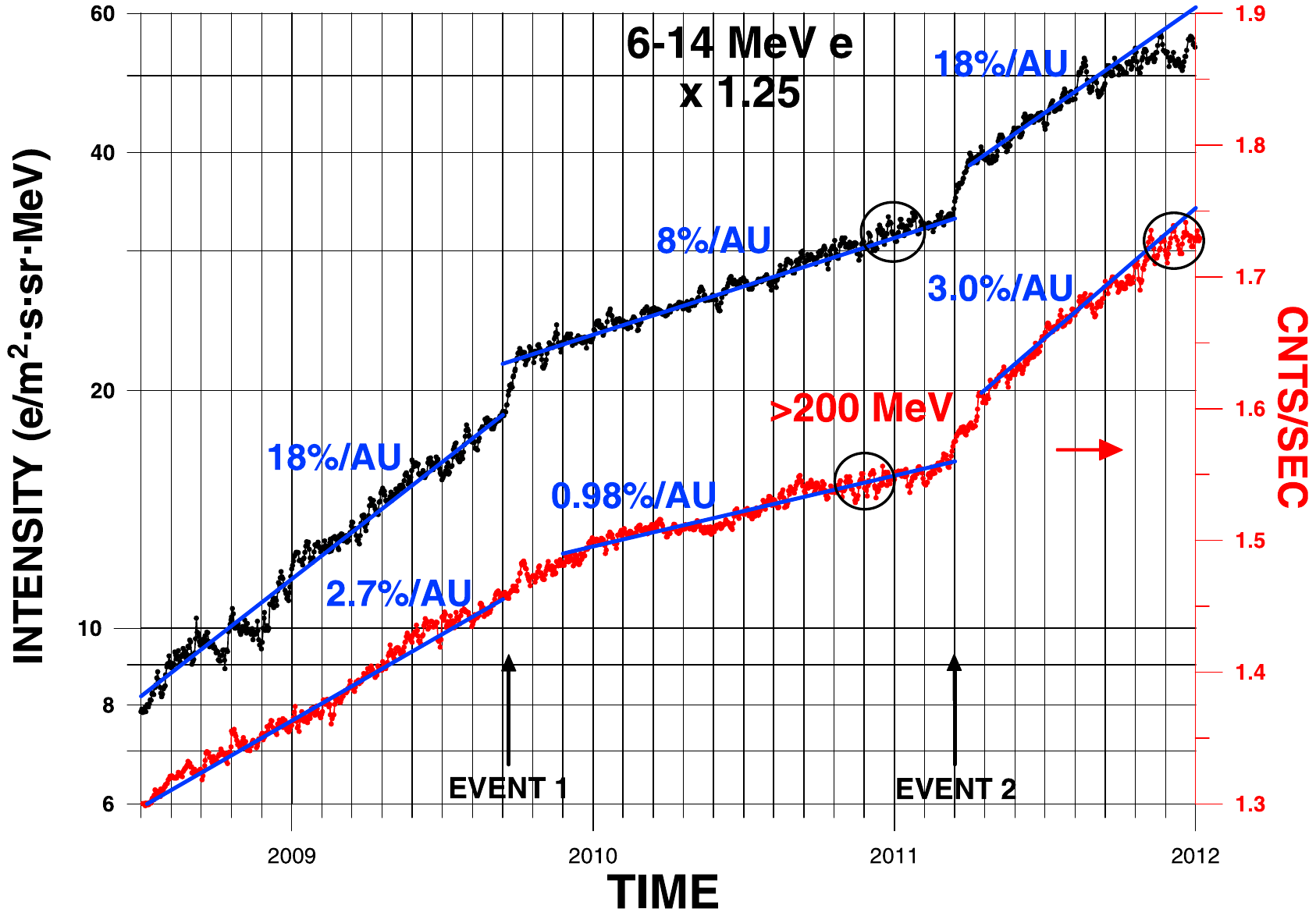}}
  \caption{The 5-day running average of the 6\,--\,14~MeV electrons
    and $E > 200$~MeV protons intensities measured at Voyager~1 from
    2008.5 to 2012. Note the first sudden intensity increase of
    electrons at 2009.70 (111.2~AU) and the change in radial gradients
    of both electrons and protons before and after this
    increase. Similar behaviour followed n the second sudden intensity
    increases of both electrons and protons at 2011.2 (116~AU). The
    continuing increase of electron and proton intensities after
    2011.3 is a notable feature. Image reproduced by permission from
    \citet{Webber-etal2012}, copyright by AGU.}
  \label{fig28}
\end{figure}}

\epubtkImage{Figure29.png}{%
\begin{figure}[htb]
 \centerline{\includegraphics[width=0.75\textwidth]{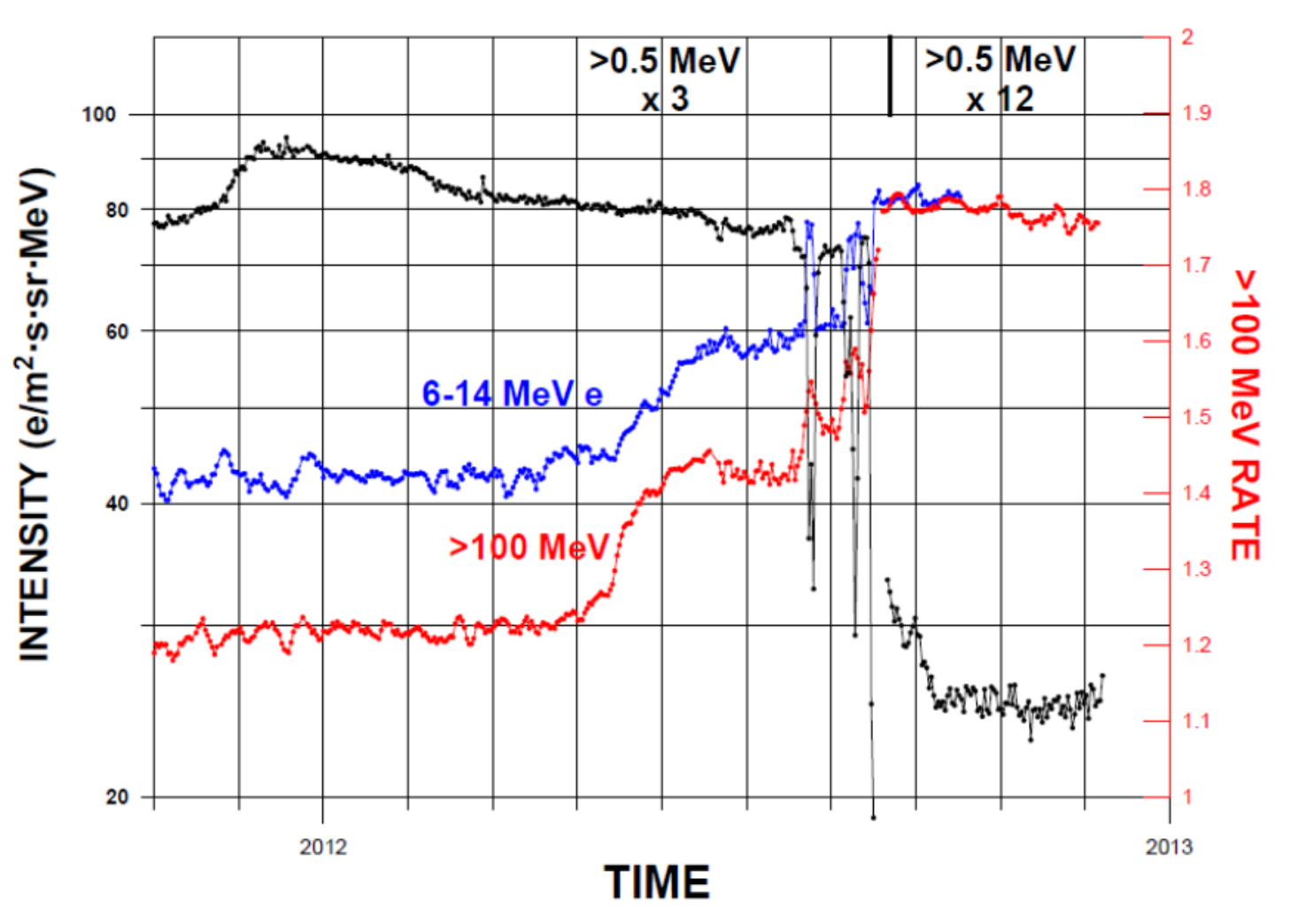}}
  \caption{Extraordinary changes were found in the 5 day running
    average intensities of 0.5~MeV protons (mainly ACR), 6\,--\,14~MeV
    galactic electrons and $E > 100$~MeV protons from 2009.0 to the end
    of the data set as observed by Voyager~1. Note how the 0.5~MeV
    proton intensity drops while the $E > 100$~MeV protons count rate
    went up. Image reproduced from the preprint of \citet{WebberMcDonald2013}.} 
  \label{fig29}
\end{figure}}

\clearpage 
\section{Constraints, Challenges, and Future Endeavours}

The modeling of the modulation of galactic CRs in the heliosphere
requires essential information, as was mentioned above, that needs to
be improved in order to make progress. In this context, a few
constraints and challenges are as follows: 

The LIS for the different galactic CR species are needed as initial
conditions at the `heliospheric boundary', presumably the HP but
probably even further out. Computed galactic CR spectra could be
different from the LIS. Surprisingly, little is known about most of
these galactic spectra at energies below a few GeV because of solar
modulation so that the uncertainties, mostly attributed to speculation
instead of solid evidence, can be rather large as illustrated by
\citet{WebberHigbie2008, WebberHigbie2009}. Adiabatic energy losses in
the heliosphere cause the galactic CR spectral shapes for ions and
anti-protons below $\sim$~10~GeV to be properly disguised,
progressively with decreasing energy and closer to the Sun. In
contrast, the intensity of~MeV-electrons is much less modified by
adiabatic energy losses and also by gradient and curvature
drifts. Unfortunately, from an electron LIS point of view, Jupiter is
a dominant source of up to $\sim$~30~MeV electrons in the inner
heliosphere, also completely concealing the electron LIS in the inner
heliosphere at these energies. This is not the case for positrons. The
spectral shape at Earth for the low energy part of the positron
spectrum is likely to be the same as for the positron LIS. There also
are indications from the PAMELA observations that the LIS for
positrons, and to some extent also electrons, may contain
contributions from local galactic sources so that the LIS may not be
as isotropic as assumed. With the two Voyager spacecraft at or close
to the HP, another milestone in the solar modulation of CRs is
foreseeable when the LIS gets observed. 

The global features, structure, and geometry of the heliosphere as
discussed above are important for modulation studies so that several
issues arise, e.g., what is the difference in the distance to the
`modulation boundary' in the polar and tail regions? How asymmetrical
is the heliosphere in the azimuthal and meridional planes and how does
it affect CR modulation? See, e.g., \citet{LangnerPotgieter2005} and
\citet{NgobeniPotgieter2011}. How much does the `modulation volume'
vary from solar minimum to maximum activity and how much is the inner
heliosheath contributing to this aspect? How much is the TS position
oscillating (moving inward and outward) with changing solar activity?
Furthermore, what is the role that the vastly extended heliospheric
tail region plays? It could also be that the alignment of the HMF and
the local interstellar magnetic field at the HP, together with the
HCS, create regions of ideal entrance for galactic CRs introducing in
the process an asymmetry in galactic CR modulation. 

The wavy HCS is a successful and well-established physical entity in
global modeling of solar modulation. The tilt angle of the HCS has
become the most useful indicator of solar activity from a
drift-modulation point of view and is widely used in data
interpretation and especially in modeling. However, the calculation of
this tilt angle is model dependent and it is not clear how the
waviness is preserved when propagating outwards away from the Sun,
especially with increasing solar activity \citep{Jiang-etal2010}. The
dynamic of the HCS has to be studied in more detail, with the first
insightful contributions being made
\citep[e.g.,][]{Borovikov-etal2011}. It is also unclear whether
particle drifts play a significant role in the heliosheath and to what
extent drift patterns are different than in the inner heliosphere
\citep{Webber-etal2008}.

Improved knowledge about the global solar wind profile and HMF remains
crucial for effective modeling. Before the Ulysses mission little was
explicitly known about the latitudinal dependence of these entities in
the inner heliosphere. Today, the solar wind profile can be specified
in detail in models, while for the HMF came the realization that it
may not be approximated in the heliospheric polar regions by the HMF
that Parker envisaged in the 1960s. This aspect is contributed to
interesting development in CR modulation. Nowadays, in numerical
models significant modifications, mostly phenomenological, to the
Parker field are applied in the polar regions, where a possible
replacement is in the form of the \citet{Fisk1996} solar magnetic
field. Unfortunately, the Fisk-type fields are too complex to handle
straight forwardly in standard numerical codes, but progress is being
made \citep[e.g.,][]{Burger-etal2008, Sternal-etal2011}. At present, a
conclusion for numerical modeling may be that Parkerian type HMFs are
mostly too simple (although the interpretation of HMF observations
keep pointing to such underlying simplicity) while Fisk-type fields
are too complex to handle numerically (and seem absent in the way that
HMF observations are interpreted). The question about the HMF geometry
is how significant are the modulation effects of the Fisk-type HMF on
CR modulation. To answer this, more study is needed to gain a better
understand of the finer details of these complex magnetic fields and
the relation to long-term CR modulation, especially how this field
changes with solar activity. An aspect that should be kept in mind is
that the study of more realistic magnetic fields requires an ever more
complex diffusion tensor and description of drifts.

Progress has been made in understanding observations of small
latitudinal gradients from Ulysses. In order to do so,
\citet{Burger-etal2000} argued that the rigidity dependence of the two
components of the diffusion tensor, which are perpendicular to the mean
HMF, should be decoupled. This approach was supported by the
investigation of~MeV electrons \citep{Ferreira-etal2001}. Although the
concept of increased polar perpendicular diffusion is well
established, no conclusive theoretical work has been published to
explain this dependence. Such an investigation may also be crucial in
deciding to what extent the `standard' Parker HMF has to be modified. 

Three-dimensional models to describe the propagation and modulation of
Jovian electrons in the inner heliosphere were applied by
\citet{Fichtner-etal2000}, \citet{Ferreira-etal2001}, and
\citet{Moeketsi-etal2005}. They studied the radial and latitudinal
transport of these particles in detail and estimated upper and lower
limits for the ratio of the parallel and perpendicular diffusion
coefficients, which in numerical model is a crucial modulation
parameter. This needs to be investigated further. They also
disentangled the galactic and Jovian contributions to these electron
observations, a process that improved understanding and the
interpretation of Ulysses data \citep[see
 also][]{Strauss-etal2012d}. They found that Jovian electrons
dominate the inner equatorial regions up to $\sim$~10\,--\,20~AU but
it is unlikely that they can dominate the low-energy galactic
electrons to heliolatitudes higher than $\sim$~30\textdegree\ off the
equatorial plane. This is determined however by how large polar
perpendicular diffusion is made and needs further study. Recently,
\citet{PotgieterNndanganeni2013a} offered computations of what can be
considered lower and upper limits for galactic electrons at Earth at
energies below a 100~MeV. 

The acceleration and propagation of CRs and the ACRs at the TS and
beyond are presently highly controversial and need further
studies. The standing paradigm of the ACRs being accelerated at the TS
was severely questioned when the Voyager spacecraft crossed the TS and
observed unexpected results. The spectrum of ACRs did not unfold to an
anticipated power-law, and the ACR fluxes simply continued to increase
as the two Voyagers moved away from the TS. Magnetic reconnection
\citep{Drake-etal2010, LazarianOpher2009} and different forms of
stochastic acceleration \citep{FiskGloeckler2009, Strauss-etal2010a}
have been invoked as causing this phenomena, or it could simply be due
to the so-called bluntness of the TS \citep[e.g.,][]{McComasSchwadron2006,
KotaJokipii2008}. The answer to this unresolved question and the
full grasp of where and how ACRs are accelerated should have important
impacts on the transport and reacceleration of CRs in the heliosphere
and elsewhere in the interstellar medium. Other interesting aspects of
CRs beyond the TS were discussed by \citet{Potgieter2008}.

In the field of the long-term (11-years and longer) CR modulation in the 
heliosphere, several issues need further investigation and research:
(1)~Despite the apparent success of the compound numerical model
described above the amount of merging taking place beyond 20~AU needs
to be studied with MHD models, especially the relation between
interplanetary CMEs and GMIRs, and how these large barriers may modify
the TS, the inner heliosheath and perhaps also the HP. (2)~The full
rigidity dependence of the compound model is as yet not well described
because the underlying effects of turbulence on the diffusion
coefficients is poorly observed and understood. (3)~A major issue with
time-dependent modeling, apart from global dynamic features such as
the wavy HCS, is what to use for the time dependence of all the
diffusion coefficients in Equation~(\ref{eq:parker_tpe}), on top of
the already complex issue of what their steady-state energy (rigidity)
and spatial dependence are in the inner heliosphere
\citep[e.g.,][]{Manuel-etal2011a, Manuel-etal2011b}. It has now also
become vital to understand the diffusion tensor beyond the
TS. Equation~(\ref{eq:parker_tpe}) is probably more complex for this
region and needs to be extended to include processes (perhaps even
completely new ones) that are otherwise considered to be
negligible. (4)~Fundamentally, from first principles, it is not yet
well understood how gradient and curvature drifts reduce with solar
activity. This aspect needs now also to be investigated for the region
beyond the TS. For example, what happens to the wavy HCS in the
heliosheath and how will the strong latitudinal and azimuthal
components of the solar wind velocity and the associated HMF influence
particle drifts and CR modulation in the heliosheath? The question
arises if the modulation in this region is really fundamentally
different from the rest of the heliosphere? (5)~Another important
question is how the heliospheric modulation volume varies with time on
the scale of thousands to millions of years? The Sun encounters
different interstellar environments during its passage through the
galaxy, and hence the outer structure and size of heliosphere should
change over such periods.

Other interesting aspects that will hopefully be addressed in future are: 
(1)~The solar modulation of many CR species and their isotopes has not been 
properly modeled. The first obstacle, and therefore prime objective, will be 
to establish LIS for these isotopes. (2)~The anisotropy of CRs at various 
energies as caused inside the heliosphere, also related to what a very 
extensive heliotail may contribute, is another specialized topic. In this 
context, see the review by \citet{Kota2012}. (3)~At what rigidity is the 
modulation of CRs beginning? This is important for the interpretation of 
observations of CRs, also for anisotropy studies. In modulation modeling it 
is simply assumed to begin at $\sim$~30~GeV. (4)~The extended solar cycle 
23/24 minimum provides an opportunity to investigate conditions that were 
different than before, perhaps similar to those in the early 1900s. This 
may reveal a 100\,--\,120~year cycle in CR modulation. (5)~Over the
last decade, there has been noteworthy progress in obtaining,
analyzing and interpreting Be\super{10} and other isotopes produced by CRs in
the atmosphere, which reveals a history of CRs going back to millions
of years \citep[e.g.,][]{Wieler-etal2011}. This certainly contributes
to the growing interest in the topic of space climate \citep[see][and
  reference therein]{Scherer-etal2006}.

\citet{Mewaldt2012} noted from an experimental point of view that there are 
presently a dozen spacecraft measuring galactic CRs and ACRs over energies 
from below 1~MeV to $\sim$~1~TeV and in situ from 1 to 122~AU (see his
Figure~1 for current missions). These space missions are supplemented
by ground-based detectors such as NMs and by balloon-flight
instruments. He concluded that the challenge during the next decade
will be to maintain enough of these missions and facilities to
support the Voyagers' approach and passage through the HP into
interstellar space, and to record in detail the evolving story of how
CRs in the heliosphere respond to the Sun's new direction in solar
activity.

\newpage 
\section{Summary}

The heliosphere shields the solar system and all living creatures from 
galactic CRs. Since the Sun is a variable star it produces significant 
modulation of these charged particles in a variety of cycles. Cosmic rays 
are excellent indicators of the various solar cycle variations and studying 
them enlighten us about the characteristics of the Sun's electrodynamic 
influence sphere. Such studies also inform us about the structural features 
and geometry of the heliosphere, including the solar wind termination
shock, the heliosheath and the heliopause because they all influence
the flux of CRs in the heliosphere, up to Earth.

Tremendous progress has been made since the beginning of the space age with 
the deployment of several space missions for detecting CRs over a wide range 
of energies, which stimulated theoretical and modeling research. Ten years 
ago, the size and geometry of the heliosphere and the importance of shock 
acceleration were all major unknown issues. Now that the Voyager spacecraft 
had crossed the TS and Voyager~1 is close to the heliopause, some of
these entities and mechanisms are becoming reasonably known, at least,
they cannot be considered `free' parameters any longer.

The present understanding of the mechanisms of the global solar modulation 
of galactic CRs in the heliosphere is considered essentially correct, an 
amazing accomplishment for Parker's theory that was developed in the early 
1960s. The main obstacles and challenges are insufficient knowledge of the 
spatial, rigidity and especially the temporal dependence of the diffusion 
coefficients, covering the underlying features of solar wind and magnetic 
field's turbulence. 

Evidently, this field of research is alive-and-well with many aspects 
remaining a work in progress.

\section{Acknowledgements}

The author acknowledges the partial financial support of the South African 
National Research Foundation (NRF), the SA High Performance Computing 
Centre (CHPC), and the SA Space Agency's (SANSA) Space Science Division. He 
also acknowledges many discussions with his MSc and PhD students, colleagues, 
and collaborators in the experimental, modeling, and theoretical fields of 
research.

\newpage



\bibliography{article}

\end{document}